\documentclass[twocolumn]{aastex63}
\usepackage{gensymb}
\usepackage{amsmath}
\usepackage{lipsum}
\usepackage[figuresright]{rotating}
\usepackage{float}

\begin{document}

\title{A Bayesian Approach to Inferring Accretion Signatures in Young Stellar Objects: A Case Study with VIRUS\footnote{Based on observations obtained with the Hobby-Eberly Telescope (HET), which is a joint project of the University of Texas at Austin, the Pennsylvania State University, Ludwig-Maximilians-Universitaet Muenchen, and Georg-August Universitaet Goettingen. The HET is named in honor of its principal benefactors, William P. Hobby and Robert E. Eberly.}}

\author[0009-0001-9962-1565]{Lauren Halstead Willett}
\affiliation{Dipartimento di Fisica e Astronomia, Università degli Studi di Padova, Via Marzolo 8, 35131, Padova, Italy}
\affiliation{Department of Astronomy \& Astrophysics, 525 Davey Laboratory, The Pennsylvania State University, University Park, PA 16802, USA}

\author[0000-0001-8720-5612]{Joe P. Ninan}
\affiliation{Department of Astronomy and Astrophysics, Tata Institute of Fundamental Research, Homi Bhabha Road, Colaba, Mumbai 400005, India}
\affiliation{Department of Astronomy \& Astrophysics, 525 Davey Laboratory, The Pennsylvania State University, University Park, PA 16802, USA}

\author[0000-0001-9596-7983]{Suvrath Mahadevan}
\affiliation{Department of Astronomy \& Astrophysics, 525 Davey Laboratory, The Pennsylvania State University, University Park, PA 16802, USA}
\affiliation{Center for Exoplanets \& Habitable Worlds, 525 Davey Laboratory, The Pennsylvania State University, University Park, PA 16802, USA}

\author[0000-0003-2307-0629]{Gregory R. Zeimann}
\author[0000-0001-9165-8905]{Steven Janowiecki}
\affiliation{Hobby-Eberly Telescope, McDonald Observatory, University of Texas at Austin, Austin, TX 78712, USA}

\author[0000-0001-6717-7685]{Gary J. Hill}
\affiliation{McDonald Observatory, The University of Texas at Austin, 2515 Speedway Boulevard, Austin, TX 78712, USA}
\affiliation{Department of Astronomy, The University of Texas at Austin, 2515 Speedway Boulevard, Austin, TX 78712, USA}

\begin{abstract}
The mass accretion rates of young stellar objects (YSOs) are key to understanding how stars form, how their circumstellar disks evolve, and even how planets form. We develop a Bayesian framework to determine the accretion rates of a sample of 15 YSOs using archival data from the VIRUS spectrograph ($R \sim 800$, 3500-5500\r{A}) on the Hobby-Eberly Telescope. 
We are publicly releasing our developed tool, dubbed \texttt{nuts-for-ysos}, as a Python package which can also be applied to other spectroscopic datasets\footnote{\url{https://github.com/laurenwillett/nuts-for-ysos}}. The \texttt{nuts-for-ysos} code fits a simple accretion model to the near-UV and optical continuum of each VIRUS spectrum. Our Bayesian approach aims to identify correlations between model parameters using the No U-Turn Sampler (NUTS). Moreover, this approach self-consistently incorporates all parameter uncertainties, allowing for a thorough estimation of the probability distribution for accretion rate not accomplished in previous works. Using \texttt{nuts-for-ysos}, we derive accretion rates of each YSO. We then verify the reliability of our method by comparing to results separately derived from only the spectral emission lines, and to results from earlier studies of the Lupus, Chamaeleon I, and NGC1333 regions. Finally, we discuss what qualitative trends, covariances, and degeneracies were found among model parameters. The technique developed in this paper is a useful improvement that can be applied in the future to larger samples of YSOs observed by VIRUS or other spectrographs.
\end{abstract}


\section{Introduction}
Understanding how quickly a young star accretes mass is important for constraining models of circumstellar disk evolution and planet formation. Over the first few million years of their life, low mass stars ($\lesssim 2M_{\odot}$) possess a circumstellar disk that is dissipated mainly through photoevaporation, stellar winds, and accretion onto the star \citep{2014prpl.conf..475A}. One commonly accepted paradigm is that young stellar objects (YSOs) in the class II phase accrete matter via magnetospheric accretion \citep{1985PASJ...37..515U, 1991ApJ...370L..39K, 1994ApJ...429..781S, 2016ARA&A..54..135H}. In this model, the stellar magnetic field couples to material from the inner circumstellar disk, and then guides gas along magnetic field lines onto the stellar surface. The gravitational energy released from this free-falling gas consequently heats up the gas and causes broad emission lines. The infalling material produces a ‘shock’ when it encounters the stellar surface, heating up patches of the surface to about $10^{4}$ K and creating UV continuum emission \citep{2000ApJ...544..927G}. Class III YSOs can still exhibit line emission like the younger accreting class II stars, but often to a lesser degree, and the emission is attributed mainly to the chromosphere rather than to accretion processes \citep{2013AandA...551A.107M}.

Accretion directly ‘uses up’ some of the circumstellar disk material. Moreover, the energy released from the accretion drives dissipation of material through outflows and photoevaporation of the disc. Finally, accretion influences the temperature and chemical composition of the inner disk. Through these channels, accretion therefore augments what material in the protoplanetary disk is available for planet formation. Deriving the mass accretion rate ($M_{acc}$) of YSOs is therefore key in understanding both low-mass star formation and consequentially how planets may form around the star \citep{2019A&A...631L...2M}. In particular, the $M_{acc}-M_{\star}$ empirical relation from class II YSOs has been heavily studied in order to probe changes in mass accretion over time (e.g. \citet{2006MNRAS.370L..10C}) and the related evolution in disks (e.g. \citet{2014MNRAS.439..256E, 2018ppd..confE..23M}, and references within). 

$L_{acc}$ is the total accretion luminosity: the luminosity from the YSO which can be attributed purely to the accretion process. $L_{acc}$ can be treated as the release of gravitational potential energy during accretion, and therefore be converted into a mass accretion rate, $M_{acc}$ (see equation \ref{eq:mass_accretion} in Section \ref{sec:estimating_mass_accretion}). Accreting YSO spectra present continuum emission in excess of a main sequence stellar photosphere, most notably through a ‘jump’ in emission in the Balmer continuum (a 'Balmer jump') for wavelengths shortwards of $\sim$3600\r{A}. The strength of this excess continuum emission can be used to estimate $L_{acc}$. Although not the only valuable wavelength range (for example, see the study of near-IR lines in \citet{2021A&A...650A..43F, 2023ApJ...944..135F}), the optical and near-UV wavelength range of a YSO spectrum therefore contains particularly useful information for estimating the total accretion luminosity $L_{acc}$. The typically high extinction of the YSO can make this a challenging regime for targeted observations however, compared to the near-IR.

The ‘direct’ method for determining $L_{acc}$ involves fitting a model of the excess continuum emission to the YSO spectrum over a wide wavelength range. This direct method has been applied to numerous YSOs (e.g. \citet{1993AJ....106.2024V, 1998ApJ...492..323G, 2008ApJ...681..594H, 2012A&A...548A..56R, 2016A&A...585A.136M, 2014A&A...561A...2A, 2017A&A...600A..20A}). However, the techniques from previous literature have not been able to thoroughly examine degeneracy among model parameters. For example, the procedure described in \citet{2013A&A...558A.114M} uses a discrete grid of parameter values and then finds the best fit by minimizing a likelihood function similar to a $\chi^2$ distribution (see Equation \ref{eq:chi_square_like}). While this approach self-consistently determines $L_{acc}$, stellar spectral type, and extinction $A_V$, it does not allow one to examine in detail the covariance of the parameters. Moreover, the uncertainty in $L_{acc}$ has generally only been roughly estimated. For example in \citet{2017A&A...600A..20A}, they state that the approximate uncertainty in $L_{acc}$ is about $\sim$ 0.25 dex for log($L_{acc}$/$L_\odot$), and they qualitatively describe the contributors to this uncertainty. However, they do not include a precise quantitative accounting of the uncertainty in $L_{acc}$ for each individual YSO. 
We aim to address these deficiencies using the same model as in \citet{2013A&A...558A.114M}, but in a Bayesian framework, to our collection of YSO spectra in Section \ref{sec:continuum_fit}. While we study a relatively small sample of YSOs, our work acts as a proof of concept for how the analysis of YSO spectra can be improved, and a demonstration of the new \texttt{nuts-for-ysos} tool. 

The accretion luminosity $L_{acc}$ can also be indirectly estimated by measuring the luminosities of a variety of emission lines ($L_{line}$) and then applying known empirical relationships between $L_{line}$ and $L_{acc}$. For example, the strength of He I lines, Ca II lines, and both visible and near-infrared hydrogen recombination lines have been shown to exhibit correlation with the accretion luminosities of YSOs derived from the direct method (e.g. \citet{1998ApJ...492..743M, 2008ApJ...681..594H, 2012A&A...548A..56R}) Therefore determining accretion luminosities from such emission lines is a reliable alternative, having yielded results that are consistent with the direct method, albeit with a higher scatter \citep{2008ApJ...681..594H, 2012A&A...548A..56R, 2014A&A...561A...2A}. Additional proxies besides emission line fluxes include excess U-band emission \citep{1998ApJ...492..323G}, and the H-alpha $10\%$ line width \citep{2004A&A...424..603N}. These various ‘indirect’ methods are useful for YSOs in which directly fitting a model of excess continuum emission near the Balmer jump is too difficult because the spectrum is low-signal-to-noise (due to high extinction in star forming regions, or low throughput in transmission optics of the spectrograph, for example) or because the spectrum is only available over a narrow wavelength range.

Using either the direct method or indirect method, accretion rates have been acquired for YSOs in various star forming regions including Taurus \citep{1998ApJ...492..323G, 2005ApJ...625..906M}, TWA \citep{2019A&A...632A..46V}, $\rho$ Oph \citep{2006A&A...452..245N}, Chamaeleon I \citep{2005ApJ...625..906M, 2016A&A...585A.136M, 2017A&A...604A.127M}, $\eta$-Chamaeleon \citep{2018A&A...609A..70R}, Lupus \citep{2014A&A...561A...2A, 2017A&A...600A..20A}, Upper Scorpius \citep{2020A&A...639A..58M}, and NGC1333 \citep{2021A&A...650A..43F}, among others. $M_{*}$ and $M_{acc}$ for these stars have shown to exhibit a loosely positive correlation, of roughly $M_{acc}$ $\propto M_{*}^{2}$. Some stars of the same mass exhibit a discrepancy in $M_{acc}$ of more than 3 dex (\citet{1998ApJ...492..323G, 2005ApJ...625..906M, 2006A&A...452..245N, 2012A&A...547A.104B, 2014A&A...572A..62A, 2017A&A...600A..20A}, and references therein). There is currently no single explanation for why there is a relationship between stellar mass and accretion rate, or why there is a large scatter in this relationship, although several possibilities have been put forward (e.g. \citet{1998ApJ...495..385H,2006A&A...452..245N,2006ApJ...645L..69D,2008ApJ...676L.139V,2014MNRAS.439..256E}). Gathering statistics on a larger population of YSOs in the future may help to parse out the reasons for the scatter in the $M_{*}-M_{acc}$ relation \citep{2023MNRAS.524.3948A}.

Despite the value in examining the Balmer jump of YSO spectra, there has nonetheless been a lack of large spectroscopic surveys dedicated towards studying YSOs in the UV range. This is partly due to instrumental limitations, and also because the extinction within star-forming regions often prohibits a high enough UV signal. The entire UV spectrum has only been accessible from space telescopes, such as through the ULYSSES survey with HST \citep{2020RNAAS...4..205R, 2022AJ....163..114E, 2022AJ....164..201P}. However, very large ground-based instruments have recently provided access to wavelengths as small as $\sim$3000\r{A} such as X-Shooter \citep{2011A&A...536A.105V}. We utilize the Hobby-Eberly Telescope (HET), a 10-meter aperture telescope located at the McDonald Observatory in the Davis Mountains in Texas \citep{1998SPIE.3352...34R, 2021AJ....162..298H}. HET is one of the largest optical telescopes in the world, and it feeds the three spectrographs housed by the observatory: Visible Integral-field Replicable Unit Spectrograph (VIRUS, \citet{2021AJ....162..298H}), the second generation Low Resolution Spectrograph (LRS2, \citet{2016SPIE.9908E..4CC}), and the Habitable-zone Planet Finder (HPF, \citet{2012SPIE.8446E..1SM, 2014SPIE.9147E..1GM}). Spectra of a 3500 - 5500\r{A} wavelength range were obtained using VIRUS, a low resolving power $(R \sim 800)$ integral field spectrograph with a 18' diameter field of view (FOV). In this paper we present the results of our analysis of VIRUS observations that were taken in parallel with the other spectrograph observations at the HET. This parallel data consists of VIRUS exposures that take place whenever another instrument aboard the HET is exposing for more than 5 minutes. 

The \texttt{nuts-for-ysos} code automates the model-fitting process such that in the future, it could be applied to numerous stars with little human input. Especially since VIRUS is a multi-object spectrograph, future targeted observations of star-forming regions by VIRUS could provide a large collection of YSO spectra to be studied using our new approach. Moreover, the wavelength region and spectral features inputted into \texttt{nuts-for-ysos} can be customized to observations from other spectrographs. A similar analysis framework can be even used for spectra in the near-infrared range, for example with future large area near-IR surveys done by the upcoming Nancy Grace Roman Space Telescope \citep{2015arXiv150303757S, 2019arXiv190205569A}. Our study therefore offers an exciting prototype of what can be achieved with both VIRUS observations or with other YSO spectroscopic surveys in the future.

We describe the VIRUS instrument and the observations in Section \ref{sec:observations}, and the selection criteria for class II and class III YSOs in Section \ref{sec:sample selection}. In Section \ref{sec:continuum_fit}, we then describe the procedure in \texttt{nuts-for-ysos} for directly fitting an accretion model to our data, and how our procedure allows us to simultaneously derive spectral type, luminosity, and $A_{V}$ for each target. We then interpret results from the fitting procedure, calculating $L_{*}$, $L_{acc}$, $M_{*}$, and $M_{acc}$ in Section \ref{sec:bayesian_results}. In Section \ref{sec:emission_lines} we measure various emission lines for these YSO spectra, and apply the empirical $L_{line}-L_{acc}$ relationship from \citet{2017A&A...600A..20A} to the targets, so that we can separately determine $L_{acc,line}$ and compare it to $L_{acc}$. In Section \ref{sec:discussion} we then discuss covariances found within the results, and compare to results from Lupus, Chamaeleon I, and NGC1333. Finally, in Section \ref{sec:conclusion} we summarize our main points and reiterate the usefulness of \texttt{nuts-for-ysos} for future possible applications. Appendix \ref{sec:data_reduction} contains the details about the data reduction process and a table of observations, and Appendix \ref{sec:result_plots} contains plots of the accretion model fit to each YSO in our current sample, and corresponding corner plots for the model parameters.

\section{Observations}
\label{sec:observations}
At its full capacity, VIRUS has 78 identical fiber-fed integral field units (IFUs) of 448 fibers, each 51'' $\times$ 51'', arrayed in a grid pattern over a field of 18' diameter. Each IFU is connected to two spectrograph channels each with CCDs read out through two amplifiers. Each of the $\sim$ 35,000 total fibers in VIRUS has a 1.5'' diameter, with 2.2'' between each fiber (1$/$3 fill-factor). The resulting spectra have R=670 at 3900\r{A}, R=850 at 4600\r{A}, and R=990 at 5200\r{A} \citep{2006SPIE.6273E..3WK, 2018SPIE10702E..1KH, 2021AJ....162..298H}. 
VIRUS is primarily built for making extragalactic observations for the Hobby-Eberly Telescope Dark Energy Experiment (HETDEX, \citet{2021ApJ...923..217G}). However, VIRUS continues to take data when LRS2 or HPF is making targeted observations for more than five minutes. These 'parallel' VIRUS pointings occasionally coincide with the galactic plane, in which case they may capture YSOs within the 18' VIRUS field of view. 
Parallel VIRUS observations are not dithered to fill in the fiber pattern, so the field has a fill factor of only $\approx 33\%$. Both the location and exposure time of each parallel observation are solely determined by the observations of whichever other spectrograph at the HET is being used for the primary science program at that time. The spectra extracted from these parallel observations between 2019 January 1 and 2023 March 31 have recently been released within the ongoing Hobby–Eberly Telescope VIRUS Parallel Survey (HETVIPS, \citet{2024ApJ...966...14Z}). The details of the data-reduction process are presented in Appendix \ref{sec:data_reduction}.

\section{Sample Selection}
\label{sec:sample selection}
VIRUS made 4,269 parallel observations between 2019-01-01 and 2021-05-30, with each of these observations typically containing hundreds to thousands of extracted spectra. We isolated potential YSO spectra from this large collection using the catalog presented in \citet{2019MNRAS.487.2522M}. This catalog adopts machine learning methods to assign a probability of a star being a YSO (as well as main sequence star, evolved star, or extragalactic object) based on Gaia DR2 \citep{2016A&A...595A...1G, 2018A&A...616A...1G, Gaia_DR2} and on AllWISE photometry \citep{2014yCat.2328....0C,irsa1}. 
\citet{2019MNRAS.487.2522M} only examines regions where the two-dimensional, 353 GHz R2.01 Planck dust opacity map \citep{2016A&A...594A..10P} yields a value of at least $\tau=1.3\times10^{-5}$, to purposely focus on dusty regions more likely to contain YSOs; \citet{2019MNRAS.487.2522M} found that 99\% YSOs known from literature occupy such regions with $\tau > 1.3\times10^{-5}$. They also do not consider any objects with multiple Gaia IDs or missing photometric bands within Gaia DR2 or AllWISE.
In building our sample, we first crossmatch VIRUS-observed objects within 2 arcseconds of \citet{2019MNRAS.487.2522M} catalog entries having an assigned YSO probability LY $> 70\%$. We then retain only the VIRUS spectra with an average signal-to-noise ratio (SNR) $>5$ over the entire spectrum, and an average SNR of at least 1.5 between 3500 and 4000\r{A}. This is because the signal covering the Balmer jump is especially important for the continuum fitting we perform in Section \ref{sec:continuum_fit}.
We then crossmatch this set of observations with the catalog presented in \citet{2021AJ....161..147B} within 2 arcseconds, to determine the distance to each object. Within this catalog we use the 'geometric distances' derived from GAIA EDR3 parallaxes \citep{2021A&A...649A...1G, Gaia_EDR3}.

This resulting sample has VIRUS spectra with sufficiently high SNR, known distances, and photometry available in the AllWISE catalog. However, it is important to note that stars on the Galactic plane can often possess spurious WISE photometry. The WISE mission was mainly designed for studying near-Earth asteroids, infrared galaxies, and brown dwarfs. The process for identifying sources from WISE images (described in their online explanatory supplement \footnote{\url{http://wise2.ipac.caltech.edu/docs/release/allwise/expsup/}} and in \citet{2012PASA...29..269M}) is not well-suited for regions containing dust or clouds bright in the mid-infrared. As a result, stars can often have W1 and W2 band photometry associated with a legitimate point source, but spurious W3 and W4 band photometry. In identifying YSOs by their infrared excess, WISE photometry must be used with great caution.

\citet{2014ApJ...791..131K} investigates this issue with WISE photometry, and finds that real detections in AllWISE can be well-isolated by making cuts on four specific parameters within each of the W1 - W4 bands:\\ \\
1. $wBsnr$: The SNR in the band B $\in$ [1,2,3,4]\\
2. $wBrchi2$: the reduced chi2 of the profile fit\\
3. $wBm$: the number of exposures over which a profile-fit flux measurement for the source could be performed\\
4. $wBnm$: the number of profile-fit flux measurements for which the source was detected with $wBsnr$ $>$ 3\\

\citet{2019MNRAS.487.2522M} addresses the problem of spurious AllWISE photometry by adopting the same general strategy of examining these AllWISE catalog parameters. However, instead of making hard cuts they take a probabilistic approach. They make a training sample of 500 real and 500 spurious sources from visually inspecting WISE W3 and W4 images. They then use the AllWISE parameters from this training sample and apply the Random Forest method to assign to other sources a probability of being real, R.

We find that although this approach generally works, many objects have a probability R which hovers around 0.5, such that requiring R $\geq$ 0.5 still results in apparent false positives.
We decide to impose additional requirements on AllWISE photometry which we find ultimately isolates true YSOs more effectively. Following Section 3.1.1 in \citet{2014ApJ...791..131K}, we require that for WISE bands W1-W4, the photometric uncertainty $wBsigmpro$ be non-null, and require that the signal-to-noise $wBsnr$ and reduced chi-squared $wBrchi2$ meet the following conditions:
\begin{equation}
    \begin{cases}
        w1rchi2 < (w1snr-3)/7\\
        w2rchi2 < 0.1 \times w2snr-0.3\\
        w3rchi2 < 0.125 \times w3snr-1\\
        w4rchi2 < 0.2 \times w4snr-2\\
    \end{cases}
\label{eq:wise_cuts}
\end{equation}
Though this set of equations is very effective in suppressing contamination from fake AllWISE detections, \citet{2014ApJ...791..131K} notes that it also eliminates around two-thirds of real detections in bands W3 and W4. \citet{2014ApJ...791..131K} thus changed their criteria to raise the retrieval rate in the W3 and W4 bands, at the cost of allowing more fake detections. We decide that instead of adopting this changed criteria, we select sources that either:\\ \\
1. Have AllWISE photometry with non-null photometric uncertainty, R $\geq$ 0.5 from \citet{2019MNRAS.487.2522M}, and meet the criteria presented in Equation \ref{eq:wise_cuts}.\\
2. Have existing Spitzer photometry (IRAC 1-4 and MIPS1 bands with non-null photometric uncertainty) that can be used as an alternative.\\

We have thus only used AllWISE photometry when catalog parameters meet a number of strict requirements, and opted to search for Spitzer photometry in cases where the requirements are not met. The Spitzer photometry is acquired using the VizieR Photometry viewer tool \footnote{\url{http://vizier.unistra.fr/vizier/sed/}} with a Python interface \footnote{\url{https://gist.github.com/mfouesneau/\\6caaae8651a926516a0ada4f85742c95}}, using a search radius of 2''.

We ultimately select 16 YSOs for our current sample. Nine YSOs have useable AllWISE photometry determined with our criteria, and fifteen of them have published Spitzer photometry. We use photometry from the c2d (Cores to Planet-forming Disks) Spitzer Legacy Program when available \citep{2003PASP..115..965E, C2D_YSOs}, and use alternatives otherwise, listed in Table \ref{tab:photometry}.
To classify the YSOs in our sample, we use the 4-class system introduced by \citet{1994ApJ...434..614G}, which classifies via the spectral index $\alpha$; essentially the slope of the SED between $\sim$2 and 20 micrometers:\\
\begin{equation}
 \alpha=\frac{d \log \lambda F_{\lambda} }{d \log \lambda}
 \label{eq:alpha}
\end{equation}
To estimate $\alpha$, we fit a line by least squares fitting to the available AllWISE and/or Spitzer photometry. We then classify $\alpha$ with the cutoffs presented in \citet{1994ApJ...434..614G}:\\
\begin{equation}
 \begin{cases}
    \text{Class I}: 0.3 \leq \alpha\\
    \text{Flat}: -0.3 \leq \alpha < 0.3\\
    \text{Class II}: -1.6 \leq \alpha < -0.3\\
    \text{Class III}: \alpha < -1.6\\
 \end{cases}
\label{eq:classify}
\end{equation}
\\
Information related to the VIRUS observations of each star is presented in Table \ref{tab:obs_log} in Appendix \ref{sec:data_reduction}. Table \ref{tab:photometry} contains the YSO likelihood LY and the AllWISE validity likelihood R obtained from \citet{2019MNRAS.487.2522M}, the classifications for each object, and the photometry used in determining each classification. We find twelve of the stars are class II, and four are class III. Among them, Object 10 (EM* LkHA 351) is classified as class II but its spectral index $\alpha$ very nearly places it in the class III category. Object 3 (2MASS J20580138+4345201) is also considered class II, but is almost in the 'Flat' category. Though it has a class II SED, the spectrum of Object 16 (ATO J052.3580+31.4444) could not be fit by a model of an accreting YSO, nor did it exhibit the Balmer emission lines expected of an accreting YSO. These two issues are discussed in Section \ref{sec:continuum_fit} and Section \ref{sec:emission_lines} respectively. Ultimately this object was not included in the results (ie. in Tables \ref{tab:bayesianresults} and \ref{tab:M_Macc_info}). By visual inspection, none of the objects had SEDs which resembled transition disk YSOs.

The four class III objects are 6, 7, 12, and 13. Two of these objects, 12 and 13 (2MASS J03283651+3119289 and 2MASS J03292815+3116285), have Spitzer IRAC and MIPS1 photometry obtained from separate catalogs since their MIPS1 photometry could only be found exclusively in Table 2 of \citet{2017ApJ...836...34M}. For both these objects, the MIPS1 photometry is labeled as 'Not Individually Detected' by \citet{2017ApJ...836...34M}. Despite the dubious nature of their MIPS1 detections, we nonetheless include these two objects because their spectra confirm that they are young stars and because the MIPS1 photometry is not used for anything beyond categorizing the YSOs as class III.

\begin{deluxetable*}{llllllllll}
\tablecaption{Photometry and classification information for the total sample.\label{tab:photometry}}
\tablewidth{0pt}
\tablehead{
    \colhead{ID} & \colhead{SIMBAD Name} & \colhead{LY$^a$} & \colhead{R$^b$} & \colhead{AllWISE ID} & \colhead{Spitzer source} & \colhead{$\alpha$} & \colhead{Class} & \colhead{Region} & \colhead{Note}
}
\startdata
1 & 2MASS J21523325+4710505 & 0.9354 & 0.456 &  & 2 & -0.84 & II & IC 5146 & \\
2 & 2MASS J21533310+4716092 & 0.9798 & 0.422 &  & 2 & -1.09 & II & IC 5146 & \\
3 & 2MASS J20580138+4345201 & 0.9966 & 0.56 & J205801.37+434520.1 & 3 & -0.32 & II & NGC 7000 & \\
4 & EM* LkHA 188 & 0.9972 & 0.568 & J205823.80+435311.3 & 3 & -0.8 & II & NGC 7000 & \\
5 & 2MASS J18300610+0106170 & 0.9996 & 0.5 & J183006.10+010616.8 & 1 & -0.71 & II & Serpens & \\
6 & 2MASS J18295618+0110574 & 0.8058 & 0.448 &  & 1 & -2.55 & III & Serpens & \\
7 & V* V776 Ori & 0.8752 & 0.388 &  & 5 & -2.22 & III & Orion & \\
8 & CVSO 1897 & 0.9888 & 0.51 & J054015.13-005726.6 & & -0.87 & II & Orion & \\
9 & [HL2013] 052.17673+30.49810 & 0.9996 & 0.55 & J032842.43+302953.0 & 1 & -0.82 & II & Perseus & \\
10 & EM* LkHA 351 & 0.997 & 0.596 &  & 1 & -1.56 & II & NGC 1333 & \\
11 & 2MASS J03285101+3118184 & 0.9996 & 0.644 & J032851.03+311818.3 & 1 & -0.61 & II & NGC 1333 & \\
12 & 2MASS J03283651+3119289 & 0.9004 & 0.21 &  & 1, 4 & -2.58 & III & NGC 1333 & ND\\
13 & 2MASS J03292815+3116285 & 0.9346 & 0.264 &  & 1, 4 & -2.5 & III & NGC 1333 & ND\\
14 & 2MASS J03284782+3116552 & 0.9956 & 0.638 & J032847.83+311655.0 & 1 & -0.91 & II & NGC 1333 & \\
15 & 2MASS J03285216+3122453 & 1.0 & 0.506 & J032852.16+312245.1 & 1 & -1.04 & II & NGC 1333 & \\
16 & ATO J052.3580+31.4444 & 0.9972 & 0.55 & J032925.92+312640.0 & 1 & -0.69 & II & NGC 1333 & \\
\enddata
\tablecomments{ND: Object has MIPS1 photometry from Table 2 of \citet{2017ApJ...836...34M} with the flag 'Not Individually Detected'.\\
$^{(a)}$ LY: the YSO likelihood in \citet{2019MNRAS.487.2522M}.\\
$^{(b)}$ R: the AllWISE validity likelihood in \citet{2019MNRAS.487.2522M}.}
\tablerefs{(1) \cite{2003PASP..115..965E}; (2) \cite{2008ApJ...680..495H}; (3) \cite{2011ApJS..193...25R}; (4) \cite{2017ApJ...836...34M}; (5) \cite{2021AandA...647A.116C}}
\end{deluxetable*}

\section{Analysis: Bayesian Fitting to the Continuum}
\label{sec:continuum_fit}
In Section \ref{sec:sample selection} we identified a sample set of YSO spectra, for which we now develop a procedure to fit a multi-component YSO model. 
The goal is to acquire accretion luminosities $L_{acc}$ of our class II sample using the ‘direct’ method (fitting a model of excess continuum emission to the spectrum) with a Bayesian approach. We have written our procedure into a publicly-available Python code called \texttt{nuts-for-ysos}. We first outline the methodology of \texttt{nuts-for-ysos} in Sections \ref{sec:the_model}, \ref{sec:criteria}, and \ref{sec:initializing_and_priors}. We then discuss in Section \ref{sec:nuts-for-ysos_details} the specific requirements of \texttt{nuts-for-ysos} for inputted YSO spectra, and which parts of the analysis can be customized by the user.

\subsection{The Model}
\label{sec:the_model}
The continuum of a YSO spectrum in the UV and optical range is mainly affected by two different phenomena, which need to be modeled simultaneously. Accretion from its circumstellar disk causes a YSO to display stronger continuum emission in the blue end of its spectrum, along with emission lines and a veiling of its photospheric absorption lines. On the other hand, flux from the YSO may be extinguished by foreground material, local material within the surrounding molecular cloud, and material within its own circumstellar disk. This extinction subdues the blue part of the spectrum. The two phenomena thus have opposite effects on the perceived temperature of the star, making them difficult to disentangle \citep{2013A&A...558A.114M}. We therefore use a model that incorporates both the accretion and extinction simultaneously so that these dual effects can be considered when determining the stellar properties. Since we are fitting the U band region of the spectrum, any contribution from the protoplanetary disk itself is negligible and we can ignore that component in the modeling.

\emph{Accretion Spectrum:} We use a slab of isothermal hydrogen in local thermodynamic equilibrium (LTE), including emission from both H and H$^-$, to model the excess continuum emission from accretion. This approach has been used numerous times in the past to derive accretion luminosities (e.g. \citet{1993AJ....106.2024V, 2008ApJ...681..594H, 2012A&A...548A..56R, 2013A&A...558A.114M,
2014A&A...561A...2A,
2017A&A...600A..20A}). We take the equations for the slab model from Section 2.2 of \citet{2014PhDT.......477M}. The slab model by itself has three parameters: electron temperature ($T_{slab}$), electron density ($n_e$), and optical depth at $\lambda$ = 300 nm ($\tau_0$). This model was originally developed to describe particles transversing a boundary layer between the disk and stellar surface \citep{1974MNRAS.168..603L}. This is contrary to the more current paradigm of magnetospheric accretion producing shocks on the stellar surface, as in models like that of \citet{1998ApJ...509..802C}. Nonetheless, the slab model is still oftentimes trusted as a relatively simple empirical way to determine a bolometric correction for the accretion luminosity $L_{acc}$, even if the three parameters $T_{slab}$, $n_e$, and $\tau_0$ themselves do not have a physical basis. Further justification for using the slab model is discussed in Section \ref{sec:param_correlations} and can also be found in Section 2.2 of \citet{2014PhDT.......477M}.

\emph{Photospheric Templates:} While the slab model represents the excess emission due to accretion, the photospheric contribution to a YSO spectrum can be represented by a class III spectrum of the same spectral type. Modeling the photosphere of an accreting YSO (the ‘target’) using a non-accreting YSO (the ‘template’) is considered a better approach than using a main sequence star. This is because it better captures the effects of elevated chromospheric activity and the altered surface gravity of YSOs compared to field dwarfs, which makes for an overall more accurate representation of the photosphere \citep{2013A&A...558A.114M}.
We use a total of 23 class III photospheric templates taken from \citet{2013AandA...551A.107M} and \citet{2017AandA...605A..86M}. These class III YSOs were chosen by the authors because they have an $A_{V}$ $\approx 0$. The spectra were acquired with the ESO VLT/X-shooter spectrograph (R $\sim$ 4000-17000, depending on wavelength and slit width) \citep{2011A&A...536A.105V}, and have been convolved with a Gaussian to match the lower VIRUS resolution.
The templates range in spectral type from G5 - M9.5, with the most thorough sample (at least one template per SpT) ranging from G8 - M6.5.
Each of these templates has a $T_{eff}$ estimated from its SpT, following the same SpT-$T_{eff}$ scale as in \citet{2013AandA...551A.107M} and \citet{2017AandA...605A..86M}. For the earlier templates up to M0, the scale uses the relation from \citet{1995ApJS..101..117K} and for later templates, the relation from \citet{2003ApJ...593.1093L}.
The corresponding distances, luminosities, and uncertainty on the luminosity (estimated 0.2 dex) for the templates we take directly from \citet{2013AandA...551A.107M} and \citet{2017AandA...605A..86M}. As discussed in Section \ref{sec:interp_templates}, the flux for each template is rescaled such that the NUTS sampler can easily interpolate between them.
The complete list of photospheric templates used in this paper is provided in Table \ref{tab:template_info}. Recently, 19 more de-reddened class III photospheric templates were introduced by \citet{2024A&A...690A.122C} to make a grid of 57 templates in total. \citet{2024A&A...690A.122C} also includes updated estimates for the luminosity $L_*$ and the effective temperature $T_{eff}$, using the SpT-$T_{eff}$ scale from \citet{2014ApJ...786...97H}. These values for $L_*$ and $T_{eff}$ are different from the values we had taken from \citet{2013AandA...551A.107M} and \citet{2017AandA...605A..86M}. Moreover, \citet{2024A&A...690A.122C} includes uncertainties on SpT, luminosity, and distance individual to each object. While these new templates were not involved in our analysis, \texttt{nuts-for-ysos} is capable of using the updated template library, which can then be utilized in future works. \citet{2024A&A...690A.122C} also included an interpolation procedure between the templates of this enhanced grid which we discuss in Section \ref{sec:interp_templates}.

\begin{deluxetable*}{llllll}
\tablecaption{Photospheric Templates and corresponding data taken from \citet{2013AandA...551A.107M} and \citet{2017AandA...605A..86M}\label{tab:template_info}.}
\tablewidth{0pt}
\tablehead{
    \colhead{Name} & \colhead{SpT} & \colhead{$T_{eff}$} & \colhead{log($L_{\star}/Lsun$)} & \colhead{Distance (pc)} & \colhead{Source} 
}
\startdata
RXJ0445.8+1556 & G5 & 5770 & 0.485 & 140 & 2\\
RXJ1508.6-4423 & G8 & 5520 & 0.043 & 150 & 2\\
RXJ1526.0-4501 & G9 & 5410 & -0.061 & 150 & 2\\
RXJ1515.8-3331 & K0.5 & 5050 & 0.098 & 150 & 2\\
RXJ0457.5+2014 & K1 & 5000 & -0.15 & 140 & 2\\
RXJ0438.6+1546 & K2 & 4900 & -0.024 & 140 & 2\\
RXJ1547.7-4018 & K3 & 4730 & -0.081 & 150 & 2\\
RXJ1538.6-3916 & K4 & 4590 & -0.217 & 150 & 2\\
TWA9A & K5 & 4350 & -0.61 & 68 & 1\\
RXJ1540.7-3756 & K6 & 4205 & -0.405 & 150 & 2\\
TWA6 & K7 & 4060 & -0.96 & 51 & 1\\
TWA25 & M0 & 3850 & -0.61 & 54 & 1\\
TWA14 & M0.5 & 3780 & -0.83 & 96 & 1\\
TWA13B & M1 & 3705 & -0.7 & 59 & 1\\
TWA2A & M2 & 3560 & -0.48 & 47 & 1\\
TWA7 & M3 & 3415 & -1.14 & 28 & 1\\
Sz121 & M4 & 3270 & -0.34 & 200 & 1\\
SO797 & M4.5 & 3200 & -1.26 & 360 & 1\\
SO641 & M5 & 3125 & -1.53 & 360 & 1\\
SO999 & M5.5 & 3060 & -1.28 & 360 & 1\\
Par-Lup3-1 & M6.5 & 2935 & -1.18 & 200 & 1\\
CHSM17173 & M8 & 2710 & -1.993 & 160 & 2\\
TWA26 & M9 & 2400 & -2.7 & 42 & 1\\
\enddata
\tablerefs{(1) \citet{2013AandA...551A.107M}; (2) \citet{2017AandA...605A..86M}}
\end{deluxetable*}

\emph{Combined Model:} The slab and the photospheric template are separately scaled and then added together. The reason for the scaling is to match the raw flux of the model to the flux of each target, which has its own inherent distance and luminosity. Then, the entire model spectrum is reddened to match the extinction of the target. We use the reddening law presented in \citet{1989ApJ...345..245C} with $R_V$ = 3.1. There are thus a total of seven parameters involved in the model:
the three slab parameters ($T_{slab}$, $n_e$, and $\tau_0$), the scale factor of the slab ($K_{slab}$), the effective temperature $T_{eff}$ of the photospheric template, the scaling of the photospheric template ($K_{phot}$), and the overall extinction of the model ($A_V$).

\subsection{Criteria Used in Fitting}
\label{sec:criteria}
As in \citet{2013A&A...558A.114M}, we do not explicitly fit the entire target YSO spectrum but instead choose certain features of the spectrum which the model fitting attempts to match. The \texttt{nuts-for-ysos} code allows the user to customize which features of the spectrum to use. The tool is by default capable of computing several different 'types' of spectral features: individual flux values, slopes, ratios, and photometric magnitudes. The number, types, and wavelength ranges of these features can be changed by the user as desired. 
For our particular work with VIRUS spectra, we chose features which sample both bluer parts of the spectrum dominated by accretion emission (around the Balmer Jump) and redder parts of the spectrum dominated by photospheric emission. The chosen features deliberately avoid wavelengths with strong emission lines. The emission lines of an accreting YSO are not replicated by the simple model we use, and separately modeling these emission lines would be a more complicated process. 
The chosen features include the slope of the Balmer continuum between $\sim$3500 and $\sim$3600\r{A}, the slope of the Paschen continuum between $\sim$3980 and $\sim$4790\r{A}, the slope of the continuum between $\sim$5060 and $\sim$5420\r{A}, and the value of the continuum at several locations ($\sim$3600\r{A}, $\sim$3860\r{A}, $\sim$4020\r{A}, $\sim$4610\r{A}, $\sim$5480\r{A}). 
Overall, these features are similar to those used in \citet{2013A&A...558A.114M}. However, \citet{2013A&A...558A.114M} uses targets from the ESO VLT/X-shooter spectrograph which has a larger wavelength region available for fitting spectral features. Instead, our spectrum occupies the shorter wavelength region of VIRUS. We attempt to make up for the lack of VIRUS coverage from $\sim$5500 to $\sim$7150\r{A} by instead using photometry (an approach originally suggested in \citet{2013A&A...558A.114M}). We compute synthetic Pan-STARRS r and i magnitudes for models. These synthetic magnitudes are included as features to be compared to the actual Pan-STARRS DR1 r and i magnitudes of the targets. 
Because this photometry is non-simultaneous with the spectrum, and because of the inherent variability of YSOs, we assign conservative uncertainties of 0.2 mag to the Pan-STARRS r and i magnitudes when performing the fitting. 
The exact wavelength ranges for these 11 features are reported in Table \ref{table:fit_features}.
Only these features are used when fitting the model to the target spectrum. It should be emphasized that the process for fitting the model attempts to fit only the continuum and not the emission lines, a practice consistent with previous similar works. This is most evident near the higher-level Balmer lines, where the spacing between lines becomes small and the superimposed lines form a pseudo-continuum for which the model does not account (e.g. see Object 11 in Figure \ref{fig:continuum_fit_examples}).

\begin{deluxetable*}{ll}
\tablecaption{Spectral features used in fitting the total model to each target spectrum.\label{table:fit_features}}
\tablewidth{0pt}
\tablehead{
    \colhead{Name} & \colhead{Wavelength Range (\r{A})}
}
\startdata
 Slope of Balmer continuum &  mean[3580:3600]-mean[3504:3524]\\
 Slope of Paschen continuum &  mean[4770:4790]-mean[3980:4000]\\
 Slope  $\sim$ 508nm to $\sim$ 541nm & mean[5390:5424]-mean[5060:5100]\\
 Continuum at $\sim$ 360nm & mean[3580:3620]\\
 Continuum at $\sim$ 386nm & mean[3850:3870]\\
 Continuum at $\sim$ 402nm & mean[4000:4030]\\
 Continuum at $\sim$ 461nm & mean[4596:4624]\\ 
 Continuum at $\sim$ 511nm & mean[5090:5130]\\ 
 Continuum at $\sim$ 548nm & mean[5470:5490]\\ 
 Pan-STARRS r magnitude &  Pan-STARRS r filter [5200:7100] \\
 Pan-STARRS i magnitude & Pan-STARRS i filter [6700:8400] \\
\enddata
\end{deluxetable*}

\subsection{Initializing the Bayesian Fit}
\label{sec:initializing_and_priors}
We use the Python package \texttt{PyMC} \citep{pymc3} to implement a Bayesian process for fitting the total model to each target spectrum. Within \texttt{PyMC} we use the No U-Turn Sampler (NUTS) with 16 chains, each having a length of 2000. We set \texttt{target\_accept} = 0.99, so that the sampler takes very small steps. We used NUTS because for a system with significant parameter correlations such as ours, a Hamiltonian-based Monte-Carlo algorithm like NUTS is more efficient than the Metropolis-Hastings algorithm or Gibbs sampling \citep{Hoffman2014TheNS}.

To reduce the burn-in time, a starting point for the sampler is determined using a least-squares optimization. In finding the starting point we attempt to match the spectral features listed in Table \ref{table:fit_features} by minimizing a likelihood function similar to a $\chi^2$ function. This $\chi_{like}^2$ function was first introduced in \citet{2013A&A...558A.114M} and is as follows:

\begin{equation}
\chi_{like}^2 = \sum_{features}\left( \,\frac{f_{obs}-f_{mod}}{\sigma_{obs}}\right) \,^2
\label{eq:chi_square_like}
\end{equation}

where $f_{obs}$ is the feature from the VIRUS spectrum, $\sigma_{obs}$ is the respective uncertainty, and $f_{mod}$ is the same feature predicted the composite model. After the starting point for each parameter is found, it is used as the initial point for the chains of the NUTS sampler.

\subsubsection{Interpolating Between Photospheric Templates}\label{sec:interp_templates}
NUTS requires each parameter in the model to have a continuous range, because it is a gradient-based sampler. NUTS takes the gradient of the likelihood with respect to each parameter, to reach convergence faster than other sampling methods \citep{Hoffman2014TheNS}. 
Six out of the seven total model parameters are inherently continuous, but $T_{eff}$ is not, because we rely only on a set of 23 discrete photospheric templates of various spectral types. We thus made $T_{eff}$ a continuous parameter by having the sampler linearly interpolate between photospheric templates. Each of the class III photospheric templates comes from real observations rather than simulation, and so each has an (unspecified) uncertainty on its stellar distance provided in \citet{2013AandA...551A.107M} or \citet{2017AandA...605A..86M}. This introduces scatter in the perceived relative brightnesses between templates. For the sampler to smoothly interpolate between templates, we decided to scale the flux of each template so the templates are at the same distance, and then slightly rescale both the templates and their respective luminosities $L_{*,phot}$ so that the brightness from each template would smoothly and monotonically increase with $T_{eff}$. We do so by forcing the median fluxes from 4500-5500\r{A} of each template to follow a fourth-degree polynomial fitted to the original medians. These adjustments ultimately give the sampler an easier set of templates to interpolate between without fundamentally changing the nature of the results. Without this cleaning of the templates, we would not have been able to make $T_{eff}$ a continuous parameter in the model and instead would have needed to make an ad hoc selection of a template before the rest of the model fitting process. 

A similar methodology has been used by \citet{2024A&A...690A.122C} to interpolate between their grid of 57 class III photospheric templates. They first normalized their class III spectra to factor out the individual distances to each star. They then computed the median fluxes of the spectra within multiple wavelength ranges. Next, they performed non-parametric local polynomial fits to the medians as a function of SpT. The smooth curve resulting from each fit then served as a basis for an interpolable set, such that the grid of templates could be made as well-sampled as needed. Within the polynomial fitting procedure, they also account for uncertainties in the median fluxes, the extinction, and spectral type of each template. They do this by repeating the polynomial fit over 1000 Monte-Carlo iterations with the error terms sampled from Gaussian distributions. The result of this process is that each interpolated template has an uncertainty in the spectral flux. This inclusion of uncertainty within the interpolation is an important addition which was not in our work. In our work, we simply incorporated uncertainties of $A_V$ and $T_{eff}$ directly into the Bayesian inference of the parameters. However, propagating uncertainty during the template interpolation stage could be particularly relevant when using the late-M spectral type class III templates, which generally have the lowest SNRs in the grid. The interpolation procedure of \citet{2024A&A...690A.122C} is available through their \texttt{FRAPPE} tool on Github \footnote{\url{https://github.com/RikClaes/FRAPPE}}. While our own interpolation did not use this approach, \texttt{nuts-for-ysos} has recently been updated with a new version that gives the NUTS sampler the ability to include uncertainties in the spectral flux of an inputted template grid, such as those acquired from \texttt{FRAPPE}.

\begin{figure} 
\centering
\includegraphics[scale=0.45]{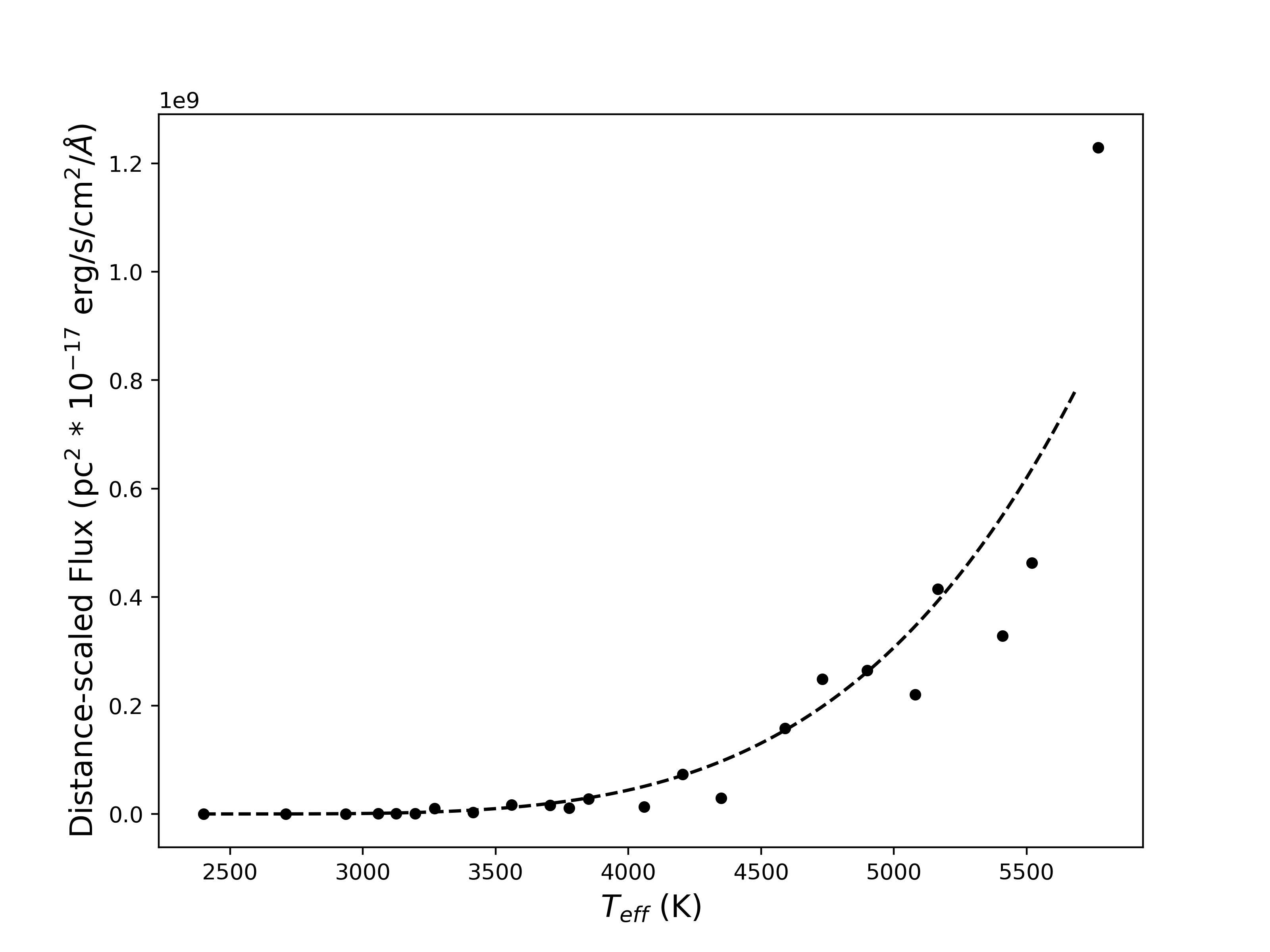}
\caption{The median values of each photospheric template flux, after being multiplied by its squared distance. A fourth-degree polynomial is fit to the medians, and the template fluxes and respective luminosities are then rescaled to match this polynomial.}
\label{fig:rescaled_flux}
\end{figure}
\subsubsection{Priors}\label{sec:priors}
\begin{figure*} 
\centering
\includegraphics[scale=0.85]{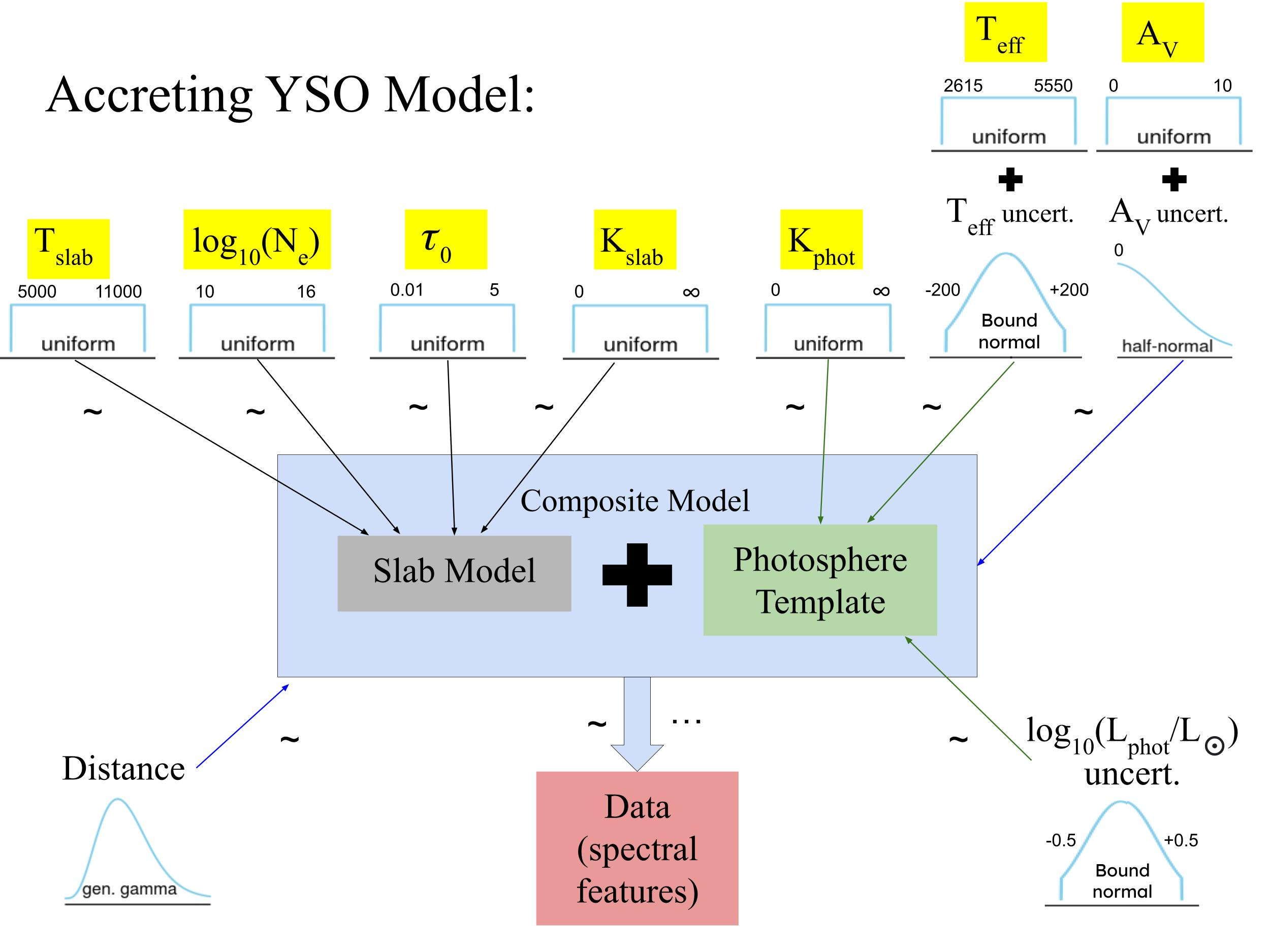}
\caption{A diagram illustrating the components of the YSO model. Parameters which are fitted for are highlighted in yellow. The shape of each prior is shown using Kruschke-style plots. Each prior has a color-coded arrow indicating whether it belongs to the slab portion, the photospheric portion, or to the entire composite model (black, green, and blue, respectively).}
\label{fig:prior_diagram}
\end{figure*}

After making the adjustment so the $T_{eff}$ parameter is continuous rather than discrete, we are then able to create priors for every component of the YSO model. A schematic diagram is presented in Figure \ref{fig:prior_diagram} showing how each prior contributes to the model. The priors associated with slab model are bounded within a mostly typical range of parameter space: $T_{slab}$ ranges from 5000 to 11,000K, $n_e$ ranges from $10^{10}$ to $10^{16}$ cm$^{-3}$, and $\tau_0$ ranges from 0.01 to 5.0. The $T_{slab}$ and $\tau_0$ priors are uniform distributions, and $n_e$ is a uniform distribution in logarithmic space. The priors for $K_{slab}$ and $K_{phot}$ are set as HalfFlat distributions, constraining them to be above zero. The $A_{V}$ prior is a uniform distribution bounded between 0 and 10.

The benefit of a Bayesian model for the fitting is that systemic uncertainties in the photospheric templates can be seamlessly integrated into the model. For example, the templates have an uncertainty in their spectral type of roughly 1 sub-class for K type stars, and half a sub-class for M type stars \citep{2013AandA...551A.107M, 2017AandA...605A..86M}, which we translate into an uncertainty in $T_{eff}$ of 100K. The prior for $T_{eff}$ by itself is set to be a Uniform prior, bound between 2615 and 5550 K (roughly SpTs M8.5 to G8), but a Normal prior centered on 0K with $\sigma = 100$K is then added to this, to reflect the uncertainty in $T_{eff}$ of the photospheric templates. However, the $T_{eff}$ uncertainty can only extend to the extrema of available templates (SpTs M9 and G5), beyond which we are unable to generate a model. We therefore bound the $T_{eff}$ uncertainty prior to be within $\pm$ 200K. 

Additionally, the photospheric templates are uncertain in their extinction; though the templates were selected by \citet{2013AandA...551A.107M} and \citet{2017AandA...605A..86M} to have $A_{V}$ close to zero, it is more accurate to include a possibility of non-zero extinction. According to \citet{2017AandA...605A..86M}, all templates compared to BT-Settl models \citep{2011ASPC..448...91A} have an $A_{V}$ $<$ 0.5 mag, with the possibility of $A_{V}$ = 0 mag being within $3\sigma$ at most. We
add a Half-Normal prior centered on 0 with $\sigma= 0.5/3$ to the uniform $A_{V}$ prior, to represent this uncertainty on $A_{V}$. 

We also include other uncertainties not directly related to a fitted parameter. The photospheric templates have an uncertainty in their luminosity $L_{*,phot}$, given by \citet{2013AandA...551A.107M} to be 0.2 dex. We represent this uncertainty with a Normal prior in log space, centered on 0 with $\sigma = 0.2$. 
For distance to each target YSO, the catalog associated with \citet{2021AJ....161..147B} provides the median, 16th percentile, and 84th percentile of a posterior probability distribution.
\citet{2021AJ....161..147B} used a generalized gamma distribution (GGD) as a prior when computing the 'geometric distances' we used. We thus choose to represent the distance to each target YSO as a GGD, defined using the median, 16th percentile, and 84th percentile.

\begin{deluxetable*}{llll}
\tablecaption{Priors used\label{table:priors}}
\tablewidth{0pt}
\tablehead{
    \colhead{Name of Prior} & \colhead{Distribution Type} & \colhead{Lower Bound} & \colhead{Upper Bound}
}
\startdata
$T_{slab}$ & Uniform & 5000K & 11000K\\
log($n_e$) & Uniform & $10^{10} $cm$^{-3}$ & $10^{16} $cm$^{-3}$ \\
$\tau_0$ & Uniform & 0.01 & 5.0\\
$K_{slab}$ & Half-Flat & 0 & $\infty$\\
$K_{phot}$ & Half-Flat & 0 & $\infty$\\
$T_{eff}$ & Uniform & 2615 K & 5550 K\\
$T_{eff}$ Uncert & Bounded Normal ($\mu=0K$, $\sigma= 100$K) & -200K & 200K\\
$A_{V}$ & Uniform & 0 & 10\\
Distance $d_{obs}$ & GGD* & 0 pc & $\infty$ pc\\
Phot. template $A_{V}$ & Half-Normal ($\sigma= 0.167$) & 0 & $\infty$\\
Phot. template log($L_{*,phot}/L_\odot$) Uncert & Bounded Normal ($\mu=0$, $\sigma= 0.2$) & -0.5 & 0.5 \\
\enddata
\tablecomments{*Median, 16th percentile, and 84th percentile of generalized gamma distribution defined for each target using the distance values in Table \ref{tab:bayesianresults}.}
\end{deluxetable*}

\subsection{The \texttt{nuts-for-ysos} Workflow}
\label{sec:nuts-for-ysos_details}
\begin{figure*} 
\centering
\includegraphics[scale=0.5]{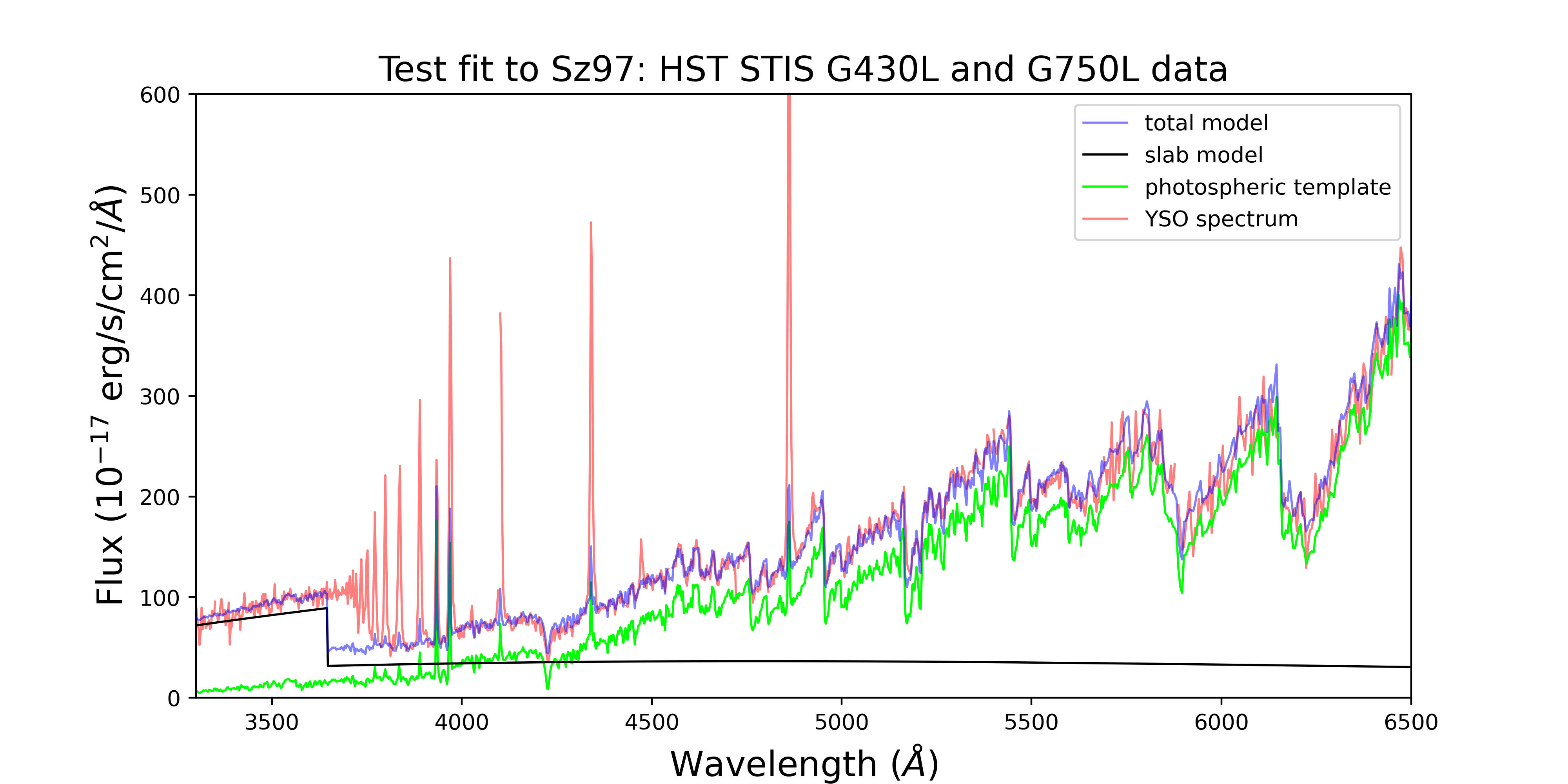}
\caption{The median model fit resulting from testing \texttt{nuts-for-ysos} on HST STIS G430L and G750L data of Sz97, considering 14 different continuum features between $\sim 3300$\r{A} and $\sim 6000$\r{A}.\\}
\label{fig:HST_test}
\end{figure*}

The \texttt{nuts-for-ysos} package incorporates the Bayesian analysis described throughout Section \ref{sec:continuum_fit}. The code has been written to also be compatible with YSO spectra obtained by other instruments besides VIRUS. The inputted YSO spectrum can be of a higher-resolution than that of VIRUS, but it must be of the same or lower resolution than the Class III templates, and must occupy a wavelength range covered by the Class III templates (for example, between 3300.0\r{A} and 10189.0\r{A} if using the UV and optical Class III templates from X-Shooter).
The user can customize several aspects of the analysis. The grid of interpolable Class III templates can be changed from the default grid of templates listed in Table \ref{tab:template_info}. For example, the new grid of templates presented in \citet{2024A&A...690A.122C} is better sampled in SpT and can be substituted within \texttt{nuts-for-ysos}. The only requirement for the templates themselves is that the wavelengths of the template spectra are no lower than 500.0\r{A} or higher than 25000.0\r{A}. As mentioned in Section \ref{sec:criteria}, the features for which the model is evaluated can also be altered within \texttt{nuts-for-ysos}. The tool is by default capable of computing individual flux values, slopes, ratios, and photometric magnitudes from spectra. The number, types, and wavelength ranges of these features can be changed by the user as desired.
As an example, we tested \texttt{nuts-for-ysos} with several low-resolution class II YSO spectra among the HST ULYSSES survey \citep{2020RNAAS...4..205R}.\footnote{Data from HST ULYSSES can be found in MAST: \dataset[https://doi.org/10.17909/t9-jzeh-xy14]{https://doi.org/10.17909/t9-jzeh-xy14}.} Specifically, we fit the model to data from the $R \sim 500$ STIS G430L and G750L gratings. We did so between $3300$\r{A} and $6000$\r{A}, using the same Class III templates in Table \ref{tab:template_info} and using the same features as in Table \ref{table:fit_features}, with three additional features included at $\sim 3310$\r{A}, $\sim 5850$\r{A}, and $\sim 6000$\r{A}. We found the code runs successfully and we show in Figure \ref{fig:HST_test} an example fit for the Class II YSO Sz97. We performed compatibility tests of \texttt{nuts-for-ysos} with X-Shooter data of several Class II YSOs from \citet{2014A&A...561A...2A} and \citet{2017A&A...600A..20A} as well. Version 1.1 of \texttt{nuts-for-ysos} is available on GitHub\footnote{\url{https://github.com/laurenwillett/nuts-for-ysos} and on Zenodo \citep{nuts-for-ysos-zenodo}}.

\section{Accretion Model Fit Results}
\label{sec:bayesian_results}

We check for convergence of each fit using the Gelman-Rubin
statistic, by requiring \^{R} $\leq$ 1.1 \citep{1992StaSc...7..457G} for every parameter. To mitigate autocorrelation, we only use every 20th value in each parameter posterior outputted by \texttt{PyMC} to compute further results. Since we ran 16 chains with length 2000 each, the original posterior has a length of 32000 and the new thinned posterior has length of 1600.

The composite model can have a $T_{eff}$ of at least 2615K and at most 5550K. We encountered two cases in which the fitted model to a target from our sample had an effective temperature on the edge of these boundaries. Object 13 (2MASS J03292815+3116285), a class III YSO, was found to have a $T_{eff}$ posterior that peaks at $\sim$2600K and is cut off for lower temperatures. Similarly, Object 4 (EM* LkHA 188), a class II YSO, was found a $T_{eff}$ posterior that peaks at $\sim$5500K and is cut off for higher temperatures. Because the sampler was unable to explore the entirety of the plausible model parameter space for these two objects, we take their $T_{eff}$ to be upper and lower limits respectively. The results for their other parameters are taken to only be approximate throughout further analysis.

\begin{figure}
\centering
\includegraphics[scale=0.37]{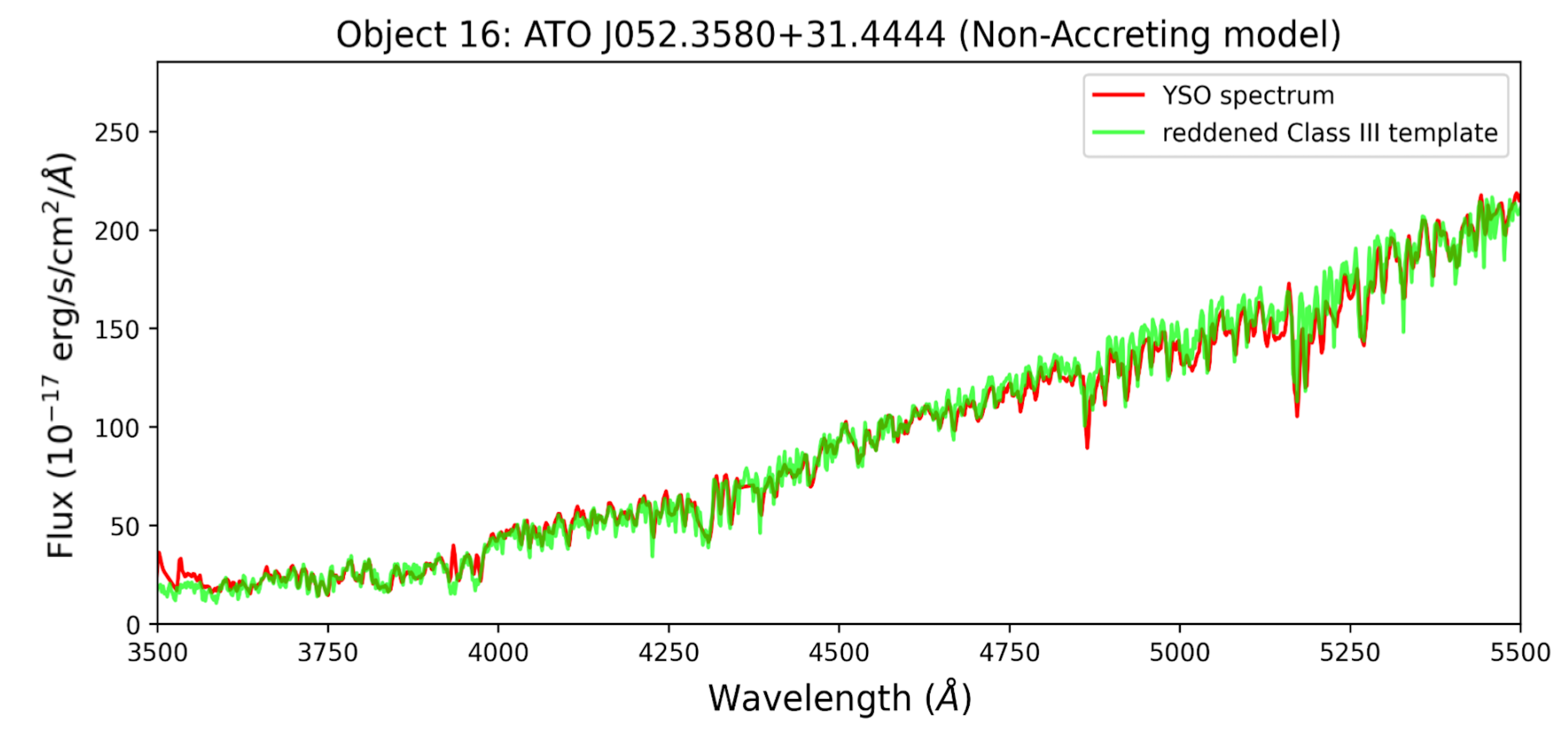}
\caption{Object 16 (ATO J052.3580+31.4444), the one YSO in our current sample for which we were unable to fit a model including accretion. Instead, we fit the spectrum (red) with only a reddened Class III template (green) having $T_{eff} \approx$ 5500K and $A_{V} \approx$ 3.2.}
\label{fig:190_classIII_fit}
\end{figure}

Object 16 (ATO J052.3580+31.4444) was the only star for which we were unable to fit an accretion model, a finding which was briefly previewed at the end of the Sample Selection section (Section \ref{sec:sample selection}). This object has seemed to be a YSO in some ways, as it belongs to NGC1333 and has a class II SED. However, we found that our process was unable to plausibly fit an accretion model to the spectrum of Object 16. The spectrum also appears to have Balmer lines which are in absorption rather than emission, as discussed in Section \ref{sec:emission_lines}. We were only able to fit a model of a reddened class III template having $T_{eff} \approx$ 5500K and $A_{V} \approx$ 3.0. A plot of the data and model is shown in Figure \ref{fig:190_classIII_fit}.

For the rest of the objects we were successful in fitting a model of an accreting YSO which converged and explored the full plausible range of $T_{eff}$. Plots of the median model fits for these 15 objects are shown in Figure \ref{fig:continuum_fit_examples}, and the full set of plots are also shown in Appendix \ref{sec:result_plots}. Underneath each of the spectra in the Appendix are also plots of probability distributions for each of the model parameters.

We find that in general, the least constrained parameters tend to be those associated with the slab portion of the model. This is especially true of the electron density $n_e$ which has a posterior sometimes occupying a flat or nearly-flat probability distribution in logarithmic space. The posterior probability distributions of slab parameters $T_{slab}$, $n_e$, and $\tau_0$ also sometimes appear to peak at or near the bounds of their uniform priors; however, extending the bounds on these priors into more extreme territory was often found not to reveal any sudden dropoff in probability. Other model parameters $K_{phot}$, $T_{eff}$, and $A_{V}$ are more constrained and tend to have posterior probability distributions that are more Gaussian in appearance. In Section \ref{sec:discussion} we discuss the posteriors and their correlations in more detail.

\begin{figure*} 
\centering
\includegraphics[scale=0.35]{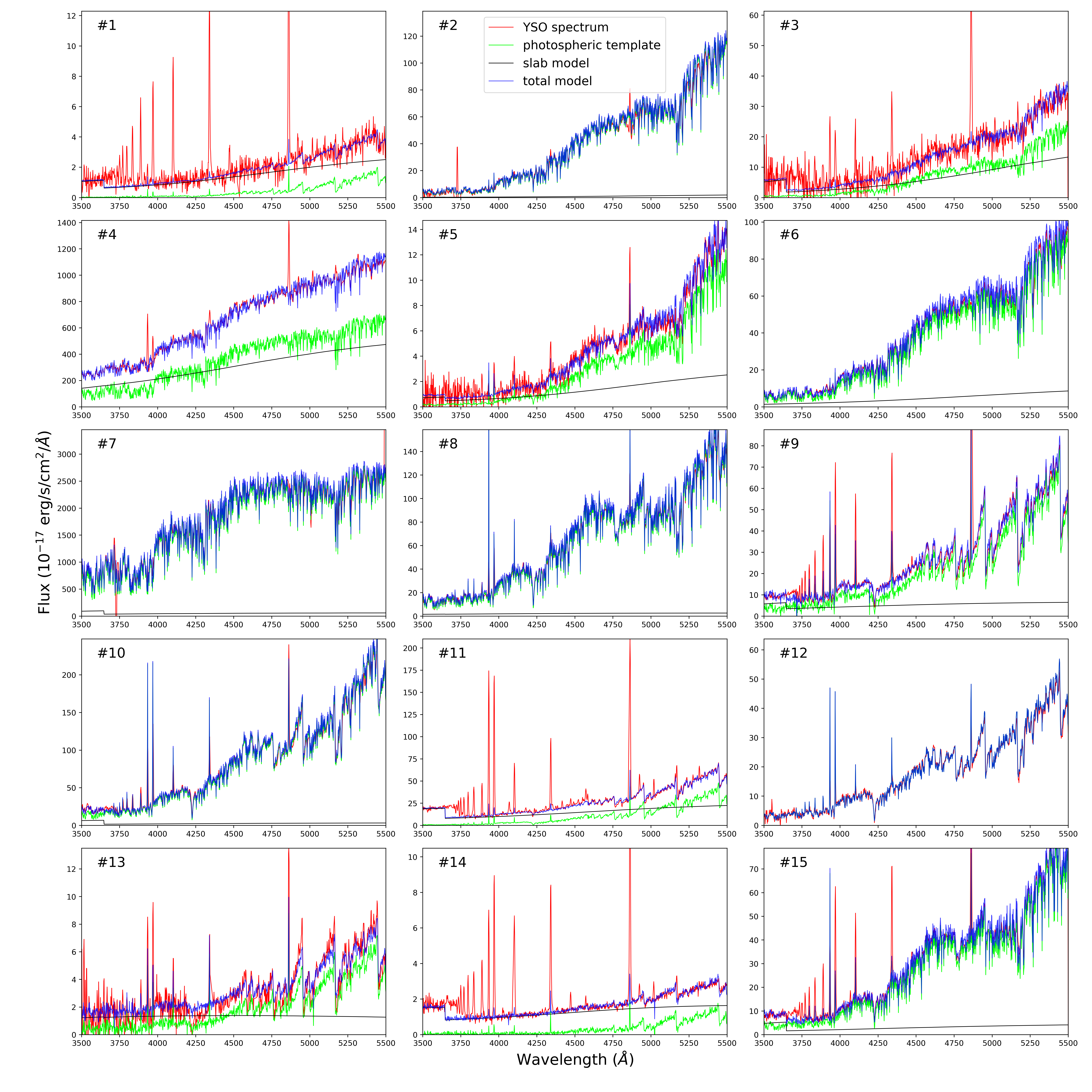}
\caption{The YSO model fit to VIRUS spectra of class II and class III YSOs. On the top left of each plot is the object ID originating in Tables \ref{tab:photometry} and \ref{tab:obs_log}. The VIRUS spectrum (red) has a model fit to its continuum (blue), which is the sum of the hydrogen accretion slab (black) and the photospheric template (green)}.
\label{fig:continuum_fit_examples}
\end{figure*}

\begin{figure}
\centering
\includegraphics[scale=0.34]{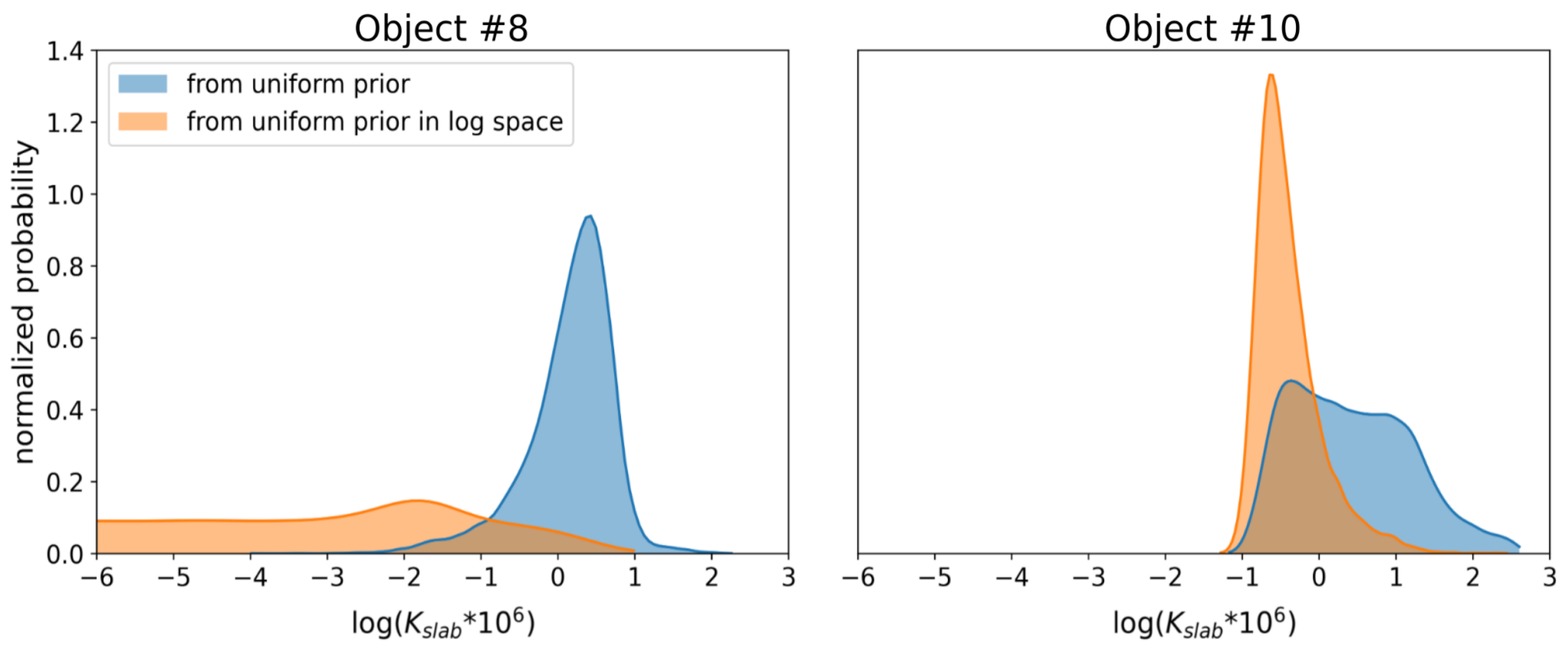}
\caption{Comparison of $K_{slab}$ posteriors (expressed as log($10^6*K_{slab})$) for Object 8 vs. Object 10. In blue is the original $K_{slab}$ posterior, using the uniform prior for $K_{slab}$ that was introduced in Section \ref{sec:priors} and Table \ref{table:priors}. In orange is the posterior resulting from an altered $K_{slab}$ prior which is uniform but in logarithmic space. From this comparison we see that Object 8 is possibly nonaccreting by virtue of having a slab component in the accretion model which approaches zero.} 
\label{fig:upperlim_comparison_plot}
\end{figure}

\subsection{Discerning Upper Limits on Accretion}\label{discern_upper_limits}
In objects that appear to have very little slab component in the model fit (e.g. Objects 8, 10, and 12, as seen in Figure \ref{fig:continuum_fit_examples} and in Appendix \ref{sec:result_plots}), there is an additional technicality that needs to be considered. The $K_{slab}$ posterior reaches very low values for these objects (for example, log($10^6*K_{slab})\sim-5$ for Object 12), but the lower bound for the uniform $K_{slab}$ prior was set exactly to 0 in Section \ref{sec:priors}. The sampler has a small but non-negligible step size which keeps it from precisely reaching this $K_{slab}$ = 0. 
We therefore must develop some way to check each object for whether the tapering of probability at low $K_{slab}$ values is a legitimate property, or if it is caused by boundary behavior.
We accomplish this by creating an altered $K_{slab}$ prior which is uniform but in logarithmic space. The lower bound of this prior is set to log($10^6*K_{slab})=-10$, which is much lower than any $K_{slab}$ resulting from the original model fits. If fitting the model with this new $K_{slab}$ prior yields a posterior that goes all the way down to -10, then this means the original model fit was affected by the lower bound of the prior. In these cases, our original model fit should only be interpreted as an upper limit on accretion. The object could be accreting, but it also might not be accreting at all.
In the opposite scenario, if the object is definitely accreting, the resulting $K_{slab}$ will be biased towards lower values but will still taper off in probability at approximately the same value as the original result.

Using this criteria, we find that five YSOs have slab components which should only be interpreted as upper limits: the class III objects 6, 7, and 12, and class II objects 2 and 8. Figure
\ref{fig:upperlim_comparison_plot} illustrates the difference between $K_{slab}$ posteriors of a YSO that is definitely accreting, albeit relatively low (Object 10) and a YSO which has an upper limit only (Object 8). For Object 10, the posterior tapers at $\approx -1$ regardless of which prior is used. For Object 8, the posterior resulting from the altered prior extends all the way to the lower bound of log($10^6*K_{slab}) = -10$.

\subsection{Calculating Luminosity and Accretion Luminosity from Fit}
\label{sec:calculate_L_Lacc}
The flux from accretion $F_{acc}$ is derived simply by integrating the hydrogen spectrum portion of the model fit (the 'slab' portion) over a very wide wavelength range; we integrate from 500\r{A} to 25000\r{A}. The integration is repeated for every possible slab created by the posterior. $F_{acc}$ is then converted into an accretion luminosity using the relation $L_{acc} = 4\pi d_{obs}^2 F_{acc}$ for a distance $d_{obs}$ from the target.
The stellar luminosity of each target is derived using the class III template portion of the model fit (the photspheric portion). Every best-fit class III template has a pre-computed luminosity $L_{*,phot}$ as discussed in Section \ref{sec:interp_templates}, which can then be combined with the determined template scaling, $K_{phot}$, to calculate the stellar luminosity $L_{*}$ of the target. The formula is:
\begin{equation}
L_{*} = K_{phot} * (d_{obs})^2 * L_{*,phot}
\end{equation}

Since every parameter involved in these calculations occupies a probability distribution defined in advance or defined by posterior outputted by \texttt{PyMC}, we are essentially building probability distributions for $L_{*}$ and $L_{acc}$.

Table \ref{tab:bayesianresults} lists averages and standard deviations for the most physically relevant parameters of the sample: $T_{eff}$, $A_{V}$, log($L_{*}$) and log($L_{acc}$), as well as the distances to each of these targets (with their 16th and 84th percentiles included). Objects 4 and 13, for which $T_{eff}$ is only a lower or upper limit, have all other model parameters listed as approximate. The objects with the lowest-SNR VIRUS spectra in Table \ref{tab:obs_log} tend to have the largest uncertainties in $T_{eff}$ and $A_{V}$, and the largest fractional uncertainty in $K_{phot}$. These uncertainties translate into wider errorbars on log($L_*$) and log($L_{acc}$). For example, the class II Objects 1 and 3 (with average SNRs of 5.6 and 5.8 respectively) both have a standard deviation in log($L_{acc}$) of $\sigma$log($L_{acc}$)$\sim$0.5, larger than the $\sigma$log($L_{acc}$)$\sim$0.15-0.30 found for other objects in the sample. In Table \ref{tab:bayesianresults} we do not list numerical results for $T_{slab}$, $n_e$, and $\tau_0$. As we briefly addressed in Section \ref{sec:the_model} (and discuss further in Section \ref{sec:discussion}, the flux of the slab model is useful for determining $L_{acc}$, but the model itself is based on simplified physical assumptions and so the numerical results for these three individual parameters should not be considered in great detail.

\begin{figure*}[p] 
\centering
\includegraphics[scale=0.909]{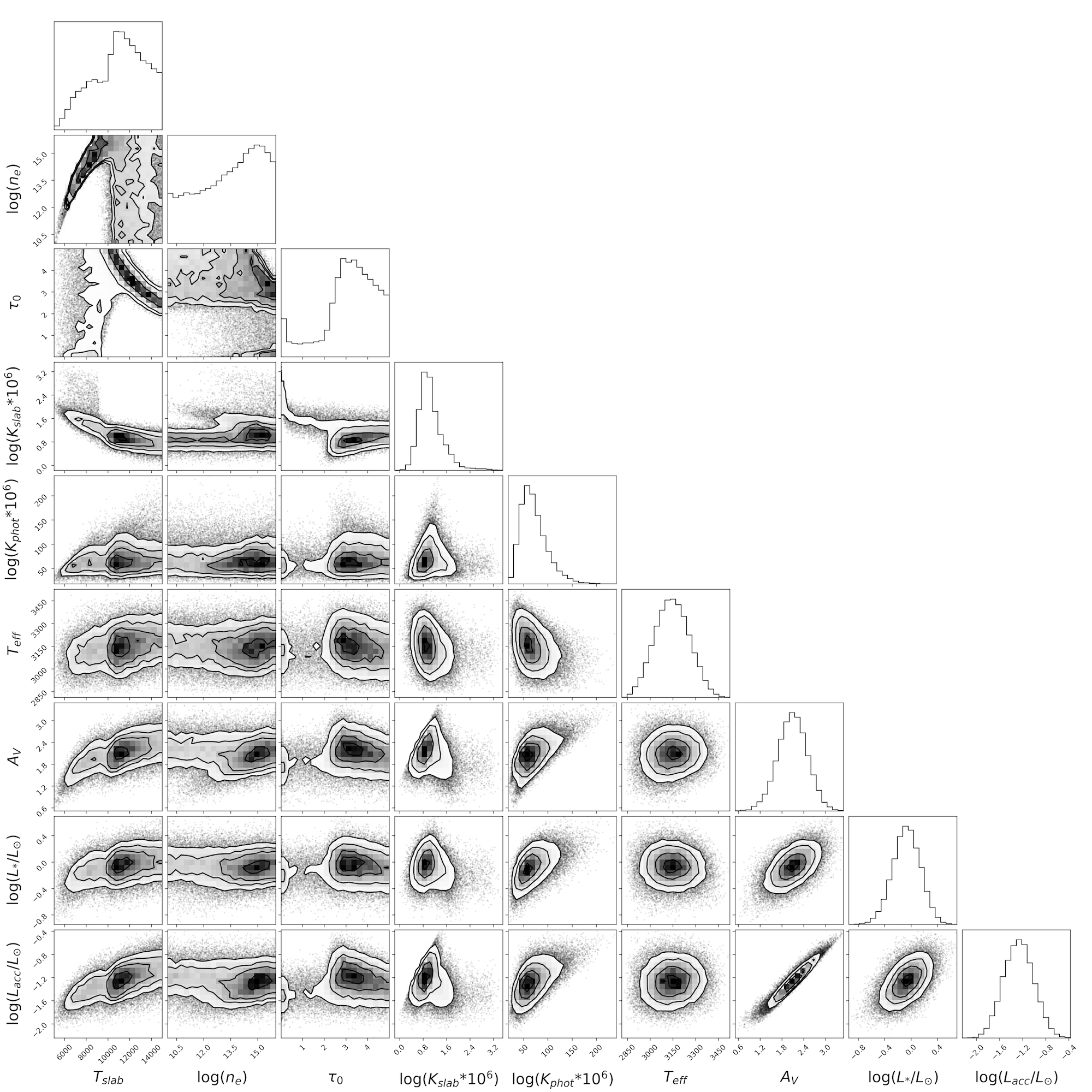}
\caption{An example corner plot for the fit of an accreting YSO model to the spectrum of Object 11 (2MASS J03285101+3118184). Included are the model parameters as well as the log($L_{*}$) and log($L_{acc}$) posteriors. The corner plots for all fits are in the in Appendix \ref{sec:result_plots}}.
\label{fig:corner_plot_example}
\end{figure*}

Appendix \ref{sec:result_plots} additionally includes corner plots of the model fit results for each object; both the parameters of the model itself, as well as log($L_{*}$) and log($L_{acc}$).
We find several correlations between parameters within the corner plots, which are discussed in Section \ref{sec:discussion}. Though all corner plots are in Appendix \ref{sec:result_plots}, we also show an example for Object 11 in Figure \ref{fig:corner_plot_example}.\\

\newpage

\begin{longrotatetable}
\begin{deluxetable*}{lllllllllllll}
\tablecaption{Both class II and class III YSOs from the sample, with information on their distances taken from \citet{2021AJ....161..147B}, and information on temperature, distance, luminosity, and accretion luminosity derived from the model fit to the continuum.\label{tab:bayesianresults}}
\tablewidth{0pt}
\tablehead{
    \colhead{ID} & \colhead{Class} & \colhead{$d_{obs}$ (pc)} & \colhead{$d_{obs,L}$ (pc)} & \colhead{$d_{obs,U}$ (pc)}  & \colhead{Mean $T_{eff}$} & \colhead{$\sigma T_{eff}$} & \colhead{Mean $A_{V}$} & \colhead{$\sigma A_{V}$} & \colhead{Mean log($L_{*}$)} & \colhead{$\sigma$log($L_{*}$)} & \colhead{Mean log($L_{acc}$)} & \colhead{$\sigma$log($L_{acc}$)}
}
\startdata
1     & II    & 694.89 & 660.88       & 746.28       & 2971       & 220      & 2.7158          & 0.9156 & -0.2056          & 0.3459   & -1.2957         & 0.5477      \\
2     & II    & 734.37 & 720.62       & 746.51       & 4291       & 107      & 2.8836          & 0.1629 & 0.5163           & 0.1657   & $\leq$-1.5617   &             \\
3     & II    & 785.71 & 754.77       & 814.22       & 3968       & 387      & 3.4723          & 0.786  & 0.241            & 0.268    & -0.1417         & 0.5203      \\
4    & II    & 746.32 & 737.9        & 753.86       & $\geq$5166 &          & $\approx$1.4186 &        & $\approx$0.5773  &          & $\approx$0.3249 & 0.2741      \\
5     & II    & 411.97 & 398.19       & 430.07       & 3761       & 116      & 3.6198          & 0.3877  & -0.3555          & 0.2052   & -1.3807         & 0.3005      \\
6     & III   & 430.93 & 424.05       & 438.52       & 4341       & 114      & 2.1424          & 0.1462 & -0.3679          & 0.1564   & $\leq$-1.5028   &             \\
7     & III   & 347.26 & 340.54       & 352.53       & 5012       & 163      & 0.1269          & 0.0994 & -0.0044          & 0.166    & $\leq$-1.6016   &             \\
8    & II    & 335.56 & 332.25       & 338.38       & 3582       & 98       & 0.1224          & 0.0904 & -0.6987          & 0.1687   & $\leq$-3.1574   &             \\
9    & II    & 282    & 277.84       & 286.37       & 3060       & 93       & 0.1863          & 0.1282 & -0.5797          & 0.1889   & -2.7998         & 0.1467      \\
10   & II    & 285.52 & 283.18   & 287.5        & 3350       & 94       & 0.5525          & 0.1467 & -0.2532          & 0.1578   & -2.813          & 0.2962      \\
11     & II    & 303.16 & 297.88       & 307.58       & 3147       & 111      & 2.0873          & 0.399  & -0.0704          & 0.2079   & -1.2693         & 0.2532      \\
12      & III   & 281.19 & 275.61       & 286.59       & 3274       & 92       & 0.0824          & 0.0707 & -1.0181          & 0.1653   & $\leq$-6.2905   &             \\
13    & III   & 297.2  & 285.6        & 308.59       & $\leq$2938 &          & $\approx$0.2679 &        & $\approx$-1.1377 &          & $\approx$-3.413 &             \\
14    & II    & 308.01 & 296.58       & 321.96       & 2828       & 114      & 2.1155          & 0.2844 & -0.7414          & 0.2095   & -2.2991         & 0.177       \\
15    & II    & 286.53 & 282.49       & 289.76       & 3676       & 106      & 1.0444          & 0.2114 & -0.8576          & 0.1712   & -2.542          & 0.2611\\
\enddata
\end{deluxetable*}
\end{longrotatetable}

\begin{figure*} 
\centering
\includegraphics[scale=0.6]{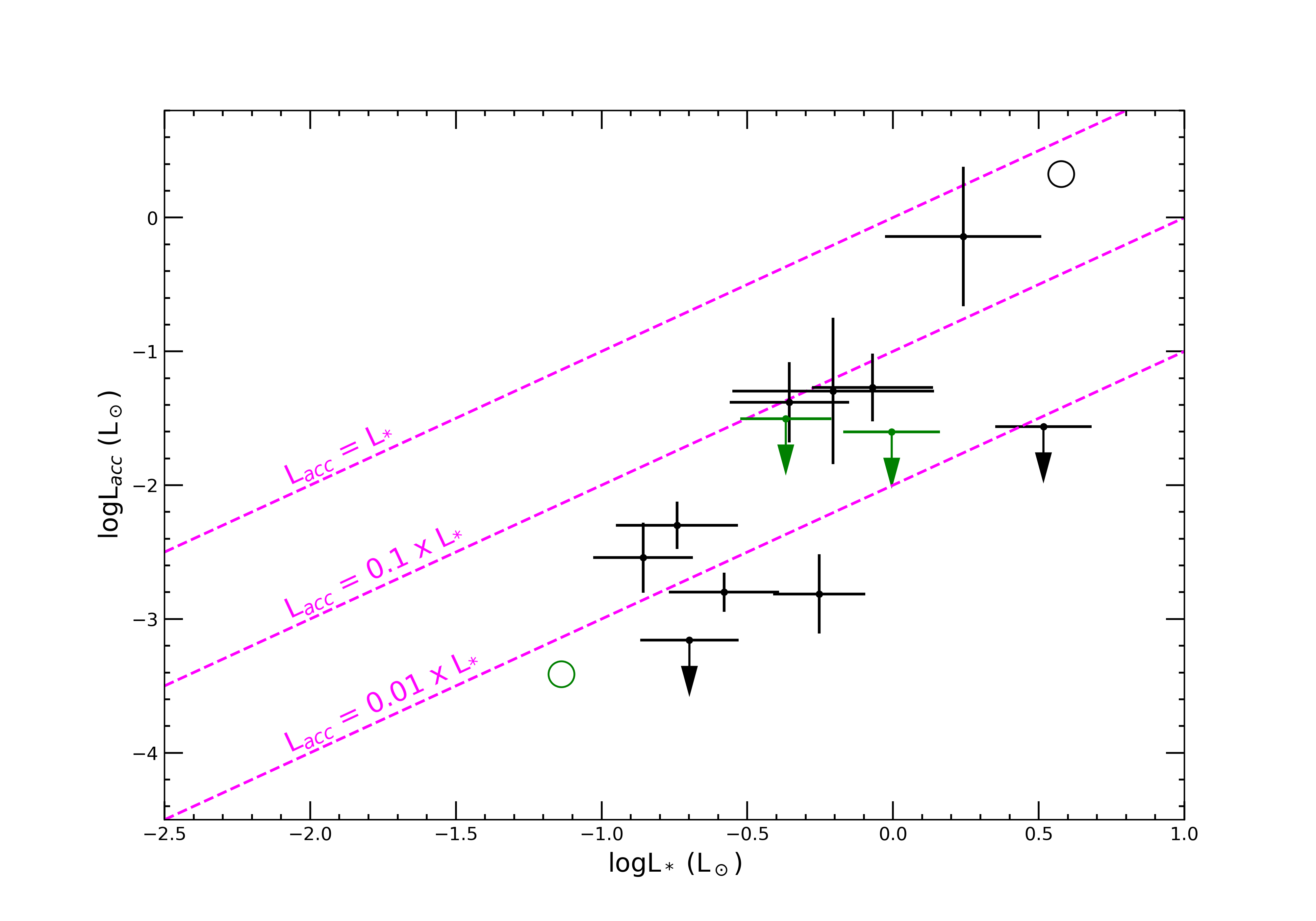}
\caption{Plot of log($L_{acc}$) vs. log($L_{*}$) for the class II and class III YSO targets. Class II YSOs are plotted in black, and the class III objects are plotted in green. Upper limits on log$L_{acc}$ are denoted with downward arrows instead of errorbars. Object 12 has an upper limit for log($L_{acc}$), but it is so low ($\approx -6.3$) that it falls below the plot limits. Object 4 is shown as a black open circle, and Object 13 as a green open circle. These two objects only have upper or lower limits on $T_{eff}$, as explained in Section \ref{sec:bayesian_results}, so their $L_{*}$ and $L_{acc}$ values are taken to be approximate.}
\label{fig:L_vs_Lacc}
\end{figure*}

\subsection{Estimating the Mass Posterior}
\label{sec:estimating_mass}
\begin{figure*} 
\centering
\includegraphics[scale=0.6]{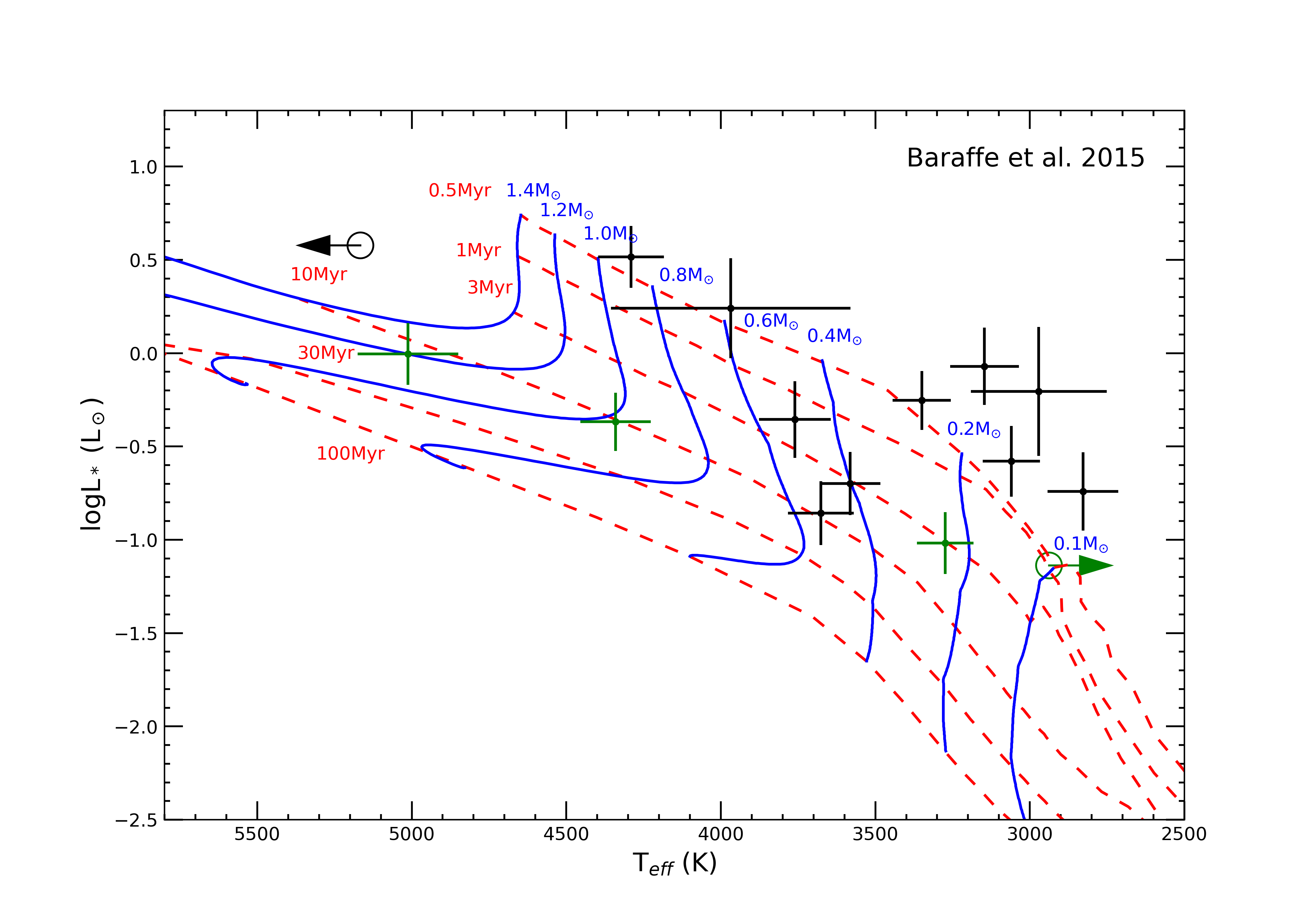}
\caption{Hertzsprung-Russell diagram for the total sample, plotted over evolutionary tracks (blue lines) and isochrones (dashed red lines) from \citet{2015AandA...577A..42B}. Class II YSOs are plotted in black, and class III YSOs in green. Objects 4 and 13 have only approximate $L_*$ and are therefore shown as open circles, with arrows to signify their respective lower and upper limits on $T_{eff}$.}
\label{fig:HR_diagram_Baraffe}
\end{figure*}

\begin{figure*} 
\centering
\includegraphics[scale=0.6]{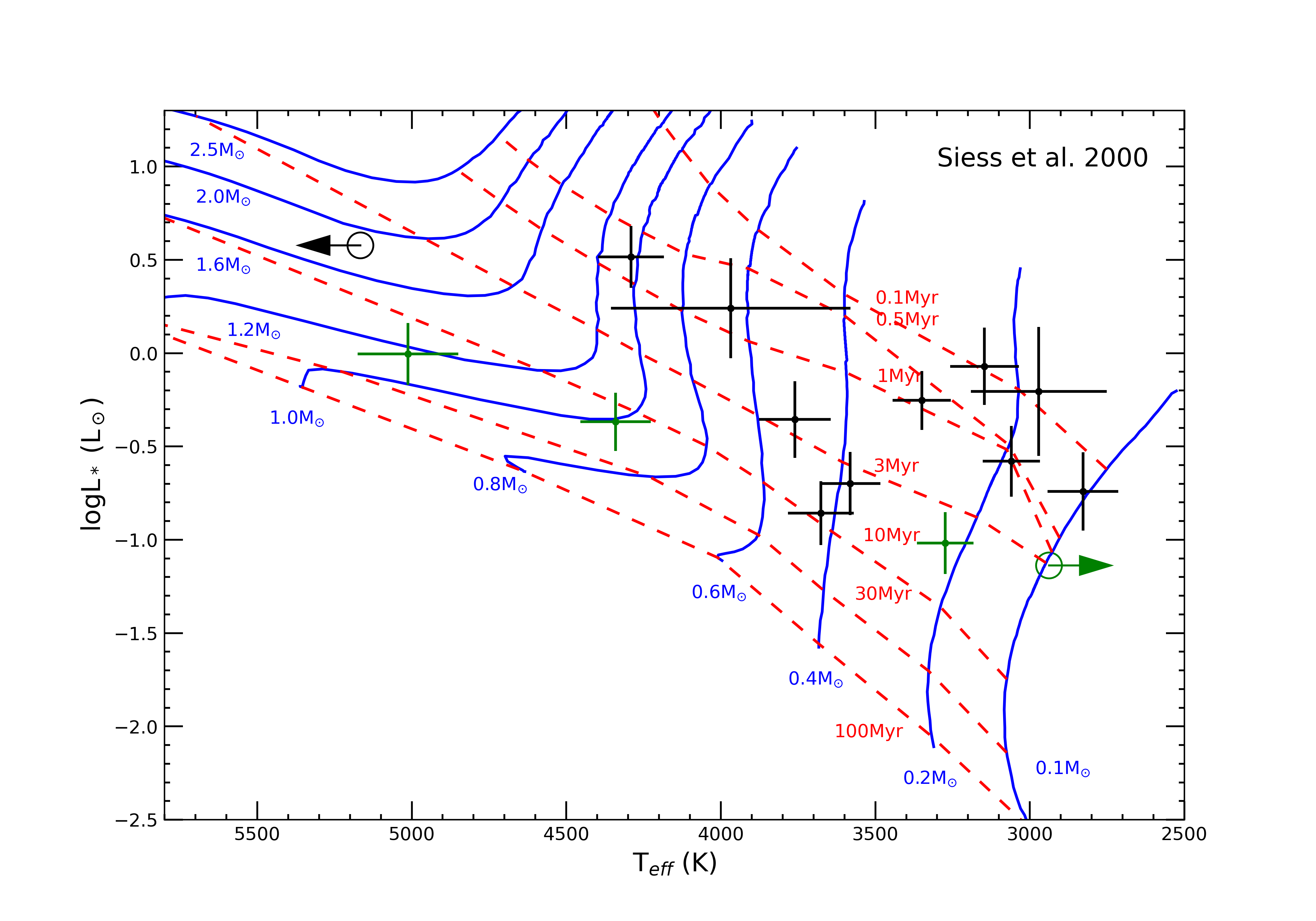}
\caption{Hertzsprung-Russell diagram for the total sample, plotted over evolutionary tracks (blue lines) and isochrones (dashed red lines) from \citet{2000AandA...358..593S}. Plotting conventions are the same as in Figure \ref{fig:HR_diagram_Baraffe}.}
\label{fig:HR_diagram_Siess}
\end{figure*}

We estimate masses for the class II and class III sample by interpolating PMS evolutionary models. Each object has a posterior outputted by \texttt{PyMC} for $T_{eff}$ and $L_{*}$, which can be fed directly into an interpolation of evolutionary models to generate a probability distribution for the stellar mass $M_\star$.
We interpolate two sets of PMS models, from \citet{2015AandA...577A..42B} and \citet{2000AandA...358..593S}. The \citet{2015AandA...577A..42B} models are available for masses $\leq 1.4 M_{\odot}$ and ages $\geq$ 0.5Myr. The \citet{2000AandA...358..593S} models extends to younger ages and to a higher mass of $7M_{\odot}$, but the lowest mass available is $0.1M_{\odot}$.
An HR diagram for all the targets is plotted in Figures \ref{fig:HR_diagram_Baraffe} and \ref{fig:HR_diagram_Siess} with the \citet{2015AandA...577A..42B} and \citet{2000AandA...358..593S} tracks, respectively.

For the \citet{2015AandA...577A..42B} tracks, there are a number of objects (all class II) which lie above the youngest isochrone. We use a nearest interpolation for these objects, as opposed to linear interpolation for the rest. When determining which mass track a data point is nearest to on the HR diagram, we create a metric where 1 dex in log($L_{*}$) scales to 1000K in $T_{eff}$. This scaling matches the ratio of the full ranges in log($L_{*}$) and $T_{eff}$ covered by the \citet{2015AandA...577A..42B} tracks.

Object 4 (EM* LkHA 188), a class II object with a SpT of roughly G8, was briefly discussed in Section \ref{sec:bayesian_results} and considered to have only a lower limit on $T_{eff}$. It is too massive to lie on any track from \citet{2015AandA...577A..42B}. For the tracks from \citet{2000AandA...358..593S} it has an estimated mass of $\approx2M_{\odot}$.
Object 13 (2MASS J03292815+3116285), a class III object with a SpT of roughly M8.5, was also briefly discussed in Section \ref{sec:bayesian_results} and considered to have only an upper limit on $T_{eff}$. This $T_{eff}$ is slightly too low-mass to lie on the $0.1M_{\odot}$ track from \citet{2000AandA...358..593S}, whereas for the tracks from \citet{2015AandA...577A..42B} it has an estimated mass of $\sim0.1M_{\odot}$ or less.

Four of the objects lying above the youngest (0.5 Myr) \citet{2015AandA...577A..42B} isochrone (Objects 10, 11, 13, and 14) are from NGC1333 which, in its entirety, is estimated to have an age of only $<$ 1Myr \citep{2021A&A...650A..43F}. NGC1333 also includes Object 13, which also demonstrates a young age of ($\sim$1Myr) according to \citet{2015AandA...577A..42B} models despite its class III SED. The other class III object from NGC 1333, Object 12, has an older estimated age of 3Myr from \citet{2015AandA...577A..42B} and 6Myr from \citet{2000AandA...358..593S}. On the other hand, Object 15 appears to be the oldest object from NGC 1333 in the sample ($\sim$10Myr), despite being a class II object, raising the possibility that it is subluminous. The rest of the class II sample have ages $\lesssim$10Myr according to both the \citet{2000AandA...358..593S} and \citet{2015AandA...577A..42B} models.

\subsection{Estimating the Mass Accretion Rate Posterior}
\label{sec:estimating_mass_accretion}
With the estimated accretion luminosities and radii from Table \ref{tab:bayesianresults} and masses estimated with the \citet{2015AandA...577A..42B} and \citet{2000AandA...358..593S} tracks, we calculate mass accretion rate $M_{acc}$ with the formula

\begin{equation}
M_{acc} = \left(1-\frac{R_\star}{R_{in}}\right)^{-1}\frac{L_{acc} R_\star}{G M_\star} \approx 1.25 \frac{L_{acc} R_\star}{G M_\star}
\label{eq:mass_accretion}
\end{equation}

where $R_\star$ is the stellar radius (calculated with $L_\star = 4\pi R_\star^2\sigma T_{eff}^4$) and $R_{in}$ is the inner radius of the circumstellar disk \citep{1998ApJ...492..323G}. $R_{in}$ is assumed to be 5$R_\star$ as done in a number of other studies (e.g. \citet{1998ApJ...492..323G}, \citet{1998ApJ...495..385H}, \citet{2014A&A...561A...2A}, \citet{2021A&A...650A..43F}).

Masses and mass accretion rates (or their upper limits) are presented in Table \ref{tab:M_Macc_info} for both the \citet{2015AandA...577A..42B} and \citet{2000AandA...358..593S} models. We find that for objects which have a mass that can be determined from both models, the masses and accretion rates from each model tend to lie within each others' errors.

\begin{deluxetable*}{ll|llll|llll}
\tablecaption{Mass and mass accretion rate information derived from both the \citet{2000AandA...358..593S} (S00) and \citet{2015AandA...577A..42B} (B15) stellar models.\label{tab:M_Macc_info}}
\tablewidth{0pt}
\tablehead{
    \colhead{ID} & \colhead{Class} & \colhead{log($M_{\bigstar}$) (S00)} & \colhead{$\sigma$ log($M_{\bigstar}$)} & \colhead{log($M_{acc}$)} & \colhead{$\sigma$ log($M_{acc}$)} & \colhead{log($M_{\bigstar}$) (B15)} & \colhead{$\sigma$ log($M_{\bigstar}$)} & \colhead{log($M_{acc}$)} & \colhead{$\sigma$ log($M_{acc}$)}\\ \colhead{} & \colhead{} & \colhead{($M_\odot$ yr$^{-1}$)} & \colhead{($M_\odot$ yr$^{-1}$)} & \colhead{($M_\odot$ yr$^{-1}$)} & \colhead{($M_\odot$ yr$^{-1}$)} & \colhead{($M_\odot$ yr$^{-1}$)} & \colhead{($M_\odot$ yr$^{-1}$)} & \colhead{($M_\odot$ yr$^{-1}$)} & \colhead{($M_\odot$ yr$^{-1}$)}
}
\startdata
1     & II    & -0.72         & 0.12      & -7.68          & 0.62         & -0.64          & 0.11        & -7.57           & 0.65           \\
2     & II    & 0.02          & 0.07      & $\leq$-8.45    &              & -0.02          & 0.06        & $\leq$-8.42     &                \\
3     & II    & -0.16         & 0.17      & -6.99          & 0.57         & -0.18          & 0.15        & -6.91           & 0.66           \\
4    & II    & $\approx$0.23 &           & $\approx$-6.93 &              &                &             &                 &                \\
5     & II    & -0.30         & 0.07      & -8.29          & 0.35         & -0.30          & 0.08        & -8.30           & 0.35           \\
6     & III   & -0.02         & 0.04      & $\leq$-8.82    &              & -0.03          & 0.04        & $\leq$-8.81     &                \\
7     & III   & 0.06          & 0.06      & $\leq$-8.94    &              & 0.07           & 0.06        & $\leq$-8.94     &                \\
8    & II    & -0.41         & 0.06      & $\leq$-10.06   &              & -0.39          & 0.08        & $\leq$-10.08    &                \\
9     & II    & -0.73         & 0.08      & -9.2           & 0.19         & -0.75          & 0.06        & -9.19           & 0.17           \\
10     & II    & -0.53         & 0.05      & -9.33          & 0.33         & -0.54          & 0.06        & -9.32           & 0.32           \\
11     & II    & -0.64         & 0.06      & -7.54          & 0.32         & -0.56          & 0.08        & -7.62           & 0.29           \\
12     & III   & -0.65         & 0.08      & $\leq$-13.06   &              & -0.62          & 0.09        & $\leq$-13.08    &                \\
13    & III   &               &           &                &              & $\approx$-0.93 &             & $\approx$-10.04 &                \\
14    & II    & -0.93         & 0.09      & -8.52          & 0.24         & -0.87          & 0.1         & -8.58           & 0.21           \\
15    & II    & -0.36         & 0.07      & -9.63          & 0.28         & -0.3           & 0.08        & -9.69           & 0.3\\         
\enddata
\end{deluxetable*}

\begin{figure} 
\centering
\includegraphics[scale=0.4]{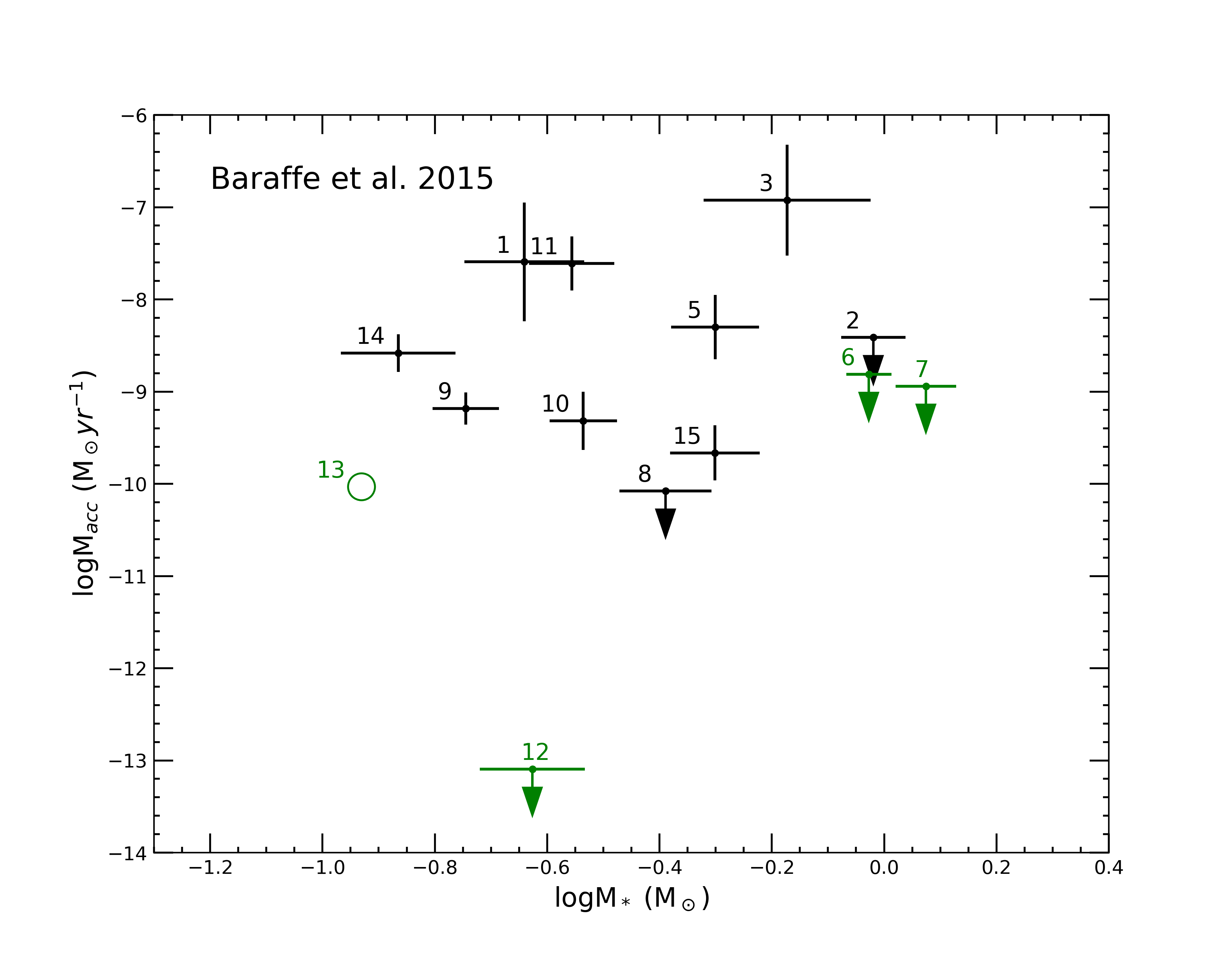}
\caption{Plot of log($M_{acc}$) vs. log($M_{*}$) for our YSO sample, using masses derived from the \citet{2015AandA...577A..42B} evolutionary model. Class II YSOs are shown in black, and class III YSOs in green. Objects are annotated with their indices according to \ref{tab:obs_log}. Upper limits on log($M_{acc}$) are denoted with downward arrows. Object 4 does not fall within the mass range of \citet{2015AandA...577A..42B} and is therefore not included in this plot. Object 13 is represented by a green open circle because its log($M_{*}$) and log($M_{acc}$) are approximate.}
\label{fig:M_vs_Macc_Baraffe}
\end{figure}

\begin{figure} 
\centering
\includegraphics[scale=0.4]{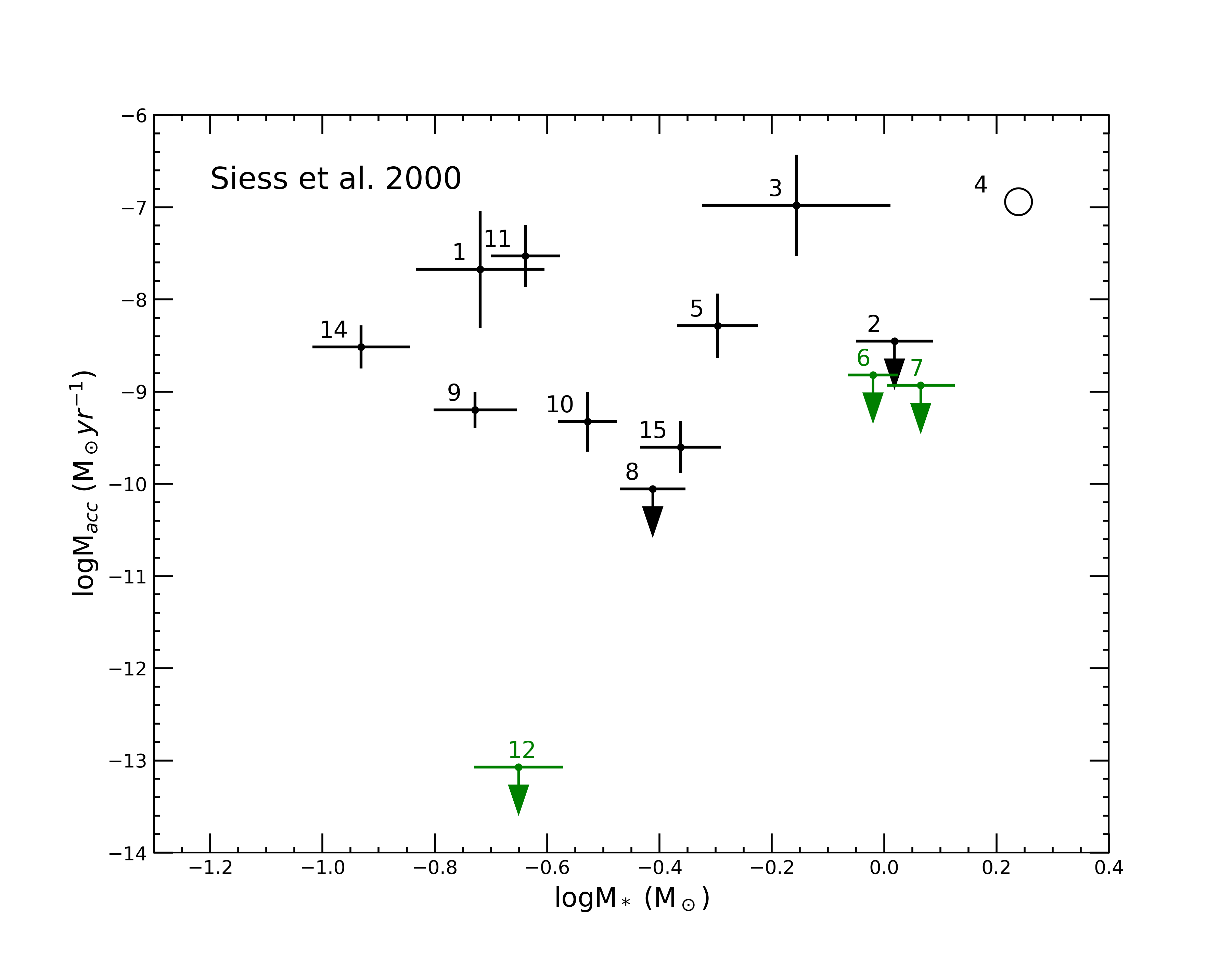}
\caption{Plot of log($M_{acc}$) vs. log($M_{*}$) for our YSO sample, using masses derived from the \citet{2000AandA...358..593S} evolutionary model. Plotting conventions are the same as \ref{fig:M_vs_Macc_Baraffe}. Object 13 does not fall within the mass range of \citet{2000AandA...358..593S} and is not included in this plot. Object 4 is represented by a black open circle because its log($M_{*}$) and log($M_{acc}$) are approximate.}
\label{fig:M_vs_Macc_Siess}
\end{figure}

\section{Emission Line Flux as a Proxy for Measuring Accretion Luminosity}
\label{sec:emission_lines}
Having determined accretion rates for the YSOs using a model fit to the continuum, we now check that they agree with results independently derived from the emission lines.
\citet{2017A&A...600A..20A} studied X-shooter spectra of 81 class II or transition disk YSOs in Lupus, using the 'direct' method on each YSO by fitting the continuum model to each spectrum. They then updated empirical linear relationships between $L_{line}$ and $L_{acc}$ to calibrate the 'indirect' method.
The 3500-5500\r{A} wavelength range of VIRUS is suited for detecting numerous emission lines characteristic to YSOs, most notably the hydrogen Balmer lines beginning with the H$\beta$ line (4860\r{A}) and higher level lines in the series. For each class II YSO, we measured fluxes (or estimated upper limits) of emission lines within the VIRUS wavelength range for which in \citet{2017A&A...600A..20A} there are empirical relations given between line luminosity $L_{line}$ and $L_{acc}$. This includes the hydrogen Balmer lines, from the H$\beta$ line up to H15, the Ca II K line, and various helium lines.

We developed a routine in Python separate from \texttt{nuts-for-ysos}, to measure these emission line fluxes from the VIRUS spectra. For each spectrum, we first subtract an approximate continuum determined via a least-squares polynomial fitting function. For each individual emission line we then estimate a baseline to match the flux of the local continuum with greater precision. In order to determine whether each emission line is detected in a spectrum, we use a 'threshold finding' function from the package \texttt{Specutils}, which looks for deviations from the continuum above a given noise factor \citep{2019ascl.soft02012A}. We deem a line ‘detected’ if the function locates an emission feature above spectrum uncertainty by a factor of at least 2, and within 3\r{A} of the expected centroid.
Finally, we measure the emission line flux using an integrating function combined with a Monte Carlo procedure. We integrate the emission line 50 times with random noise added consistent with spectrum variance at these locations. The average of these 50 integrations is taken to be the final flux measurement, and the standard deviation becomes the associated error.
If an emission line is not detected, we estimate $3\sigma$ upper limits on the flux using the same general approach as \citet{2014A&A...561A...2A}. We take the upper limit to be $3 \times F_{noise} \times \Delta \lambda$ where $F_{noise}$ is the rms flux uncertainty over a 20\r{A} region centered on the expected line centroid, and the line width $\Delta \lambda$ of the undetected line is assumed to be 5\r{A} at most. While this routine is not a part of the main \texttt{nuts-for-ysos} code, we do include it in a separate folder of the GitHub repository for the interested reader. There also exist plenty of other general line flux measurement tools in the community (eg. \texttt{Specutils}), and tools specific to YSOs such as the \texttt{STAR-MELT} package \citep{2021MNRAS.507.3331C}.

The full list of emission lines used, and the corresponding extinction-corrected flux measurements for our current sample, can be found in Tables \ref{tab:fluxes1}, \ref{tab:fluxes2}, and \ref{tab:fluxes3}.
\citet{2017A&A...600A..20A} includes $L_{line}$-$L_{acc}$ relationships for the H$\epsilon$ and Ca II H lines, but we did not attempt to measure fluxes for these emission lines. These two lines are partially blended in X-Shooter spectra, but fully blended with one another in our relatively low resolution VIRUS spectra. Therefore they could not be de-blended as done in \citet{2017A&A...600A..20A}. 
Several line fluxes were omitted within Tables \ref{tab:fluxes1}, \ref{tab:fluxes2}, and \ref{tab:fluxes3} due to data reduction-related issues over narrow wavelength regions. For example, Object 11 (2MASS J03285101+3118184) has no H$\beta$ emission line flux listed in Table \ref{tab:fluxes1}. 
We also measured line fluxes for the class III YSOs, though we found that Objects 6 and \textbf{Object 7} demonstrated very few detected emission lines.
Object 16 (ATO J052.3580+31.4444), for which we were unable to fit a continuum accretion model to its spectrum, displays almost all of the applicable lines as either unmeasurable or in absorption rather than emission. The only exception is the Ca II K line, which is in emission. We therefore did not measure line fluxes for Object 16.

We correct the emission line fluxes for extinction using the $A_V$ derived from the continuum-fitting process in Section \ref{sec:continuum_fit}. We use the same $A_V$ so that the the direct and indirect methods can be consistently compared. The extinction correction takes into account the uncertainty in $A_V$. This is accomplished by repeating the extinction correction over the entire posterior distribution of $A_V$ values. We then take the mean to be the resulting extinction-corrected flux reported in Tables \ref{tab:fluxes1}, \ref{tab:fluxes2}, and \ref{tab:fluxes3}, with the standard deviation taken to be the corresponding uncertainty.

\begin{figure*} 
\centering
\includegraphics[scale=0.49]{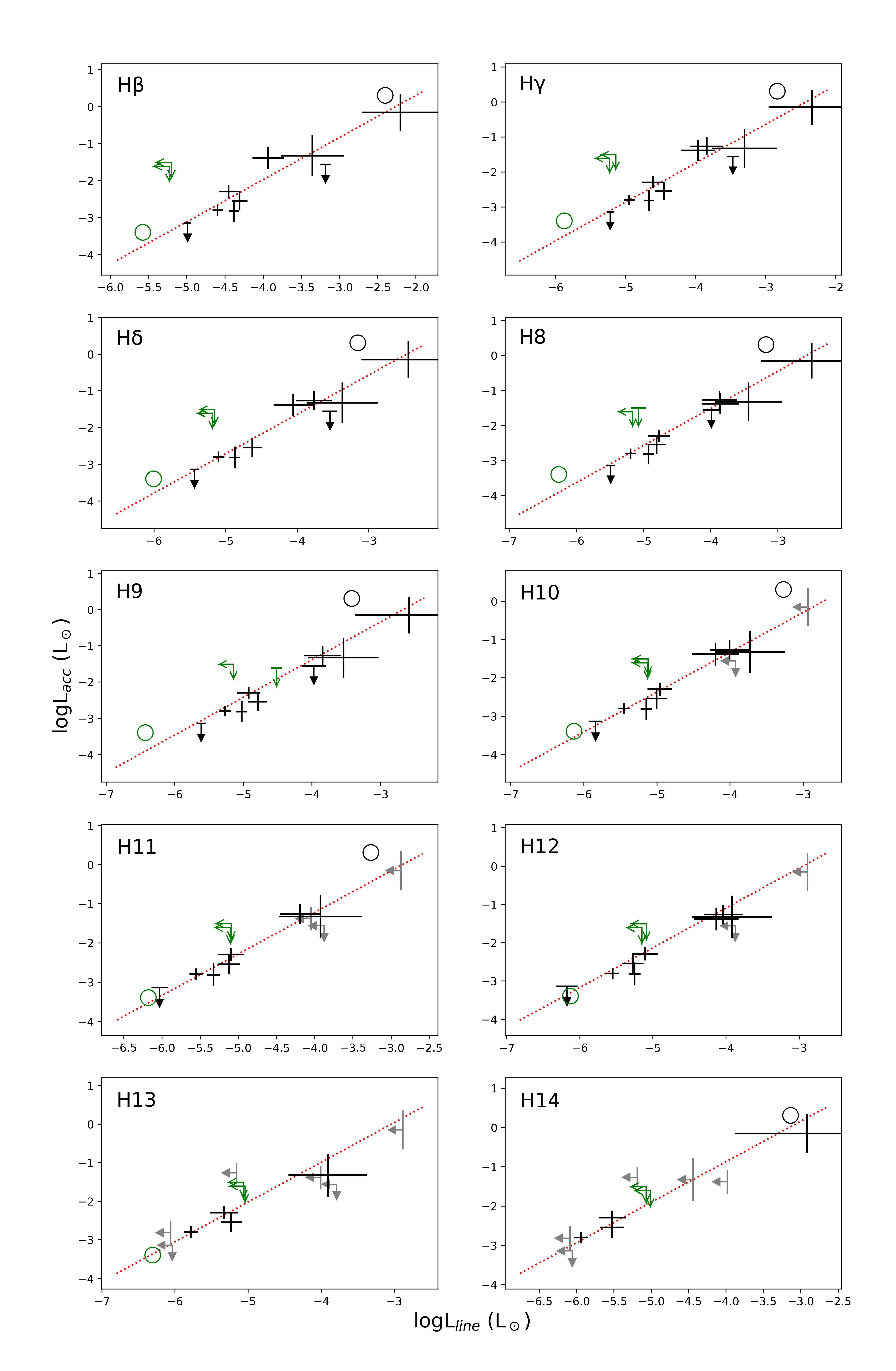}
\caption{Continuum-derived accretion luminosity $L_{acc}$ of both the class II and class III sample plotted against line luminosity for a variety of emission lines studied in \citet{2017A&A...600A..20A}. Each emission line is labeled on the top left of the plot. For the class II YSOs, detected lines are plotted in black, and lines with only upper limits in gray. For class III YSOs, all results are plotted in green with empty arrows. Measurements for objects 4 and 13, which have only an approximate $L_{acc}$, are represented with circles. The red dotted line represents the relation from \citet{2017A&A...600A..20A}, and is not a line fitted to our data.}
\label{fig:line_flux_plots_1}
\end{figure*}

\begin{figure*} 
\centering
\includegraphics[scale=0.50]{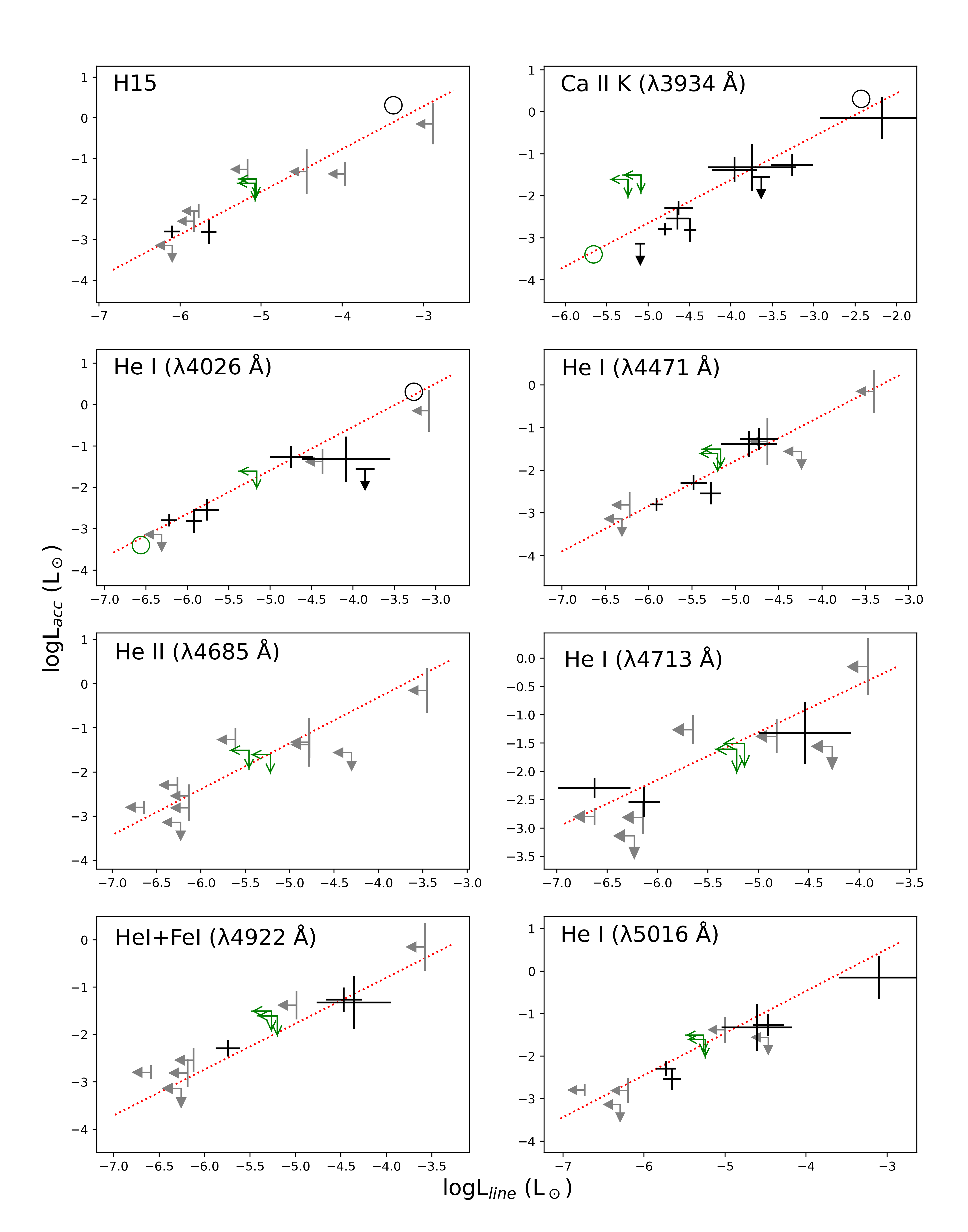}
\caption{Continuum-derived accretion luminosity $L_{acc}$ of both the class II and class III sample plotted against line luminosity for a variety of emission lines studied in \citet{2017A&A...600A..20A}. The plot conventions are the same as Figure \ref{fig:line_flux_plots_1}.}
\label{fig:line_flux_plots_2}
\end{figure*}

\begin{figure} 
\centering
\includegraphics[scale=0.35]{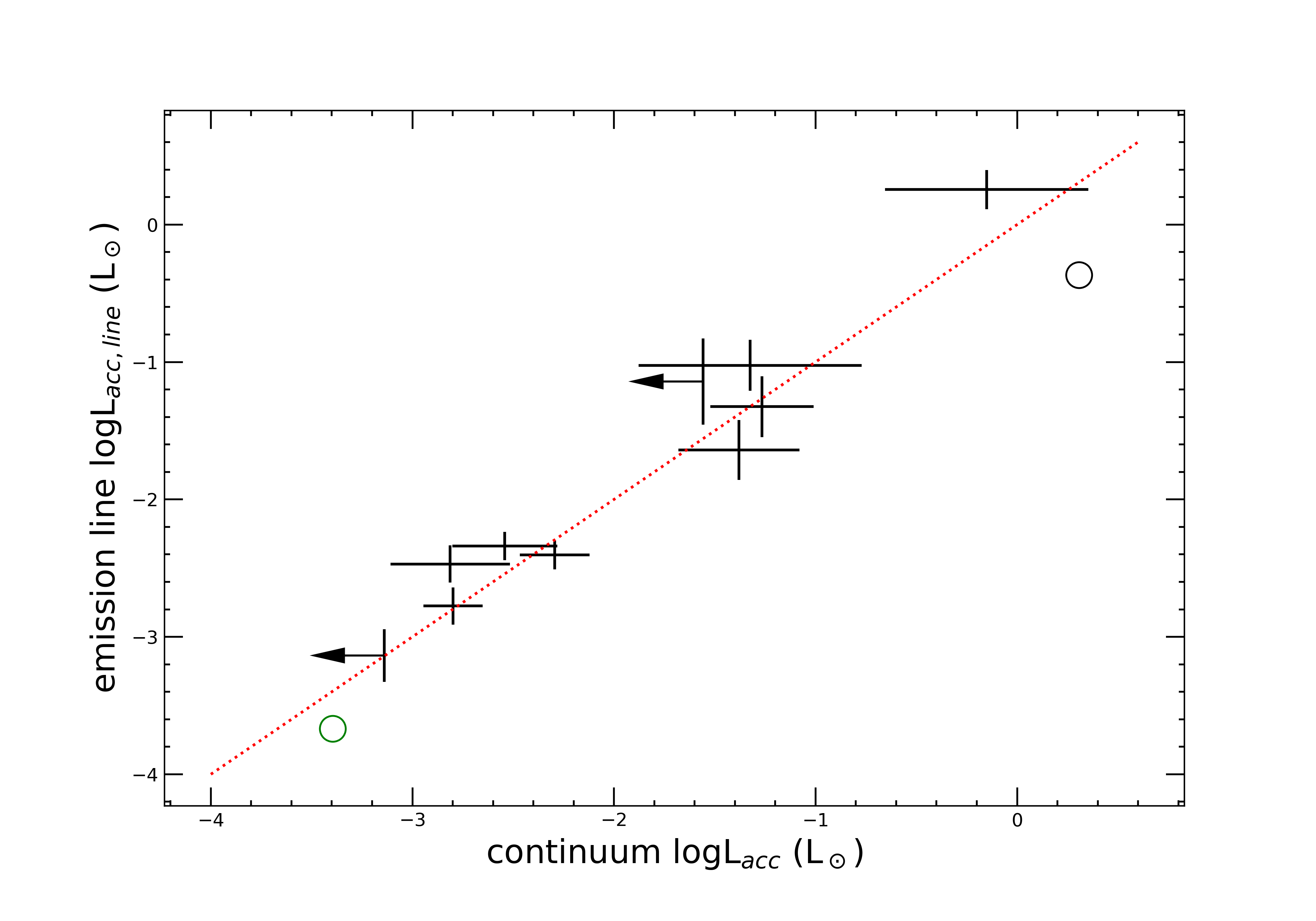}
\caption{The average $L_{acc,line}$ plotted against the continuum-derived $L_{acc}$ for the class II stars (black) as well as Object 13, a class III star (green). $L_{acc,line}$ is derived using emission line luminosities and the relations from \citet{2017A&A...600A..20A}. Objects 4 and 13, having only an approximate $L_{acc}$ and $L_{acc,line}$, are represented with open circles. The red dotted line illustrates the approximate 1:1 correlation.}
\label{fig:emission_vs_continuum_plot}
\end{figure}

The extinction-corrected emission line fluxes are converted into line luminosities using the distance to each YSO. We propagate uncertainty in distance by using the entire posterior distribution in distance of each target (described in Section \ref{sec:priors}) for each calculation. In Figures
\ref{fig:line_flux_plots_1} and \ref{fig:line_flux_plots_2} the resulting line luminosities and uncertainties are plotted against the continuum-derived $L_{acc}$ for our sample. The empirical relationship from \citet{2017A&A...600A..20A} for each line is then plotted as a dotted red line. It is important to note that each line is not fitted to our data; it is completely independently derived by \citet{2017A&A...600A..20A} but shows remarkable agreement with our data. If we apply the relationships from \citet{2017A&A...600A..20A} to each $L_{line}$ of our sample, we then acquire another estimate for the accretion luminosity. 
We then take the average of the estimates from each emission line, deriving a new quantity we call $L_{acc, line}$. We calculated $L_{acc, line}$ for every object except for the class III Objects 6 and 7, since almost all of their line fluxes are only upper limits. Ultimately we find that applying the updated relationships from \citet{2017A&A...600A..20A} to the line luminosities of our class II sample yields $L_{acc, line}$ in good agreement with each $L_{acc}$ derived in Section \ref{sec:calculate_L_Lacc}. Figure \ref{fig:emission_vs_continuum_plot} shows $L_{acc, line}$ plotted against the continuum-derived $L_{acc}$. This plot visually demonstrates an approximate 1:1 correlation, suggesting trustworthiness of our new Bayesian approach. The only exception is the class III Object 12, which has $L_{acc, line}=0.36$ and an upper limit on $L_{acc}\leq-6.28$ that lies well outside of the plot limits. This very likely non-accretor has significant emission lines (unlike Objects 6 and 7), but the lines could possibly be attributed mainly to chromospheric activity.

\begin{figure} 
\centering
\includegraphics[scale=0.5]{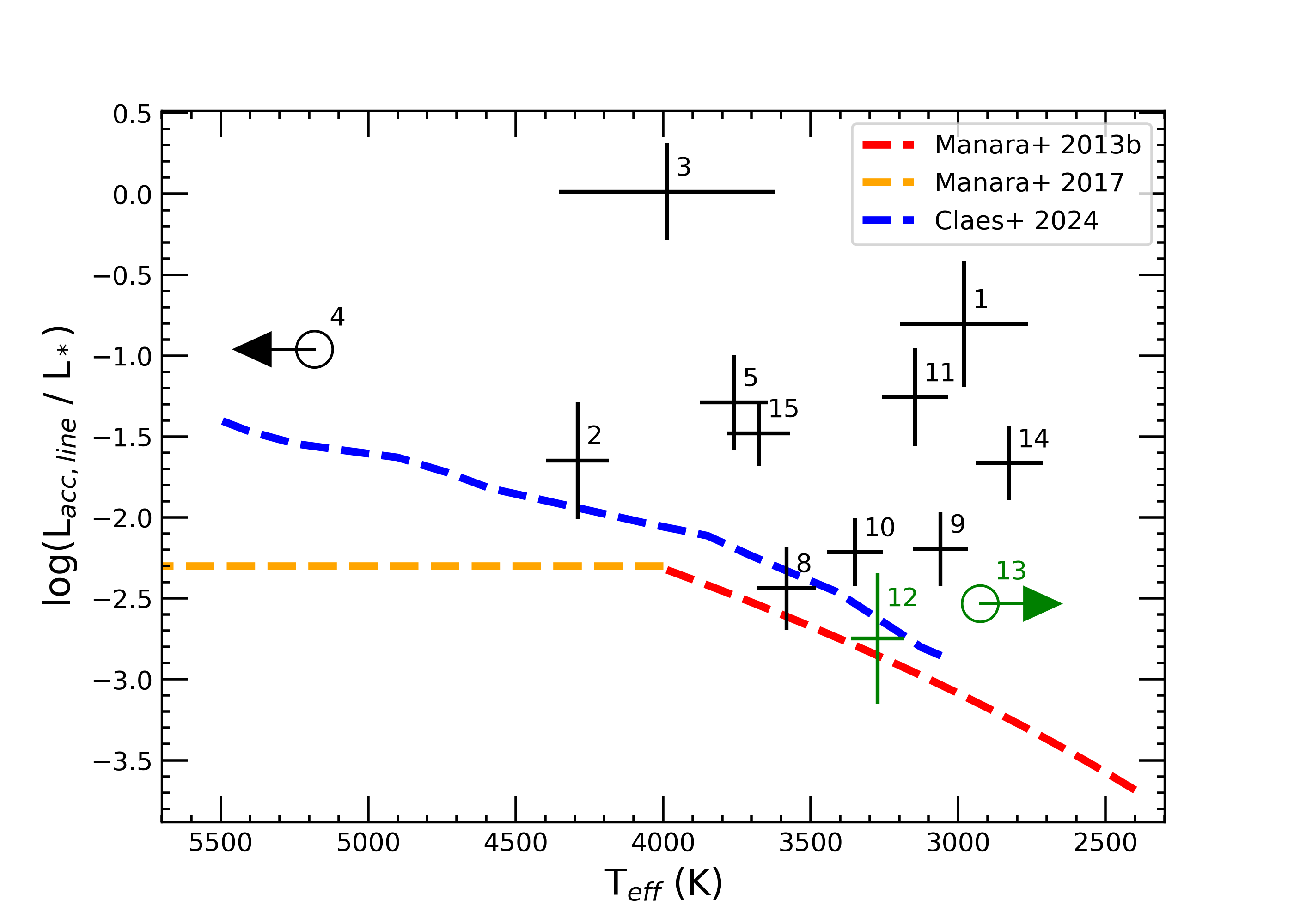}
\caption{The ratio of the average $L_{acc,line}$ with $L_{*}$ for each star for which $L_{acc,line}$ had been measured. The ratio is plotted in a logarithmic scale against $T_{eff}$. Class II YSOs are plotted in black, and the two class III YSOs are plotted in green. The red and orange dashed lines mark the approximate boundary below which chromospheric emission becomes an important contributor to $L_{acc}$, according to \citet{2013AandA...551A.107M} and \citet{2017AandA...605A..86M}. The blue dashed line marks the same but according to \citet{2024A&A...690A.122C}, with the $T_{eff}$ scale converted from \citet{2014ApJ...786...97H} to the one used in this work in Section \ref{sec:the_model}. Objects 4 and 13, having only an approximate $L_{acc,line}$, are represented with open circles.}
\label{fig:chromosphere_limit_plot}
\end{figure}

Emission lines are produced by chromospheric activity in addition to accretion processes. This adds a bias to $L_{acc,line}$ which needs to be assessed. We check the $L_{acc, line}$ of each star relative to $L_*$ to evaluate whether any of the YSOs in our current sample can be considered 'weak accretors' having emission line fluxes dominated by chromospheric activity. Figure \ref{fig:chromosphere_limit_plot} shows the ratio log($L_{acc, line}$/$L_*$) plotted against $T_{eff}$ for each object. The red dashed line shows the level at which chromospheric noise is expected to become important, as derived by \citet{2013AandA...551A.107M} using a set of class III YSOs. The orange dashed line shows the update for $4000K < T_{eff} < 5800K$ objects, made by \citet{2017AandA...605A..86M}. Finally, the blue dashed line shows the chromospheric noise level derived by \citet{2024A&A...690A.122C}, with the $T_{eff}$ scale converted from \citet{2014ApJ...786...97H} to the one used in this work (\citet{2003ApJ...593.1093L} for M-type stars, and \citet{1995ApJS..101..117K} for K-type stars).

Among the two class III YSOs for which $L_{acc, line}$ was calculated, Object 12 is the closest to the chromospheric threshold of \citet{2013AandA...551A.107M}, and is below that of \citet{2024A&A...690A.122C}. This implies that the log($L_{acc, line}$) value for this object is overestimated due to chromospheric line emission. Indeed, the upper limit of log($L_{acc}$) found by fitting to the continuum is lower than the log($L_{acc,line}$) value by approximately 2.5 dex. 
The two class II YSOs with upper limits on the $L_{acc}$ derived in the continuum fit (Objects 2 and 8) also both demonstrate emission line fluxes at or near to the chromospheric level. In Figure \ref{fig:chromosphere_limit_plot}, Object 8 has a log($L_{acc, line}$/$L_*$) which lies under the chromospheric threshold of \citet{2024A&A...690A.122C}. Object 2 has a higher log($L_{acc, line}$/$L_*$) lying slightly above the threshold. These objects are both plausible 'weak accretors' by separate consideration of both their continuum emission and their emission lines.

\begin{deluxetable*}{lllllll}[!ht]
\tablecaption{Extinction corrected fluxes of Balmer lines $\mathrm{H\beta}$ to $\mathrm{H10}$ (excluding $\mathrm{H\epsilon}$) in erg s$^{-1}$ cm$^{-2}$.\label{tab:fluxes1}}
\tablewidth{0pt}
\tablehead{
    \colhead{ID} & \colhead{$f_\mathrm{H\beta}$} & \colhead{$f_\mathrm{H\gamma}$} & \colhead{$f_\mathrm{H\delta}$} & \colhead{$f_\mathrm{H8}$} & \colhead{$f_\mathrm{H9}$} & \colhead{$f_\mathrm{H10}$}
}
\startdata
1          & 2.84($\pm$2.63)e-14 & 3.16($\pm$3.36)e-14 & 2.74($\pm$3.07)e-14 & 2.27($\pm$2.66)e-14 & 1.80($\pm$2.12)e-14 & 1.16($\pm$1.39)e-14 \\
2          & 3.87($\pm$0.67)e-14 & 2.02($\pm$0.41)e-14 & 1.71($\pm$0.40)e-14 & 6.09($\pm$1.92)e-15 & 6.43($\pm$2.44)e-15 & $<$7.08e-15         \\
3          & 3.25($\pm$3.74)e-13 & 2.40($\pm$3.40)e-13 & 1.81($\pm$2.84)e-13 & 1.65($\pm$2.81)e-13 & 1.44($\pm$2.46)e-13 & $<$6.03e-14         \\
4         & 2.27($\pm$0.70)e-13 & 8.54($\pm$3.01)e-14 & 4.06($\pm$1.53)e-14 & 3.78($\pm$1.49)e-14 & 2.19($\pm$0.88)e-14 & 3.09($\pm$1.24)e-14 \\
5          & 2.19($\pm$1.00)e-14 & 2.07($\pm$1.12)e-14 & 1.66($\pm$1.03)e-14 & 2.61($\pm$1.63)e-14 & ...                 & 1.16($\pm$0.87)e-14 \\
6          & $<$1.08e-15         & $<$1.25e-15         & $<$1.21e-15         & 1.44($\pm$0.38)e-15 & $<$1.23e-15         & $<$1.30e-15         \\
7          & $<$1.58e-15         & $<$1.59e-15         & $<$1.73e-15         & $<$1.84e-15         & 8.07($\pm$1.32)e-15 & $<$1.98e-15         \\
8         & 2.92($\pm$0.31)e-15 & 1.72($\pm$0.21)e-15 & 1.04($\pm$0.14)e-15 & 9.29($\pm$1.40)e-16 & 6.92($\pm$1.02)e-16 & 4.12($\pm$0.82)e-16 \\
9         & 1.02($\pm$0.15)e-14 & 4.54($\pm$0.78)e-15 & 3.20($\pm$0.59)e-15 & 2.60($\pm$0.50)e-15 & 2.19($\pm$0.43)e-15 & 1.43($\pm$0.28)e-15 \\
10          & 1.62($\pm$0.22)e-14 & 8.58($\pm$1.36)e-15 & 5.28($\pm$0.90)e-15 & 4.64($\pm$0.82)e-15 & 3.73($\pm$0.67)e-15 & 2.79($\pm$0.51)e-15 \\
11          & ...                 & 4.98($\pm$2.67)e-14 & 6.01($\pm$3.43)e-14 & 4.77($\pm$2.86)e-14 & 4.92($\pm$2.98)e-14 & 3.47($\pm$2.12)e-14 \\
12          & 9.90($\pm$0.89)e-16 & 2.47($\pm$0.66)e-16 & 2.38($\pm$0.35)e-16 & 1.13($\pm$0.25)e-16 & $<$1.11e-16         & $<$1.07e-16         \\
13         & 9.65($\pm$3.68)e-16 & 4.86($\pm$2.24)e-16 & 3.55($\pm$1.82)e-16 & 2.01($\pm$1.32)e-16 & 1.35($\pm$0.86)e-16 & 2.68($\pm$1.63)e-16 \\
14         & 1.18($\pm$0.35)e-14 & 8.25($\pm$2.78)e-15 & ...                 & 5.62($\pm$2.11)e-15 & 3.99($\pm$1.53)e-15 & 3.66($\pm$1.40)e-15 \\
15         & 1.91($\pm$0.46)e-14 & 1.38($\pm$0.39)e-14 & 9.28($\pm$2.78)e-15 & 6.16($\pm$1.94)e-15 & 6.37($\pm$2.04)e-15 & 3.83($\pm$1.25)e-15\\
\enddata
\end{deluxetable*}

\begin{deluxetable*}{lllllll}[!ht]
\tablecaption{Extinction corrected fluxes of Balmer lines $\mathrm{H11}$ to $\mathrm{H15}$ and $\mathrm{Ca\: II\: \lambda\: 3934}$ in erg s$^{-1}$ cm$^{-2}$.\label{tab:fluxes2}}
\tablewidth{0pt}
\tablehead{
    \colhead{ID} & \colhead{$f_\mathrm{H11}$} & \colhead{$f_\mathrm{H12}$} & \colhead{$f_\mathrm{H13}$} & \colhead{$f_\mathrm{H14}$} & \colhead{$f_\mathrm{H15}$} & \colhead{$f_{\mathrm{Ca\: II\: \lambda\: 3934}}$}
}
\startdata
1          & 7.71($\pm$9.30)e-15 & 7.86($\pm$9.86)e-15 & 7.87($\pm$9.69)e-15 & $<$2.31e-15          & $<$2.36e-15         & 1.15($\pm$1.34)e-14 \\
2          & $<$7.87e-15         & $<$7.92e-15         & $<$9.70e-15         & ...                  & ...                 & 1.37($\pm$0.35)e-14 \\
3          & $<$7.01e-14         & $<$6.75e-14         & $<$6.80e-14         & 6.37($\pm$14.16)e-14 & $<$6.85e-14         & 3.55($\pm$5.83)e-13 \\
4         & 3.12($\pm$1.26)e-14 & $<$5.22e-15         & $<$5.31e-15         & 4.20($\pm$1.70)e-14  & 2.49($\pm$1.02)e-14 & 2.16($\pm$0.84)e-13 \\
5          & $<$1.67e-14         & 1.37($\pm$0.96)e-14 & $<$1.83e-14         & $<$1.95e-14          & $<$2.02e-14         & 2.07($\pm$1.29)e-14 \\
6          & $<$1.39e-15         & $<$1.40e-15         & $<$1.49e-15         & $<$1.47e-15          & 6.00($\pm$1.33)e-15 & 5.66($\pm$1.18)e-15 \\
7          & $<$2.10e-15         & $<$1.88e-15         & $<$2.42e-15         & $<$2.58e-15          & $<$2.25e-15         & $<$1.53e-15         \\
8         & 2.66($\pm$0.64)e-16 & 1.89($\pm$0.63)e-16 & $<$2.60e-16         & $<$2.48e-16          & $<$2.28e-16         & 2.29($\pm$0.31)e-15 \\
9         & 1.13($\pm$0.23)e-15 & 1.14($\pm$0.23)e-15 & 6.64($\pm$1.40)e-16 & 4.62($\pm$0.96)e-16  & 3.23($\pm$0.71)e-16 & 6.45($\pm$1.23)e-15 \\
10          & 1.86($\pm$0.35)e-15 & 2.21($\pm$0.42)e-15 & $<$3.41e-16         & $<$3.20e-16          & 8.86($\pm$1.95)e-16 & 1.26($\pm$0.22)e-14 \\
11          & 2.24($\pm$1.38)e-14 & 3.17($\pm$1.96)e-14 & $<$2.42e-15         & $<$2.27e-15          & $<$2.38e-15         & 1.89($\pm$1.12)e-13 \\
12          & 2.09($\pm$0.41)e-16 & 1.23($\pm$0.29)e-16 & $<$1.27e-16         & 1.41($\pm$0.30)e-16  & 1.20($\pm$0.37)e-16 & 7.09($\pm$0.84)e-16 \\
13         & 2.38($\pm$1.68)e-16 & 2.65($\pm$1.81)e-16 & 1.75($\pm$1.22)e-16 & $<$2.19e-16          & $<$2.27e-16         & 8.14($\pm$4.32)e-16 \\
14         & 2.66($\pm$1.03)e-15 & 2.61($\pm$1.04)e-15 & 1.57($\pm$0.66)e-15 & 9.82($\pm$4.07)e-16  & $<$5.65e-16         & 7.90($\pm$2.93)e-15 \\
15         & 2.96($\pm$0.98)e-15 & 2.07($\pm$0.70)e-15 & 2.27($\pm$0.76)e-15 & 1.16($\pm$0.41)e-15  & $<$5.82e-16         & 8.93($\pm$2.79)e-15\\
\enddata
\end{deluxetable*}

\begin{deluxetable*}{lllllll}[!ht]
\tablecaption{Extinction corrected fluxes of helium lines in erg s$^{-1}$ cm$^{-2}$.\label{tab:fluxes3}}
\tablewidth{0pt}
\tablehead{
    \colhead{ID} & \colhead{$f_\mathrm{He\: I\: \lambda\: 4026}$} & \colhead{$f_\mathrm{He\: I\: \lambda\: 4470}$} & \colhead{$f_\mathrm{He\: II\: \lambda\: 4686}$} & \colhead{$f_\mathrm{He\: I\: \lambda \:4712}$} & \colhead{$f_\mathrm{He\: I+Fe\: I\: \lambda \:4922}$} & \colhead{$f_\mathrm{He\: I\: \lambda\: 5016}$}
}
\startdata
1          & 5.56($\pm$6.46)e-15 & $<$1.52e-15         & $<$1.07e-15         & 1.86($\pm$1.94)e-15 & 2.73($\pm$2.64)e-15 & 1.64($\pm$1.60)e-15 \\
2          & 8.36($\pm$2.20)e-15 & $<$3.44e-15         & $<$2.96e-15         & $<$3.22e-15         & ...                 & $<$2.03e-15         \\
3          & $<$4.33e-14         & $<$2.07e-14         & $<$1.82e-14         & $<$6.37e-15         & $<$1.38e-14         & 4.22($\pm$4.69)e-14 \\
4         & 3.13($\pm$1.20)e-14 & $<$3.51e-15         & $<$3.40e-15         & $<$3.40e-15         & $<$3.49e-15         & $<$3.20e-15         \\
5          & $<$8.00e-15         & 2.77($\pm$2.02)e-15 & $<$3.11e-15         & $<$2.84e-15         & $<$1.92e-15         & $<$1.85e-15         \\
6          & ...                 & $<$1.17e-15         & 2.40($\pm$0.55)e-15 & $<$1.26e-15         & $<$9.35e-16         & $<$9.35e-16         \\
7          & $<$1.84e-15         & $<$1.67e-15         & $<$1.61e-15         & $<$1.63e-15         & $<$1.68e-15         & $<$1.52e-15         \\
8         & $<$1.39e-16         & $<$1.40e-16         & $<$1.68e-16         & $<$1.67e-16         & $<$1.57e-16         & $<$1.45e-16         \\
9         & 2.43($\pm$0.53)e-16 & 4.93($\pm$0.85)e-16 & $<$9.17e-17         & $<$9.52e-17         & $<$1.04e-16         & $<$7.44e-17         \\
10          & 4.76($\pm$1.11)e-16 & $<$2.36e-16         & $<$2.87e-16         & $<$2.83e-16         & $<$2.56e-16         & $<$2.48e-16         \\
11          & 6.30($\pm$3.72)e-15 & 6.50($\pm$3.38)e-15 & $<$8.57e-16         & $<$7.90e-16         & 1.19($\pm$0.54)e-14 & 1.19($\pm$0.53)e-14 \\
12          & $<$1.03e-16         & $<$7.96e-17         & $<$8.59e-17         & 1.54($\pm$0.22)e-16 & $<$8.36e-17         & $<$8.05e-17         \\
13         & 1.04($\pm$0.73)e-16 & $<$9.10e-17         & $<$1.08e-16         & $<$1.04e-16         & $<$7.85e-17         & $<$7.25e-17         \\
14         & ...                 & 1.10($\pm$0.36)e-15 & $<$1.81e-16         & 7.95($\pm$6.39)e-17 & 6.02($\pm$1.79)e-16 & 6.21($\pm$1.78)e-16 \\
15         & 6.72($\pm$2.32)e-16 & 2.03($\pm$0.56)e-15 & $<$2.85e-16         & 2.86($\pm$1.06)e-16 & $<$2.96e-16         & 8.63($\pm$2.12)e-16\\
\enddata
\end{deluxetable*}

\section{Discussion}
\label{sec:discussion}

\subsection{Parameter Probabilities and Correlations}
\label{sec:param_correlations}
We have demonstrated an improved method to fitting an accretion model to YSO spectra by using a Bayesian framework. One main motivation in using a Bayesian approach is to understand hidden correlations among the parameters. From our small sample of YSOs we have noticed several trends in behavior. 

\begin{figure*} 
\centering
\includegraphics[scale=0.6]{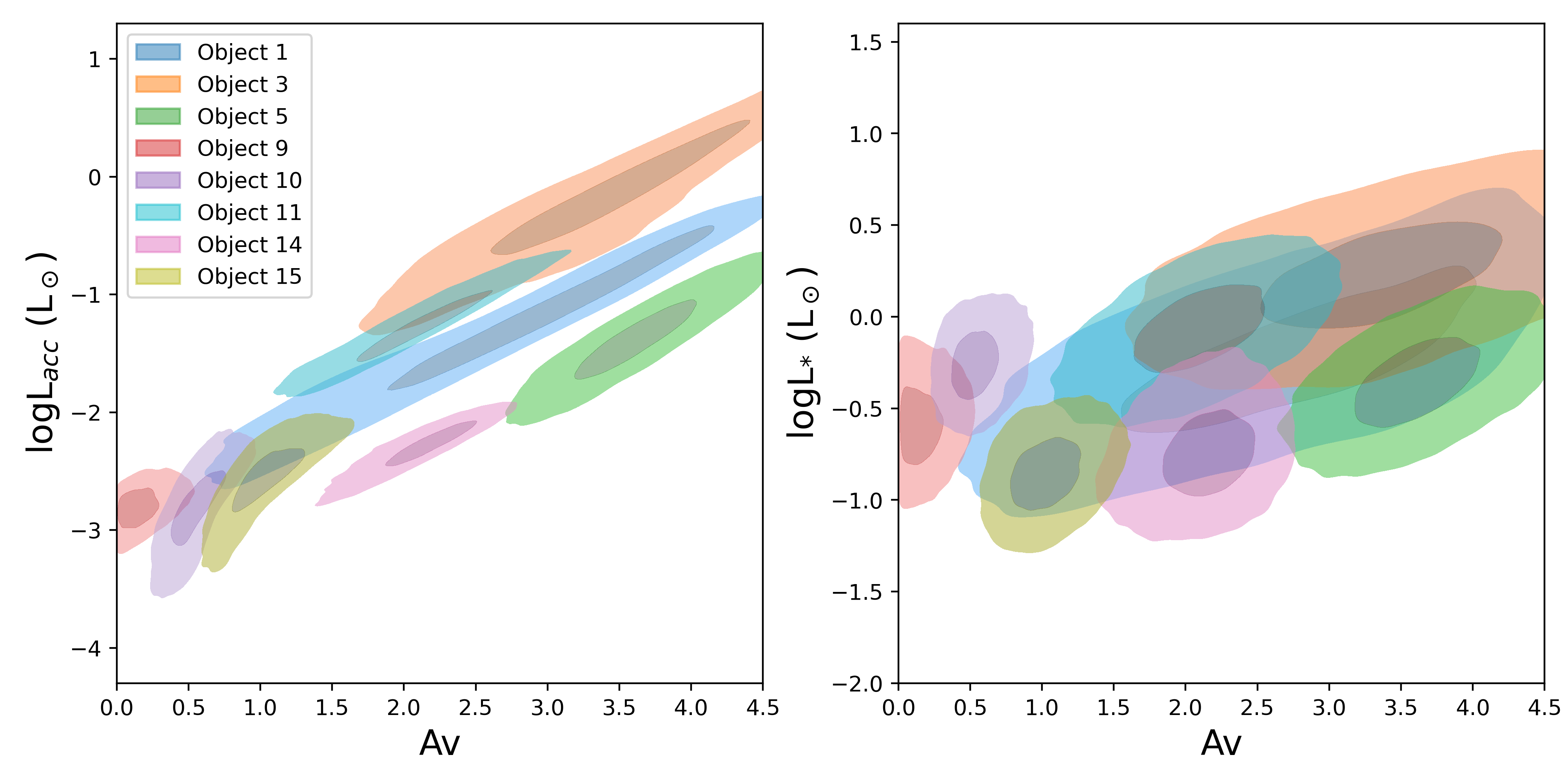}
\caption{The overlaid posteriors for log($L_{acc}$) and log($L_{*}$) versus $A_{V}$, for every object in our sample in which log($L_{acc}$) is not an upper limit or is approximated. A strong correlation can be seen between $A_{V}$ and log($L_{acc}$) in particular.}
\label{fig:L_Lacc_Av}
\end{figure*}

The quantity log($L_{acc}$) often exhibits correlation with $A_{V}$. Uncertainties in $A_{V}$ can therefore have a particularly strong influence on the determination of log($L_{acc}$). The left-hand side of Figure \ref{fig:L_Lacc_Av} shows the overlaid posteriors of log($L_{acc}$) versus $A_{V}$ for every object in the class II sample for which log($L_{acc}$) is not an upper limit. Within individual posteriors, a strong correlation is often evident between $A_{V}$ and log($L_{acc}$). Moreover, there is a scattered trend that the objects with a higher median log($L_{acc}$) value tend to have the higher median $A_{V}$ values. This degeneracy between the intrinsic luminosity of the accretion slab and the extinction could be partially improved by analyzing YSO spectra with a wider wavelength coverage. \citet{2013A&A...558A.114M} pointed out that if one fits the model to a broader wavelength range (such as that of X-Shooter) or even just includes more photometric data from other bands in addition to the spectroscopic data, the reddening effect of extinction and the blue enhancement by the accretion slab may be better disentangled, although never completely.
We also find that $K_{phot}$ and $A_{V}$ typically appear correlated. This makes sense since $K_{phot}$ and $A_{V}$ both pertain directly to the scaling of the class III photospheric template and can be adjusted somewhat interchangeably, resulting in degeneracy. This can introduce a degeneracy of log($L_{*}$) with $A_{V}$ as well, although we find that the correlation between these two is less pronounced, as shown in the right-hand side of Figure \ref{fig:L_Lacc_Av}.

Examining the corner plots in in Appendix \ref{sec:result_plots}, there is often a notable lack of correlation between log($L_{acc}$) and the parameters that make up the slab portion of the model: $T_{slab}$, $n_e$, and $\tau_0$. This is true both within individual posteriors and also over the entire sample; for example, objects with a high median $T_{slab}$ value do not demonstrate higher or lower median $L_{acc}$ values than other objects. 
As previously noted in \citet{2012A&A...548A..56R}, there are degeneracies in the slab model, in which tradeoffs of different slab parameters are able to produce similar accretion luminosities. \citet{2012A&A...548A..56R} found that reasonable slab model fits to a YSO spectrum yielded $L_{acc}$ estimates within only $\sim$ 10\% of each other, even over a wide variety of slab parameters. \citet{2014PhDT.......477M} includes a full discussion of the interplay of these three parameters, and how they each affect the shape of the slab model. The Bayesian method shows us that these are nuisance parameters that can be marginalized over. Instead, only the overall scaling of the slab model $K_{slab}$ tends to demonstrate some correlation with log($L_{acc}$) in the corner plots, alongside the extinction $A_{V}$.

Within our current sample we find only one instance in which the model fit shows clear preference for a certain $T_{slab}$ value between the upper and lower bounds: Object 13 has a $T_{slab}$ posterior that peaks at $\sim$7000K with a symmetric distribution. In other cases, we mainly find broad probability distributions for $T_{slab}$ that are either relatively flat or skewed to low temperatures. The least constrained parameter tends to be $n_e$. We find that $n_e$ often does not display a strong peak in probability at any one specific value between the imposed lower and upper bounds of $10^{10}$ cm$^{-3}$ and $10^{16}$ cm$^{-3}$. In most cases the distribution for log($n_e$) gradually rises to peak probability at $\sim$10 or $\sim$16, or remains nearly flat. 
For $\tau_0$, we find only two instances where the model settles on a specific value between the lower and upper bounds of 0.01 and 5.0. These are Objects 11 and 14, for which the most probable $\tau_0$ is $\sim$ 3.5 and $\sim$ 4.0 respectively. We more often find $\tau_0$ skewed to the lowest or highest values possible, and otherwise a flat distribution or somewhat bimodal distribution between both extremes.

Interestingly, we notice that there is sometimes a correlation between $T_{slab}$ and log($n_e$) for $T_{slab}$ values below $\sim$10000K. This can be seen most clearly in the corner plots for Objects 1, 5, 9, 10, 11, and 15.
One possible reason for the correlation is the way that $T_{slab}$ and $n_e$ jointly affect the height of the Balmer jump (ie. the ratio of flux at 3600\r{A} and 4000\r{A}). \citet{2014PhDT.......477M} shows (in Figure 2.4 of Section 2.3) that a specific value for the Balmer jump can be achieved for higher and higher $T_{slab}$, as long as $n_e$ is also increased. However, the ratio gradually becomes independent of $n_e$ at high $T_{slab}$ values. A similar phenomenon occurs when  \citet{2014PhDT.......477M} examines the ratio of Balmer continuum emission to Paschen continuum emission (Figure 2.5). 

Three of the lowest-mass class II stars (Objects 1, 11, and 14) demonstrate both high $T_{slab}$ and $\tau_0$ values, and display an apparent anticorrelation between these two parameters for $T_{slab}>$ 10000K. In their work, \citet{2014PhDT.......477M} shows in Figures 2.4, 2.5, and 2.6 how either increasing $\tau_0$ or increasing $T_{slab}$ can suppress the strength of features such as the height of the Balmer jump and the slope of the Balmer continuum. This trade-off may be one possible reason for the anticorrelation. Such a phenomenon may be particularly noticeable for these three objects because the Balmer jump is a fairly dominant component in their spectra, and because their $T_{slab}$ and $\tau_0$ posterior distributions happen to both prefer higher values.

In Section 2.2 of their work, \citet{2014PhDT.......477M} discusses the trustworthiness of the slab model for the sole purpose of determining $L_{acc}$. As we alluded in Section \ref{sec:the_model}, the slab model is based off an older paradigm for accreting YSOs, where particles cross a boundary layer between the inner circumstellar disk and the star (e.g. \citet{1974MNRAS.168..603L}). This is different from the presently more widely-accepted magnetospheric accretion model. Therefore, the three parameters $T_{slab}$, $n_e$, and $\tau_0$ of the slab model should not be interpreted as physical quantities due to their lack of basis in reality. According to \citet{2014PhDT.......477M} however, there are a number of reasons why the slab model is still useful besides its simplicity. One reason is that more complex 'shock models' based in magnetospheric accretion (e.g. that of \citet{1998ApJ...509..802C}) have not been able to fully match the veiling of spectra for accreting class II stars at long wavelengths \citep{2013ApJ...767..112I}. However, the slab model has been able to match observations (e.g. \citet{1993AJ....106.2024V, 2008ApJ...681..594H}). Moreover, for smaller wavelengths of $\lesssim 3000$\r{A}, using either model has yielded similar bolometric correction factors for the accretion luminosity $L_{acc}$ (e.g. \citet{1998ApJ...509..802C, 2008ApJ...681..594H, 2014PhDT.......477M}). Therefore the slab model, which is simple to implement with only a few parameters, is often still used. Nonetheless, it is sensible to treat the model with some caution given its dubious physical origins. Continued tests on YSO spectra at a variety of photometric wavebands can further verify the usefulness of the slab model for calculating $L_{acc}$.

\subsection{Comparison to Specific SFRs}
We have studied only a small sample of YSOs that do not all belong to one star-forming region. We therefore refrain from making physical interpretations of our results for $M_{acc}$. Instead, our study offers a prototype of what can be accomplished with optical/UVB spectroscopy of YSOs, combined with our Bayesian framework for fitting the accretion model, which can be applied to larger surveys in the future. With our small sample we can at least note that our results for the $L_{*}$-$L_{acc}$ plot (Figure \ref{fig:L_vs_Lacc}) and $M_{*}$-$M_{acc}$ plots (Figures \ref{fig:M_vs_Macc_Baraffe} and \ref{fig:M_vs_Macc_Siess}) occupy a similar range as previous studies of various star-forming regions. We briefly compare to previous results of class II stars from Lupus \citep{2017A&A...600A..20A}, Chamaeleon I \citep{2016A&A...585A.136M}, and NGC1333 \citep{2021A&A...650A..43F}.

\begin{figure*} 
\centering
\includegraphics[scale=0.6]{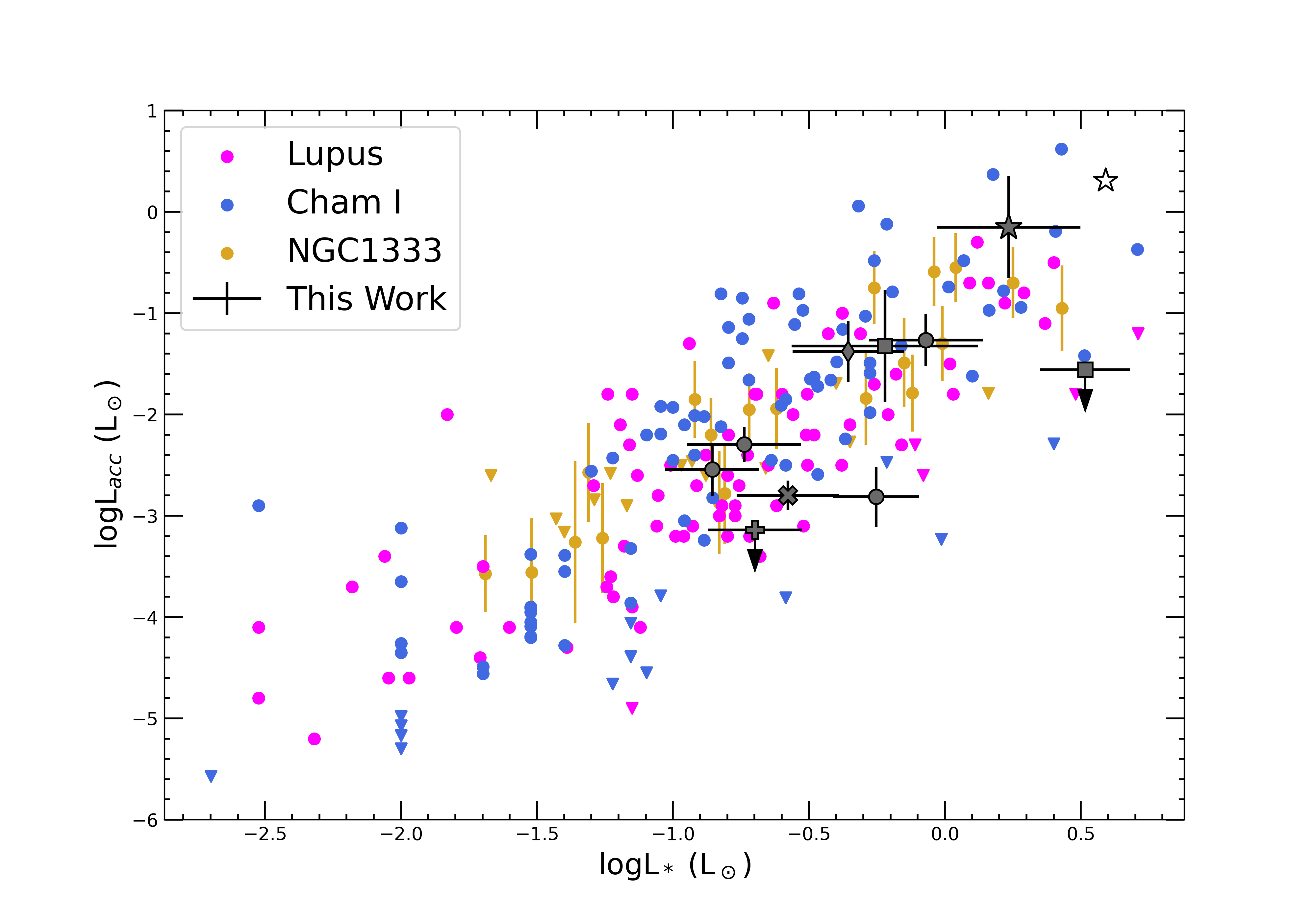}
\caption{Plot of the continuum-derived log($L_{acc}$) vs. log($L_{*}$) for the class II targets of our sample, compared to the results for class II YSOs from Lupus \citep{2017A&A...600A..20A}, Chamaeleon I \citep{2016A&A...585A.136M}, and NGC1333 \citep{2021A&A...650A..43F}. These three sets are plotted in magenta, blue, and orange, respectively, with upper limits represented by triangles. The values from this work are plotted in black and grey, with different shapes for different regions: circles (NGC1333), stars (NGC7000), squares (IC5146), a diamond (Serpens), a plus sign (Orion), and an X-symbol (Perseus). Object 4, having only an approximate $L_{acc}$, is represented with an unfilled star.}
\label{fig:Lacc_comparison}
\end{figure*}

Figure \ref{fig:Lacc_comparison} shows that the $L_{*}$ and $L_{acc}$ of our class II sample follow a loose linear relationship in agreement with these three star-forming regions. A few objects in our current sample (Objects 8, 9, and 10) have fairly low $L_{acc}$/$L_{*}$ compared to most objects, but they are still certainly plausible, as several stars from Lupus and Chamaeleon I demonstrate similarly low $L_{acc}$/$L_{*}$ values or upper limits. 

Figure \ref{fig:Macc_comparison} shows how the $M_{*}$-$M_{acc}$ relationship from our current sample of 15 YSOs compares to Lupus, Chamaeleon I, and NGC1333. Past studies have demonstrated a loose correlation between $M_{*}$ and $M_{acc}$. Given our small sample size, we do not attempt to fit a line to log($M_{*}$)-log($M_{acc}$) plot. However, we note that our sample, although scattered, occupies a space similar to these three star-forming regions. 
The highest accretion rates in our current sample are attained by objects 3 and 4 (respectively log($M_{acc}$) = -6.99 and  log($M_{acc}$) $\approx$ -6.93 in $M_\odot$/yr$^{-1}$, via the \citet{2000AandA...358..593S} evolutionary model). Although high, these accretion rates are similar to several objects from Chamaeleon I.
We also observe a relatively high $M_{acc}$/$M_{*}$ ratio for objects 1, 11, and 14. Though they lie along the high-$M_{acc}$ border of the $M_{*}$-$M_{acc}$ relationships for Lupus, Chamaeleon I, and NGC1333, they are not implausibly high. Objects 1 and 11 are both from NGC1333 and are within expectations when compared to similar-mass class II YSOs studied by \citet{2021A&A...650A..43F}.

\begin{figure} 
\centering
\includegraphics[scale=0.55]{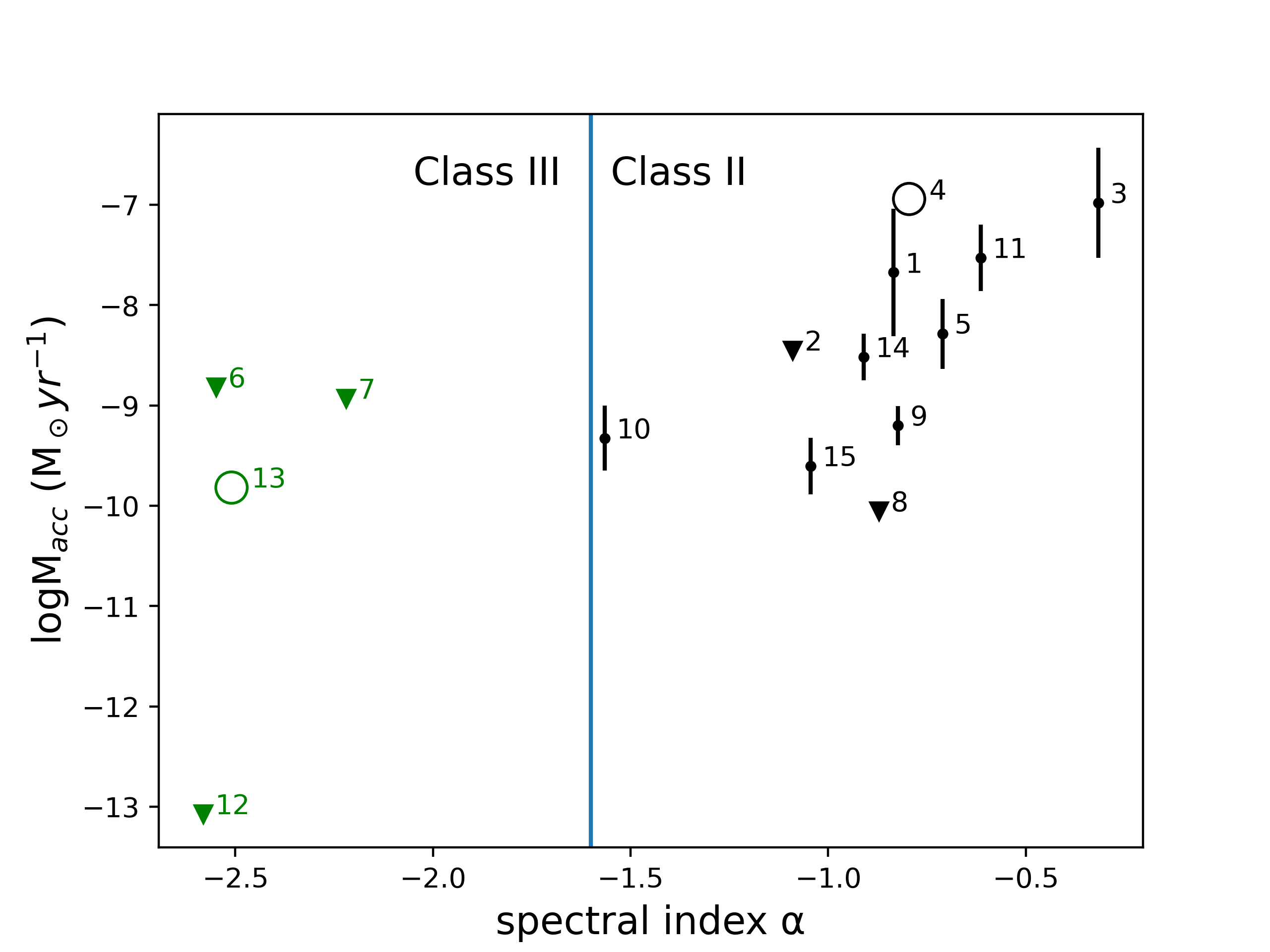}
\caption{Plot of log($M_{acc}$) vs. the spectral index $\alpha$ used for YSO classification in Section \ref{sec:sample selection}. Class II YSOs are plotted in black, and Class III YSOs are plotted in green. Upper limits in $M_{acc}$ are denoted by downward triangles. Objects 4 and 13, having only an approximate $M_{acc}$, are represented with open circles. The blue line is the dividing line between Class II and Class III in \citet{1994ApJ...434..614G}.}
\label{fig:alpha_accretion_plot}
\end{figure}

\begin{figure*} 
\centering
\includegraphics[scale=0.6]{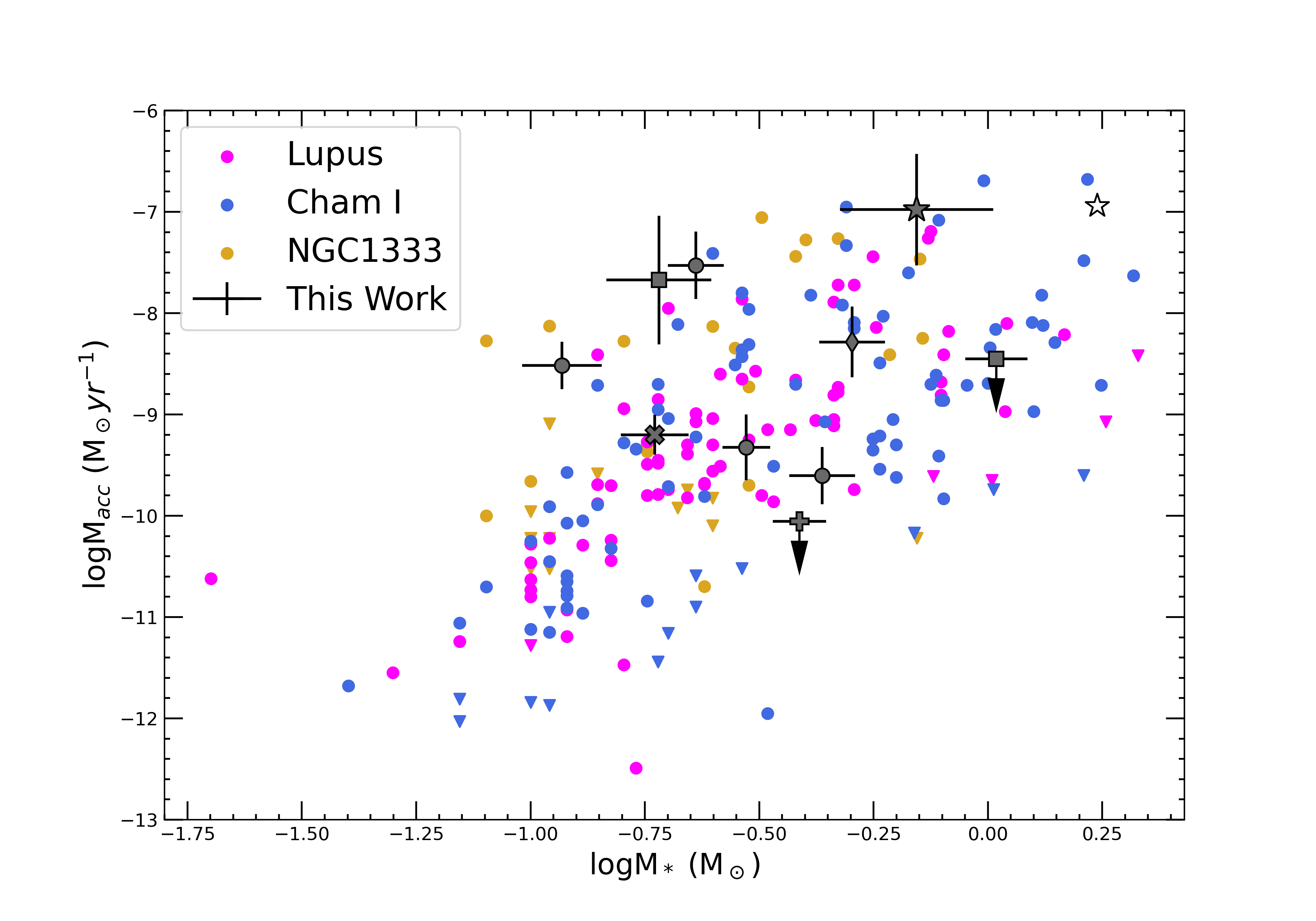}
\caption{Plot of log($M_{acc}$) vs. log($M_{*}$) for the class II targets of our current sample, compared to the results for class II YSOs from Lupus \citep{2017A&A...600A..20A}, Chamaeleon \citep{2016A&A...585A.136M}, and NGC1333 \citep{2021A&A...650A..43F}. These three sets are plotted in magenta, blue, and orange, respectively, with upper limits represented by triangles. The values from this work are plotted in black and grey, with different shapes for different regions: circles (NGC1333), stars (NGC7000), squares (IC5146), a diamond (Serpens), a plus sign (Orion), and an X-symbol (Perseus). Object 4 in NGC7000, having only an approximate $M_{acc}$, is represented with an unfilled star.}
\label{fig:Macc_comparison}
\end{figure*}

There are two class II YSOs in the sample for which we derive only upper limits on mass accretion rate. Object 8 (CVSO 1897) demonstrates the lowest mass accretion rate among all the class II stars, with an upper limit of log($M_{acc}$) $\leq$-10.0 in $M_\odot$/yr$^{-1}$. This upper limit lies along the low-$M_{acc}$ border of the $M_{*}$-$M_{acc}$ plots found for Lupus, Chamaeleon I, and NGC1333. As can be seen in Figure \ref{fig:alpha_accretion_plot}, Object 8 has a lower accretion rate than Object 10 (EM* LkHA 351), even though Object 10 has a lower spectral index $\alpha$ (being nearly in the class III category). Though Object 8 has a class II SED, its spectrum appears similar to a class III star. As shown in Figure \ref{fig:chromosphere_limit_plot} and previously discussed in Section \ref{sec:emission_lines}, Object 8 has emission line fluxes close to the expected chromospheric level for its $T_{eff}$, according to the thresholds defined by \citet{2013AandA...551A.107M} and \citet{2024A&A...690A.122C}. 
One other measurement for the accretion rate of Object 8 has been made by \citet{2020ApJ...893...56M}, and they found a log($M_{acc}$) of -9.12 in $M_\odot$/yr$^{-1}$. They derived this accretion rate using the equivalent width of the H$\alpha$ emission line. While their estimated $M_{acc}$ is above ours, their estimates for $A_{V}$, $L_{*}$ and $M_{*}$ are similar. They found an $A_{V}$ = 0.14 mag, log($L_{*}$/$L_\odot$) = -0.80, and log($M_{*}$/$M_\odot$) = -0.48, whilst we derive an $A_{V}$ = 0.12 $\pm$ 0.09 mag, log($L_{*}$/$L_\odot$) = -0.70 $\pm$ 0.17, and log($M_{*}$/$M_\odot$) = -0.41 $\pm$ 0.06 using the \citet{2000AandA...358..593S} models.
The other upper limit in $M_{acc}$ within our sample is Object 2, one of the highest mass stars in the sample with $M_{*} \approx M_{\odot}$. Object 2 demonstrates an upper limit of log($M_{acc}$) $\leq$ -8.4. 
Similarly to Object 8, Object 2 also has emission line fluxes close to the expected chromospheric level in Figure \ref{fig:chromosphere_limit_plot}.

There is one other star in our sample, Object 14 (2MASS J03284782+3116552), for which we found a mass accretion rate measurement in previous literature. This object is present in the \citet{2021A&A...650A..43F} study of NGC1333. \citet{2021A&A...650A..43F} found that the star has log($L_{*}$/$L_\odot$) = -0.72, in agreement our result of log($L_{*}$/$L_\odot$) = -0.74 $\pm$ 0.2. Using the fluxes of the Pa$\beta$ and Br$\gamma$ lines and applying the empirical line luminosity relations from \citet{2017A&A...600A..20A}, they found log($L_{acc}$/$L_\odot$) = -1.95 $\pm$ 0.36. This is larger than the log($L_{acc}$/$L_\odot$) = -2.30 $\pm$ 0.18 found in our study. Using the \citet{2000AandA...358..593S} models, we estimate $M_{*}$ = 0.12$M_\odot$, in close agreement to the $M_{*}$ = 0.11$M_\odot$ mass from \citet{2021A&A...650A..43F}. However, we find a lower accretion rate of log($M_{acc}$) = -8.81 $\pm$ 0.25 in $M_\odot$/yr$^{-1}$, as opposed to their log($M_{acc}$) = -8.13. The disagreement could potentially be ascribed to the difference in $A_V$ estimated. \citet{2021A&A...650A..43F} estimates $A_V$ = 4.0 mag by fitting the J and H bands of the class II spectrum with a class III template by hand. We estimate $A_V$ = 2.1 though, using a similar process but in the near-UV and optical range. 
The disagreement may also be due to the underlying chromospheric activity, differences in the choice of photospheric template between our work and \citet{2021A&A...650A..43F}, or accretion variability since the disagreement is small. Besides Objects 8 and 14, there were no other literature values to compare with our sample, as far as we are aware.

The scatter in the log($M_{*}$)-log($M_{acc}$) relationship for previous works suggests that $M_{acc}$ depends on more than just $M_{*}$. The aformentioned studies of Lupus, Chamaeleon I, and NGC1333 each used samples of YSOs from single star-forming regions, therefore all borne from the same local conditions at roughly the same time (although possibly with some spread in age; see the discussion in \citet{2023ASPC..534..539M} regarding age gradients within regions). As a result, scatter in the log($M_{*}$)-log($M_{acc}$) found by these studies can be more confidently attributed to individual differences in disk mass or the angular momentum of prenatal cores, or perhaps to time variability in accretion. On the other hand, the YSOs of our current sample have the additional complication of being from several SFRs rather than just one. Within singular star-forming regions we do still note this definite diversity in $M_{acc}$. A prime example is the difference in $M_{acc}$ between Objects 9 and 11. Both of these class II YSOs belong to NGC1333 and have nearly the same mass (0.23 $M_{\odot}$ and 0.19 $M_{\odot}$ respectively, via \citet{2000AandA...358..593S} models). However, they display $M_{acc}$ that differ by $\approx$ 1.5 dex. Object 11, the stronger accretor, has a spectrum which displays a visibly much more dramatic Balmer jump compared to Object 9.

\section{Conclusion}
\label{sec:conclusion}
We have developed a Python-based package called \texttt{nuts-for-ysos} and used it to study the spectra of 15 photometrically identified YSOs from VIRUS parallel data. With a wavelength range of 3500-5500\r{A}, VIRUS captures several important accretion indicators in YSOs, including the excess Balmer continuum and a variety of optical emission lines. Within \texttt{nuts-for-ysos} we applied the No U-Turn Sampler (NUTS) to fit a model consisting of a theoretical accretion spectrum added to a class III YSO template, in order to replicate the continuum of our 15 VIRUS spectra. In doing so, we were able to simultaneously derive values for several stellar parameters and accretion parameters while examining uncertainties and covariances to an extent not achieved in the past. From these parameters, we were then able to determine mass accretion rates for each YSO. One main takeaway is the strong correlation between $A_V$ and $L_{acc}$ derived from the model fits. This highlights the importance of precisely determining $A_V$ in order to constrain $L_{acc}$. This can better be achieved by fitting the model over a wide wavelength range, to differentiate the effects of extinction and accretion on the spectrum as much as possible. 
We then compared our results to those acquired using emission lines of the spectra, and found strong agreement. In comparing our study to previous studies of Lupus, Chamaeleon I, and NGC1333, we also found our results to occupy a typical range in $L_{acc}$ and $M_{acc}$.

Our results demonstrate the promise of applying a Bayesian framework like \texttt{nuts-for-ysos} for analyzing YSO spectra. The \texttt{nuts-for-ysos} code is flexible with regards to the number and types of spectral features it fits for, as well as the wavelengths of these features. It can therefore be applied to both future VIRUS observations or to observations from other spectrographs. Regardless, a future cohesive analysis of a larger collection of YSOs will be especially important for better characterizing the relationship between $M_{*}$ and $M_{acc}$. This, in turn, could help elucidate the physics behind star formation and by association the formation of planets. 

\subsection{Acknowledgements}
We thank the anonymous referee for their insightful comments that informed edits to our manuscript.
Observations were obtained with the Hobby–Eberly Telescope (HET), which is a joint project of the University of Texas at Austin, the Pennsylvania State University, Ludwig-Maximilians-Universität München, and Georg-August-Universität Göttingen. The HET is named in honor of its principal benefactors, William P. Hobby and Robert E. Eberly.

VIRUS is a joint project of the University of Texas at Austin,
Leibniz-Institut f\"ur Astrophysik Potsdam (AIP), Texas A\&M University
(TAMU), Max-Planck-Institut f\"ur Extraterrestrische Physik (MPE),
Ludwig-Maximilians-Universit\"at Muenchen, Pennsylvania State
University, Institut f\"ur Astrophysik G\"ottingen, University of Oxford,
and the Max-Planck-Institut f\"ur Astrophysik (MPA). In addition to
Institutional support, VIRUS was partially funded by the National
Science Foundation, the State of Texas, and generous support from
private individuals and foundations.

The authors acknowledge the Texas Advanced Computing Center (TACC) at The University of Texas at Austin for providing high-performance computing, visualization, and storage resources that have contributed to the research results reported within this paper.

Computations for this research were performed on the Pennsylvania State University’s Institute for Computational and Data Sciences’ Roar supercomputer. The Center for Exoplanets and Habitable Worlds is supported by the Pennsylvania State University and the Eberly College of Science.

This work has made use of data from the European Space Agency (ESA) mission
{\it Gaia} (\url{https://www.cosmos.esa.int/gaia}), processed by the {\it Gaia}
Data Processing and Analysis Consortium (DPAC,
\url{https://www.cosmos.esa.int/web/gaia/dpac/consortium}). Funding for the DPAC
has been provided by national institutions, in particular the institutions
participating in the {\it Gaia} Multilateral Agreement.

This publication makes use of data products from the Wide-field Infrared Survey Explorer, which is a joint project of the University of California, Los Angeles, and the Jet Propulsion Laboratory/California Institute of Technology, funded by the National Aeronautics and Space Administration.

The Pan-STARRS1 Surveys (PS1) and the PS1 public science archive have been made possible through contributions by the Institute for Astronomy, the University of Hawaii, the Pan-STARRS Project Office, the Max-Planck Society and its participating institutes, the Max Planck Institute for Astronomy, Heidelberg and the Max Planck Institute for Extraterrestrial Physics, Garching, The Johns Hopkins University, Durham University, the University of Edinburgh, the Queen's University Belfast, the Harvard-Smithsonian Center for Astrophysics, the Las Cumbres Observatory Global Telescope Network Incorporated, the National Central University of Taiwan, the Space Telescope Science Institute, the National Aeronautics and Space Administration under Grant No. NNX08AR22G issued through the Planetary Science Division of the NASA Science Mission Directorate, the National Science Foundation Grant No. AST-1238877, the University of Maryland, Eotvos Lorand University (ELTE), the Los Alamos National Laboratory, and the Gordon and Betty Moore Foundation. 

This research has made use of the VizieR catalogue access tool, CDS,
Strasbourg, France (DOI : 10.26093/cds/vizier). The original description 
of the VizieR service was published in \citet{2000A&AS..143...23O}.

\software{Python \citep{van2009introduction}, PyMC \citep{pymc3}, PyTensor \citep{pytensor, 2016arXiv160502688full}, Matplotlib \citep{Hunter:2007}, NumPy \citep{harris2020array}, SciPy \citep{2020SciPy-NMeth}, Astropy \citep{astropy:2013, astropy:2018, astropy:2022}, Specutils \citep{2019ascl.soft02012A}, PySpecKit \citep{2011ascl.soft09001G}}

\bibliography{ms}{}
\bibliographystyle{aasjournal}

\newpage
\appendix

\section{Data Reduction}
\label{sec:data_reduction}
The parallel observations were reduced using Remedy, a data processing pipeline for VIRUS\footnote{\url{https://github.com/grzeimann/Remedy}}. Remedy is responsible for calibrating the spectra from each individual fiber, and then combining these spectra into final ‘extracted spectra’ of bright continuum sources. A full description of parallel observation reductions is presented in \citet{2024ApJ...966...14Z}, but we present relevant details to our study below.
For each observation, Remedy first performs the gain multiplication, bias and dark frame subtractions, and masking of hot pixels identified in the master dark frame. The location of the fibers on the detector are then determined using a master twilight frame compiled over several days. Wavelength calibration for each fiber is performed using Cd and Hg arc lamps.
In order to extract complete spectra of continuum sources from the fiber spectra, a fiber normalization must be performed, which accounts for variations in CCD quantum efficiency and both fiber and spectrograph throughput variations. A master twilight frame is used to evaluate the relative throughput between each fiber and their average, and the fibers are normalized appropriately. A fiber is omitted from the extracted spectrum if its normalization factor is below 10\%. 
For each exposure, a single sky model is constructed using identified 'blank' fibers in the field. This first sky model is individually subtracted from every normalized fiber spectrum. Then a more localized residual sky model is subtracted, employing a Gaussian kernel in the fiber and wavelength direction.  This kernel has a standard deviation of seven fibers in the fiber direction and 14\r{A} in the wavelength direction.  Fibers with continuum emission greater than 2-$\sigma$ are masked in this residual sky process. This entire sky subtraction process tends to fail in crowded fields with a lack of fibers pointed on blank sky. For example, several VIRUS parallel observations contained YSOs in nebulous regions such as the Orion Nebula. These observations were omitted from our study because the sky subtraction was unreliable.
After sky subtraction, the VIRUS observations are then astrometrically calibrated using Pan-STARRS \citep{2017AAS...22922303C} Data Release 2. Remedy matches point sources from each exposure with Pan-STARRS, and then shifts and rotates the VIRUS astrometry accordingly; shifts are usually $< 5$'' and rotations $\lesssim 0.1 \degree $.

Remedy then extracts continuum sources from the normalized, sky-subtracted, and astrometrically calibrated fiber spectra. The seeing at the HET is usually around the same as the 1.5’’ diameter of the VIRUS fibers, meaning that the weight for each fiber in receiving light from a source can easily change with the presumed position of the source. Another complication is the presence of differential atmospheric refraction, which causes variation in the throughput with wavelength for a given fiber. 
Both of these phenomena are especially problematic for parallel VIRUS observations, which are not dithered, and therefore sparsely sample the sky compared to the usual 3-point dither pattern of VIRUS observations. Remedy uses the other stars in the VIRUS field of view (usually more than 20) to model the spatial PSF and the differential atmospheric refraction, so that it can address these issues and accurately extract the spectrum of the source.

Each VIRUS parallel observation can ultimately contain hundreds to thousands of extracted spectra. Flux calibration is first performed on the extracted spectra using the throughput curve for the HETDEX collaboration \citep{2012AAS...21942402G}. Then, these preliminary calibrated spectra are convolved with the Pan-STARRS g filter. For each entire parallel observation, the overall biweight-estimated offset between the Pan-STARRS1 DR2 g magnitudes and VIRUS g magnitudes is then used to normalize this preliminary calibration. The finished flux calibration has a standard deviation of usually 0.1-0.15 mag.
In this study, however, we chose to individually re-normalize extracted spectra using their Pan-STARRS1 DR2 'Mean PSF' g magnitudes \citep{2017AAS...22922303C, MAST_Panstarrs}.\footnote{Pan-STARRS1 DR2 data can be found in MAST: \dataset[https://doi.org/10.17909/s0zg-jx37]{https://doi.org/10.17909/s0zg-jx37}.} 
This was because stars in star-forming regions may be young and variable in brightness, and these variations can affect a normalization done over an entire shot at once. Normalizing using photometry one-at-a-time does not remove possibility of each target's variability affecting flux calibration, but it does allow for a more consistent approach not affected by the number and variability of neighboring stars. Moreover, we use the Pan-STARRS1 DR2 Mean PSF r and i magnitudes when fitting the accretion model to the VIRUS spectra in Section \ref{sec:continuum_fit}, so it is best that the g magnitudes of the VIRUS spectra are kept completely consistent with Pan-STARRS photometry. We are therefore assuming that the YSO spectrum has not drastically changed between the Pan-STARRS photometric observation and VIRUS observation.
While this might be an unsuitable assumption for sources significantly varying in luminosity, like erupting FUors or EXors stars, none of our targets have previously been identified with these categories. The uncertainty in the Pan-STARRS photometry is taken into account when calculating the final uncertainty for each VIRUS spectrum.

\begin{deluxetable}{llllllll}
\tablecaption{Observation Log\label{tab:obs_log}}
\tablewidth{0pt}
\tablehead{
    \colhead{ID} & \colhead{SIMBAD Name} & \colhead{Obs. date} & \colhead{RA} & \colhead{DEC} & \colhead{Exp. time} & \colhead{Avg. SNR} & \colhead{PanSTARRS ID}
}
\startdata
1 & 2MASS J21523325+4710505 & 2020-11-05 & 21:52:33.24 & 47:10:50.50 & 907.03 & 5.6 & 164613281385307509\\
2 & 2MASS J21533310+4716092 & 2020-11-05 & 21:53:33.10 & 47:16:09.20 & 907.03 & 52.1 & 164723283879153751\\
3 & 2MASS J20580138+4345201 & 2020-10-04 & 20:58:01.38 & 43:45:20.15 & 607.8 & 5.84 & 160503145057527374\\
4 & EM* LkHA 188 & 2020-12-01 & 20:58:23.81 & 43:53:11.40 & 608.15 & 138.56 & 160663145991974465\\
5 & 2MASS J18300610+0106170 & 2020-04-15 & 18:30:06.11 & 01:06:16.81 & 1207.45 & 10.93 & 109322775254456070\\
6 & 2MASS J18295618+0110574 & 2020-04-19 & 18:29:56.18 & 01:10:57.32 & 1207.8 & 58.24 & 109412774841009544\\
7 & V* V776 Ori & 2020-02-07 & 05:34:50.97 & -05:42:21.44 & 608.13 & 201.04 & 101150837123603275\\
8 & CVSO 1897 & 2020-11-15 & 05:40:15.14 & 00:57:26.71 & 1826.3 & 67.41 & 106850850630851552\\
9 & [HL2013] 052.17673+30.49810 & 2021-01-16 & 03:28:42.44 & 30:29:53.03 & 1506.9 & 53.16 & 144590521768208302\\
10 & EM* LkHA 351 & 2019-01-01 & 03:28:46.20 & 31:16:38.44 & 908.15 & 88.41 & 145530521924233432 \\
11 & 2MASS J03285101+3118184 & 2019-01-01 & 03:28:51.03 & 31:18:18.39 & 908.15 & 47.75 & 145560522125686781\\
12 & 2MASS J03283651+3119289 & 2019-01-01 & 03:28:36.53 & 31:19:28.81 & 908.15 & 31.91 & 145590521521560240\\
13 & 2MASS J03292815+3116285 & 2021-02-09 & 03:29:28.16 & 31:16:28.44 & 1508.32 & 6.54 & 145530523673090085\\
14 & 2MASS J03284782+3116552 & 2021-02-09 & 03:28:47.84 & 31:16:55.05 & 1508.32 & 16.81 & 145530521992468976\\
15 & 2MASS J03285216+3122453 & 2021-02-09 & 03:28:52.17 & 31:22:45.15 & 1508.32 & 44.76 & 145650522173355698\\
16 & ATO J052.3580+31.4444 & 2021-02-12 & 03:29:25.93 & 31:26:39.93 & 1507.25 & 61.23 & 145730523580053950\\
\enddata
\end{deluxetable}

\newpage
\section{Model fit and corner plots}
\label{sec:result_plots}
\begin{figure*}[ht!] 
\centering
\includegraphics[scale=0.22]{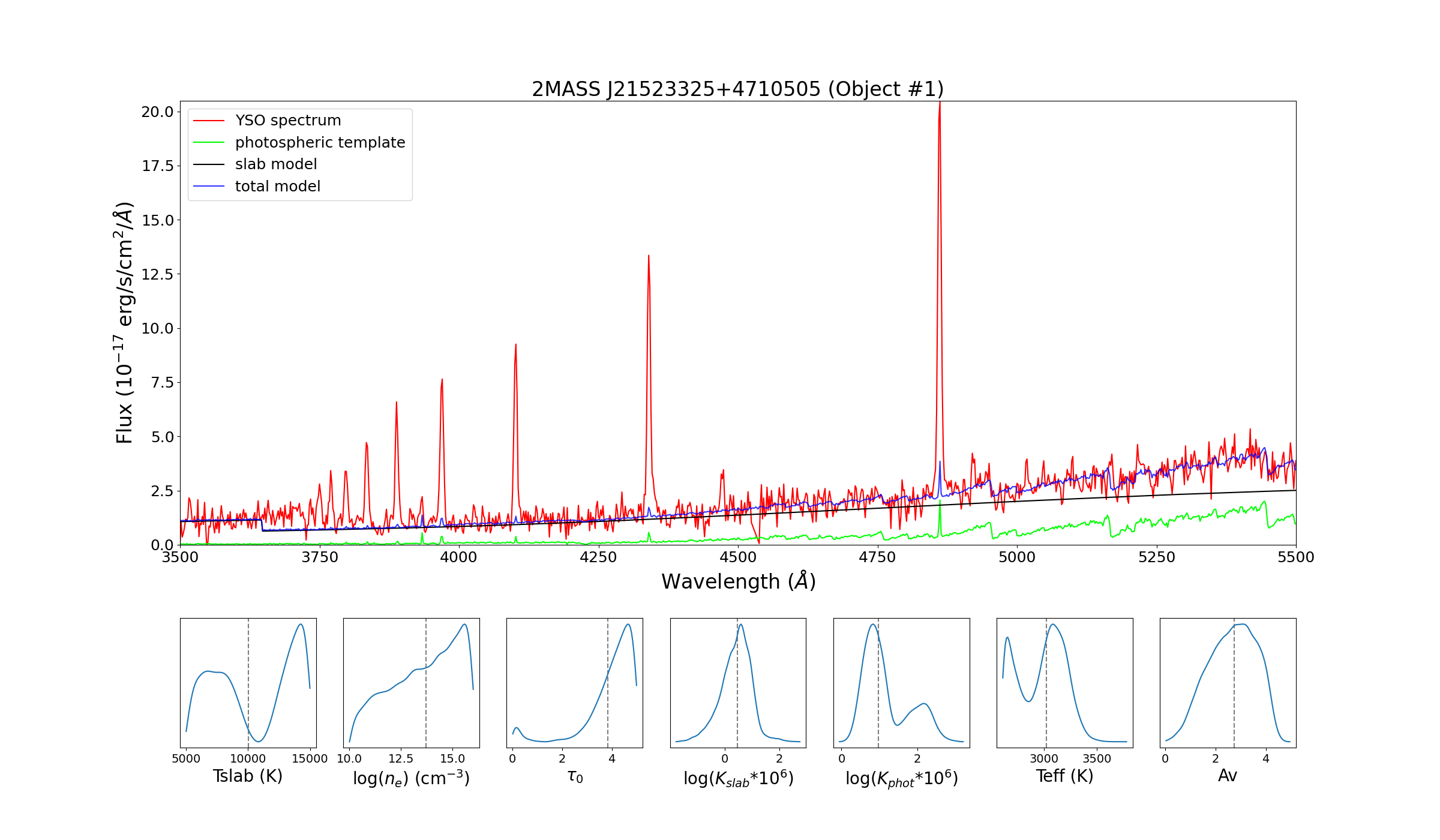}
\caption{The median model fit for Object 1 and the parameter posteriors below (the median parameters marked with a vertical gray dotted line).}
\label{fig:result_1}
\end{figure*}

\begin{figure*}[ht!]
\centering
\includegraphics[scale=0.63]{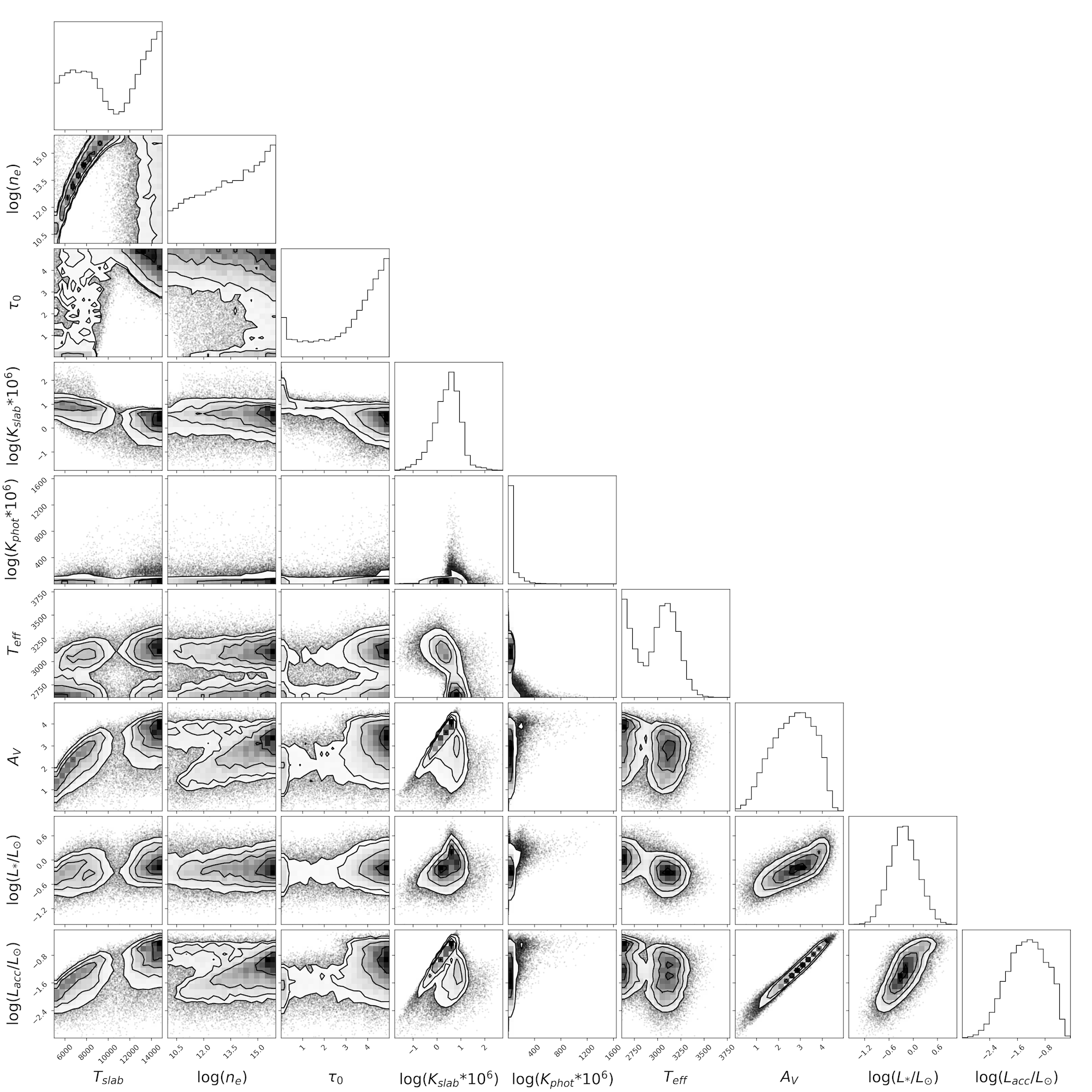}
\caption{The corner plot for Object 1, for model parameters and the log($L_{*}$) and log($L_{acc}$) posteriors.}
\end{figure*}

\begin{figure*} 
\centering
\includegraphics[scale=0.25]{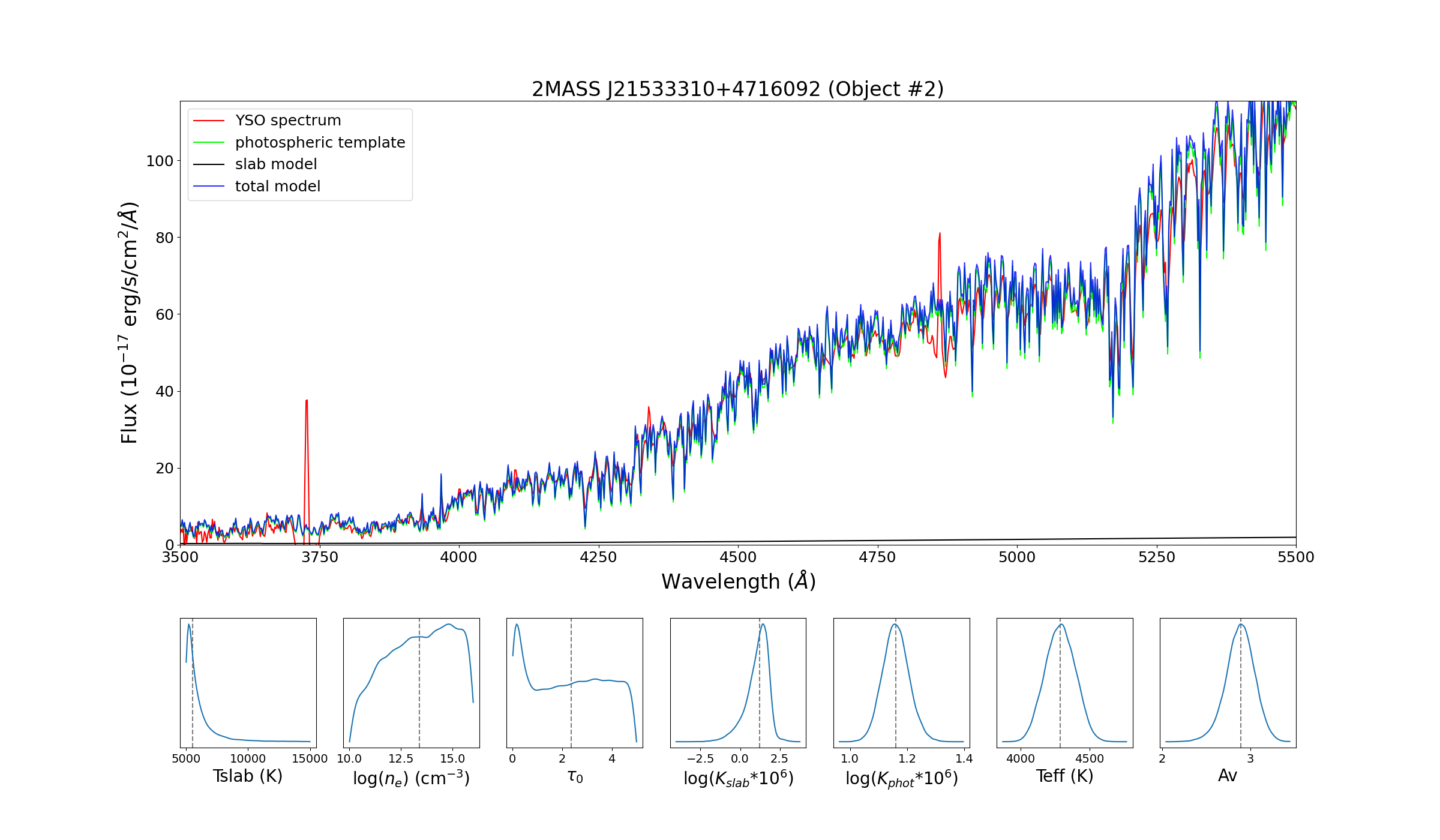}
\caption{The median model fit for Object 2 and the parameter posteriors with the same plotting convention as Figure \ref{fig:result_1}.}
\end{figure*}

\begin{figure*} 
\centering
\includegraphics[scale=0.66]{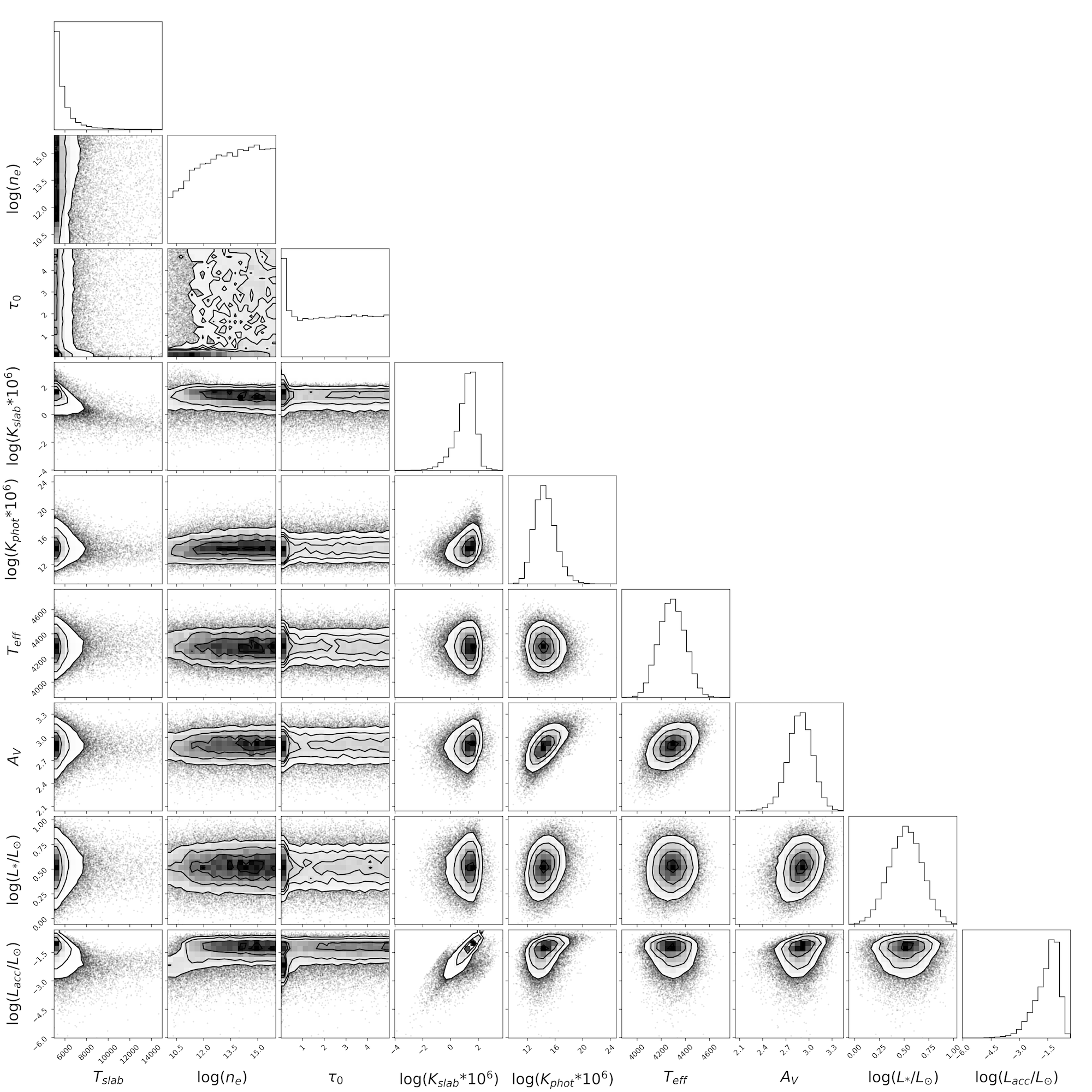}
\caption{The corner plot for Object 2, for model parameters and the log($L_{*}$) and log($L_{acc}$) posteriors.}
\end{figure*}

\begin{figure*} 
\centering
\includegraphics[scale=0.25]{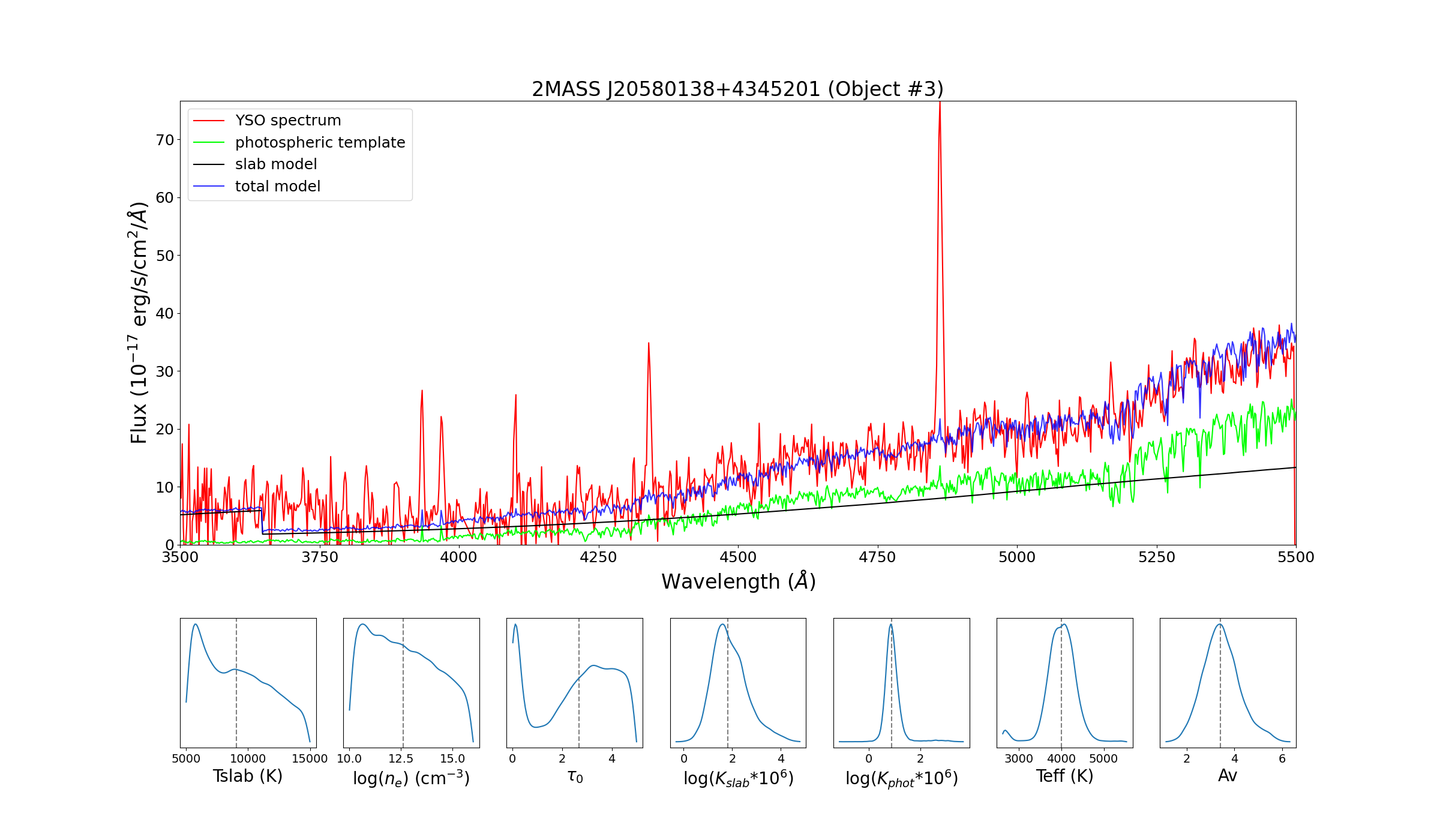}
\caption{The median model fit for Object 3 and the parameter posteriors with the same plotting convention as Figure \ref{fig:result_1}.}
\end{figure*}

\begin{figure*} 
\centering
\includegraphics[scale=0.66]{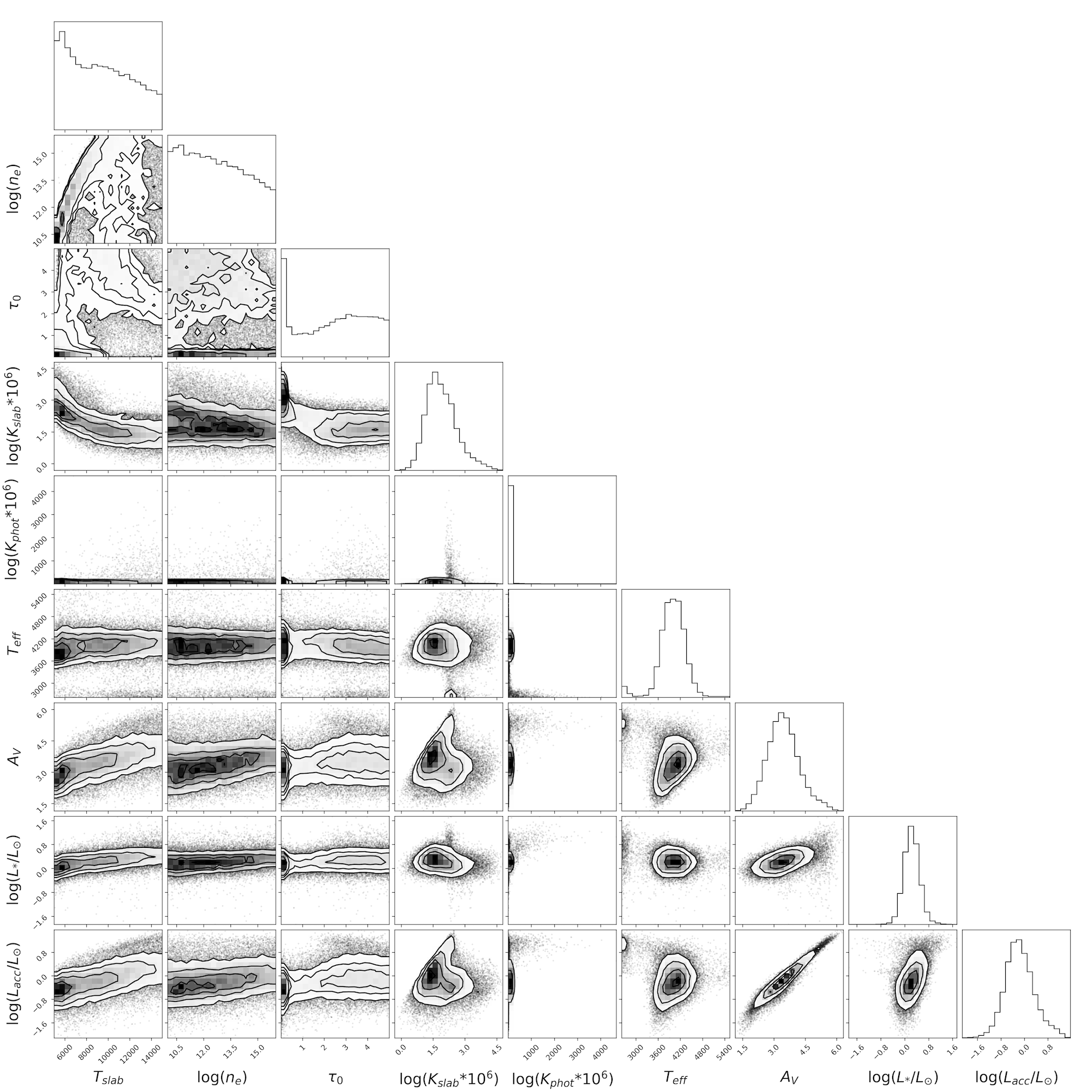}
\caption{The corner plot for Object 3, for model parameters and the log($L_{*}$) and log($L_{acc}$) posteriors.}
\end{figure*}

\begin{figure*} 
\centering
\includegraphics[scale=0.25]{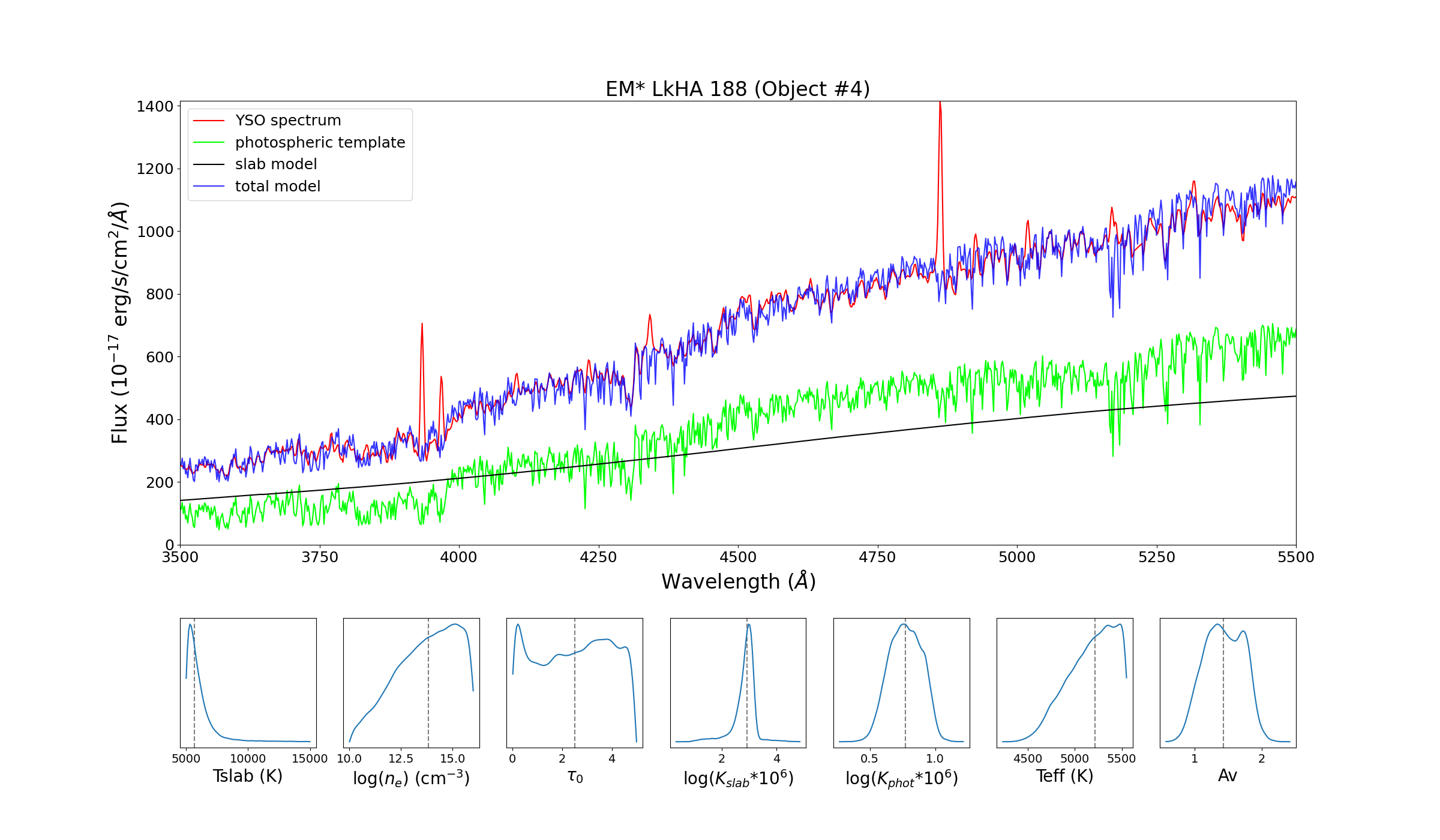}
\caption{The median model fit for Object 4 and the parameter posteriors with the same plotting convention as Figure \ref{fig:result_1}.}
\end{figure*}

\begin{figure*} 
\centering
\includegraphics[scale=0.66]{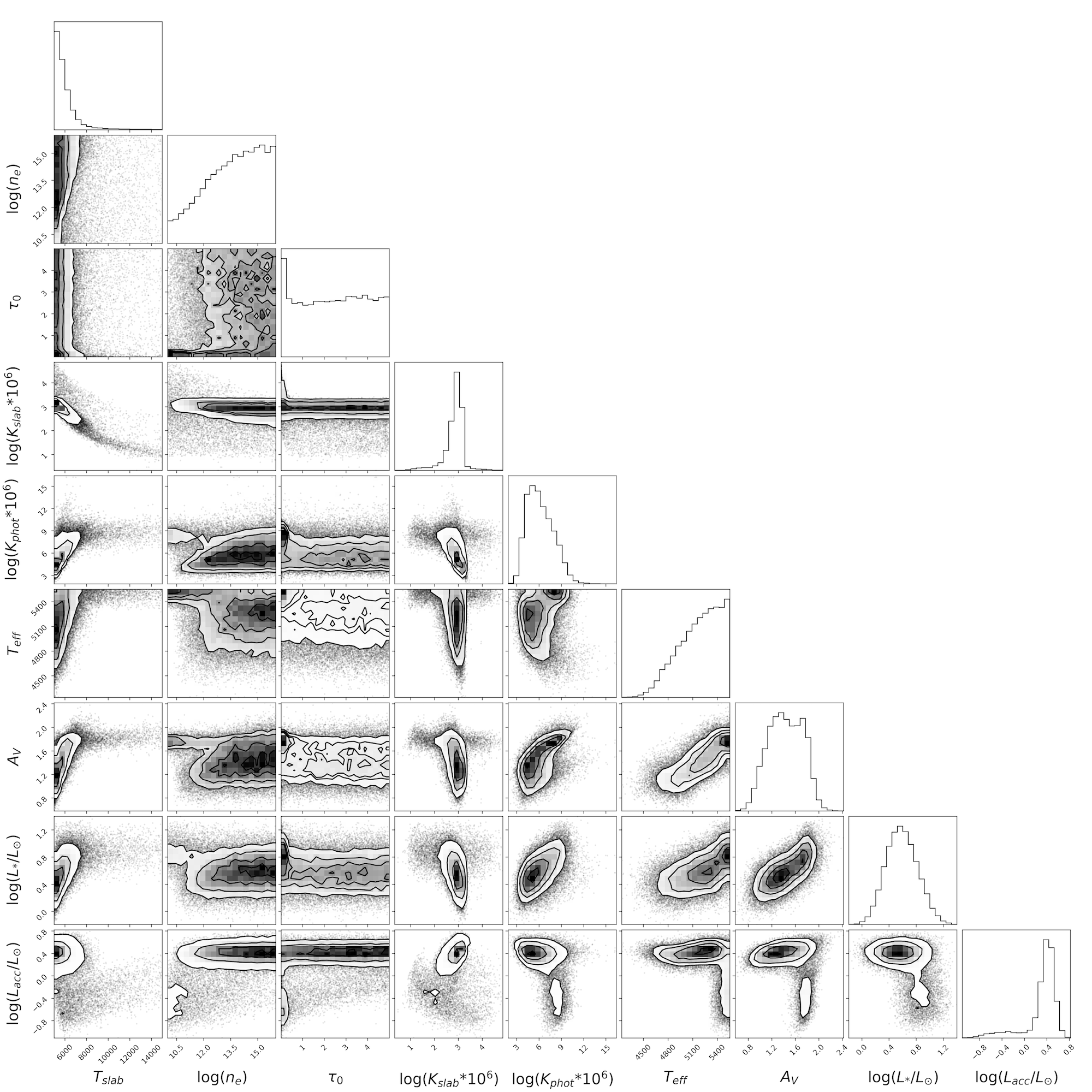}
\caption{The corner plot for Object 4, for model parameters and the log($L_{*}$) and log($L_{acc}$) posteriors.}
\end{figure*}

\begin{figure*} 
\centering
\includegraphics[scale=0.25]{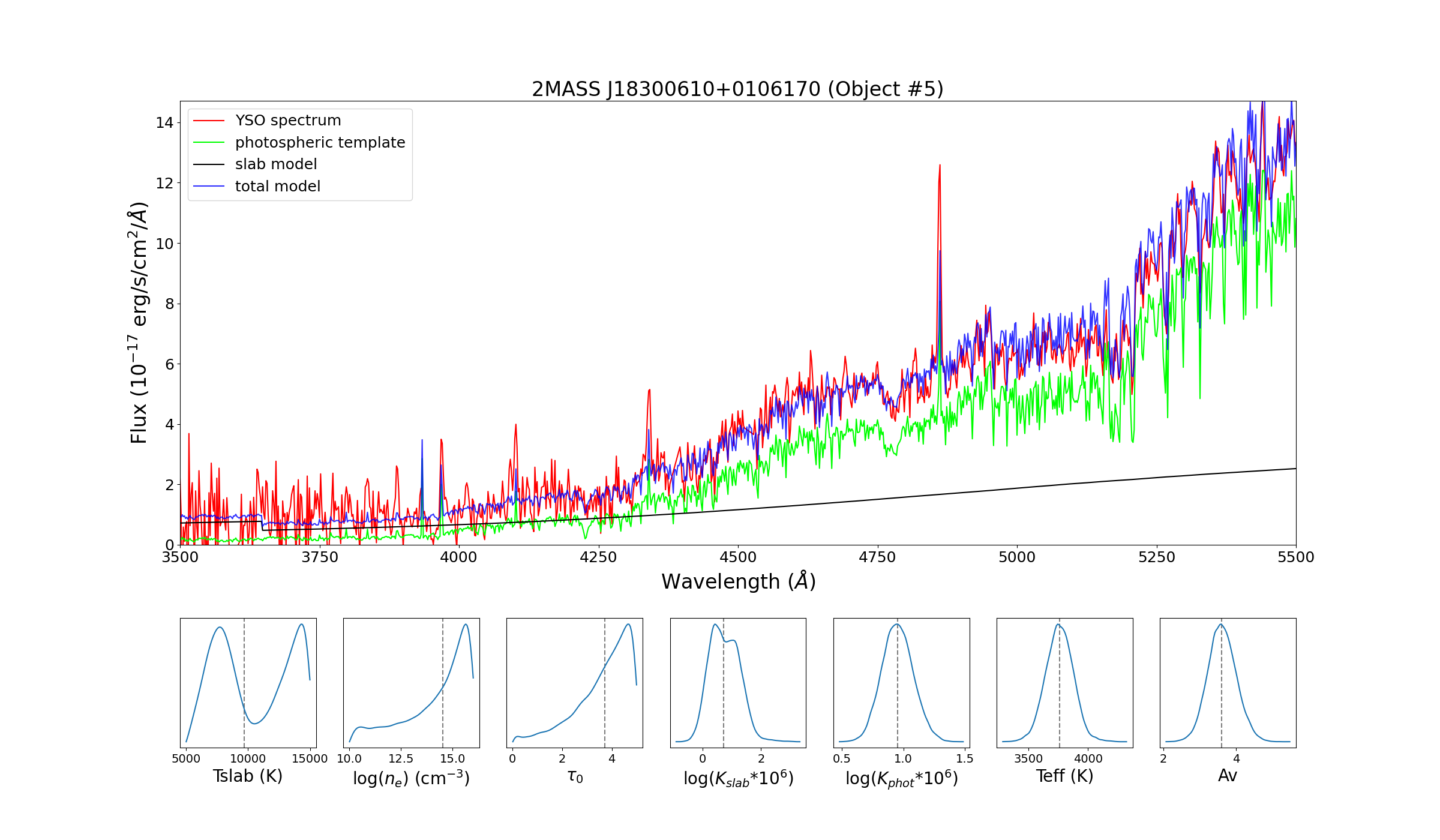}
\caption{The median model fit for Object 5 and the parameter posteriors with the same plotting convention as Figure \ref{fig:result_1}.}
\end{figure*}

\begin{figure*} 
\centering
\includegraphics[scale=0.66]{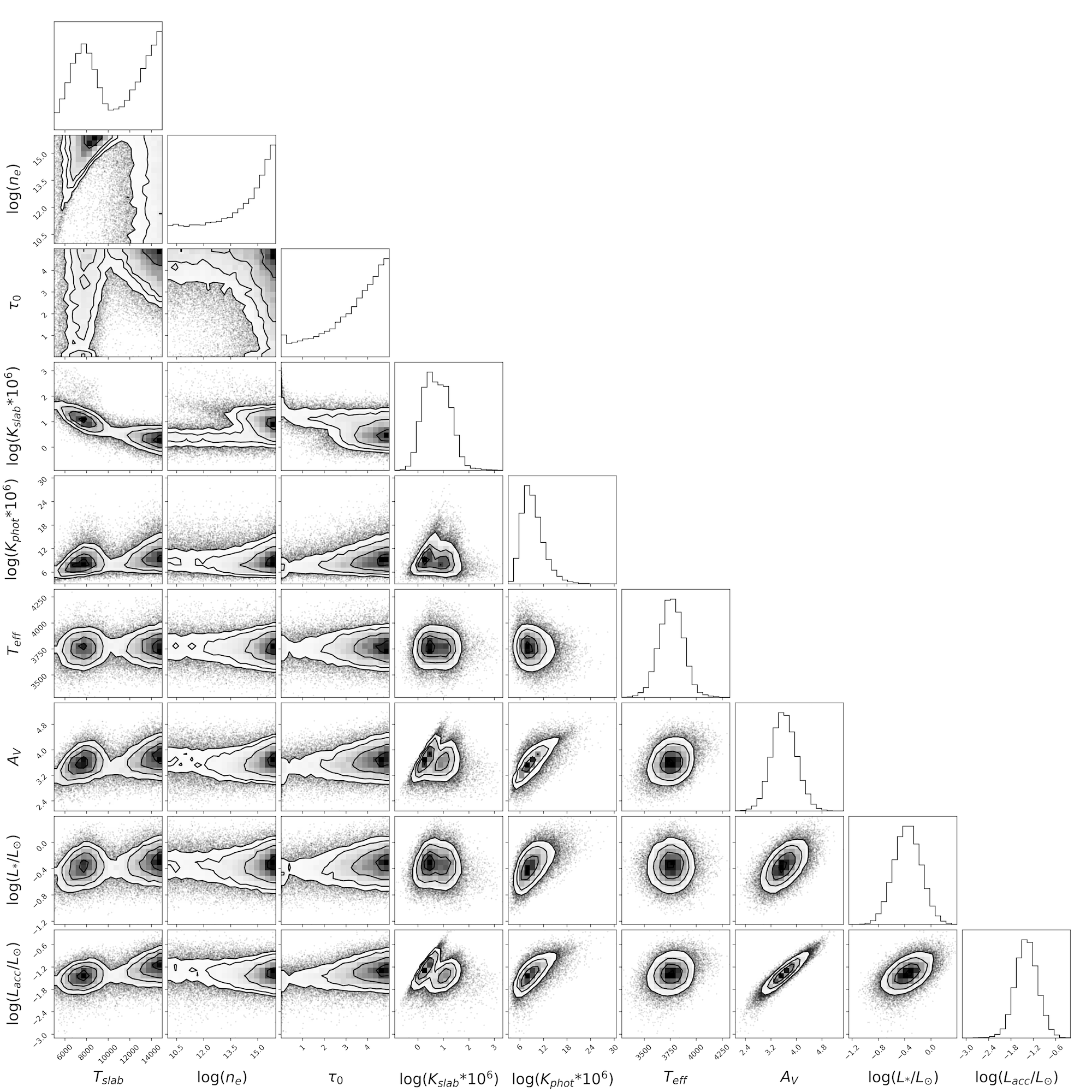}
\caption{The corner plot for Object 5, for model parameters and the log($L_{*}$) and log($L_{acc}$) posteriors.}
\end{figure*}

\begin{figure*} 
\centering
\includegraphics[scale=0.25]{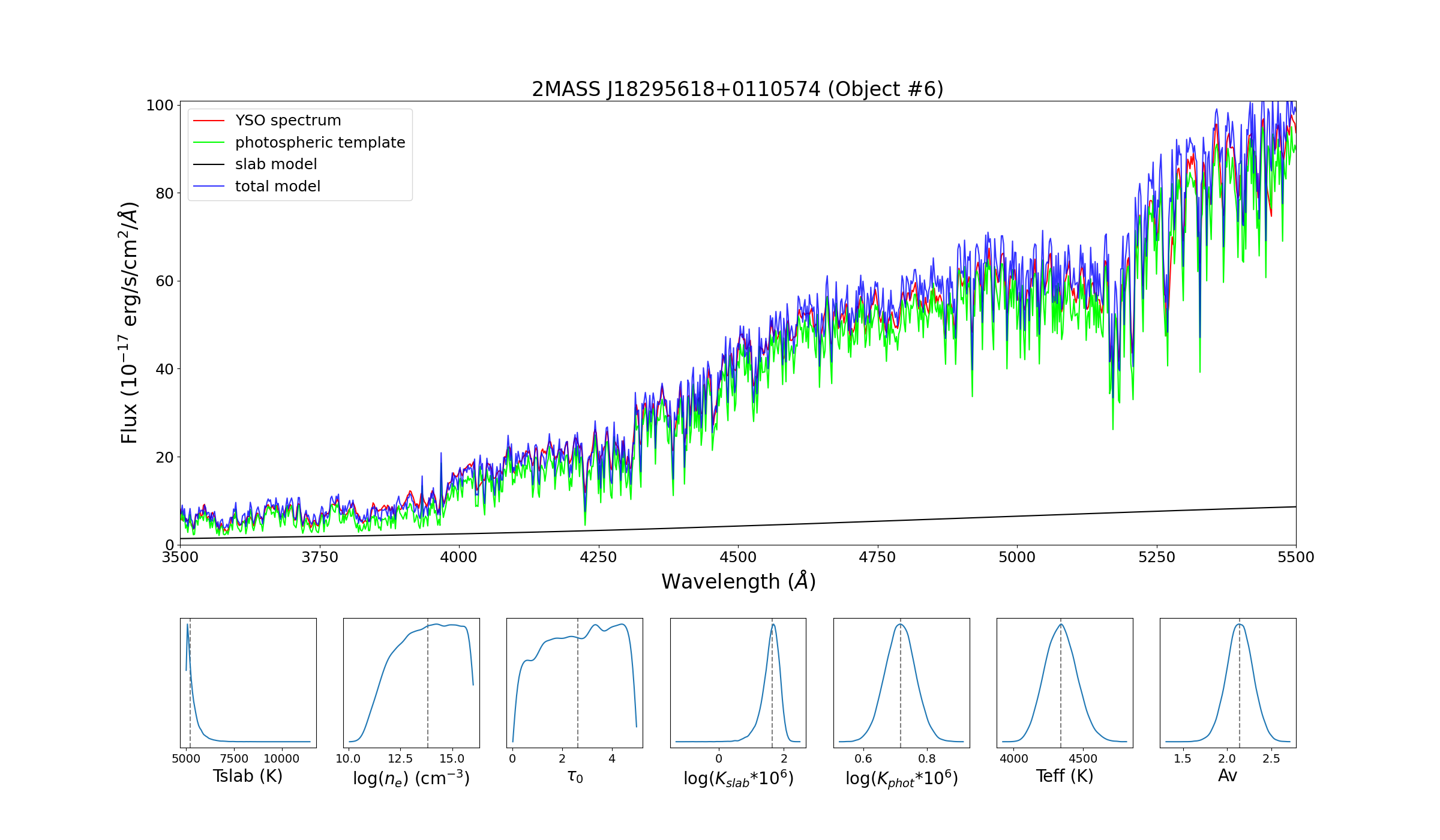}
\caption{The median model fit for Object 6 and the parameter posteriors with the same plotting convention as Figure \ref{fig:result_1}.}
\end{figure*}

\begin{figure*} 
\centering
\includegraphics[scale=0.66]{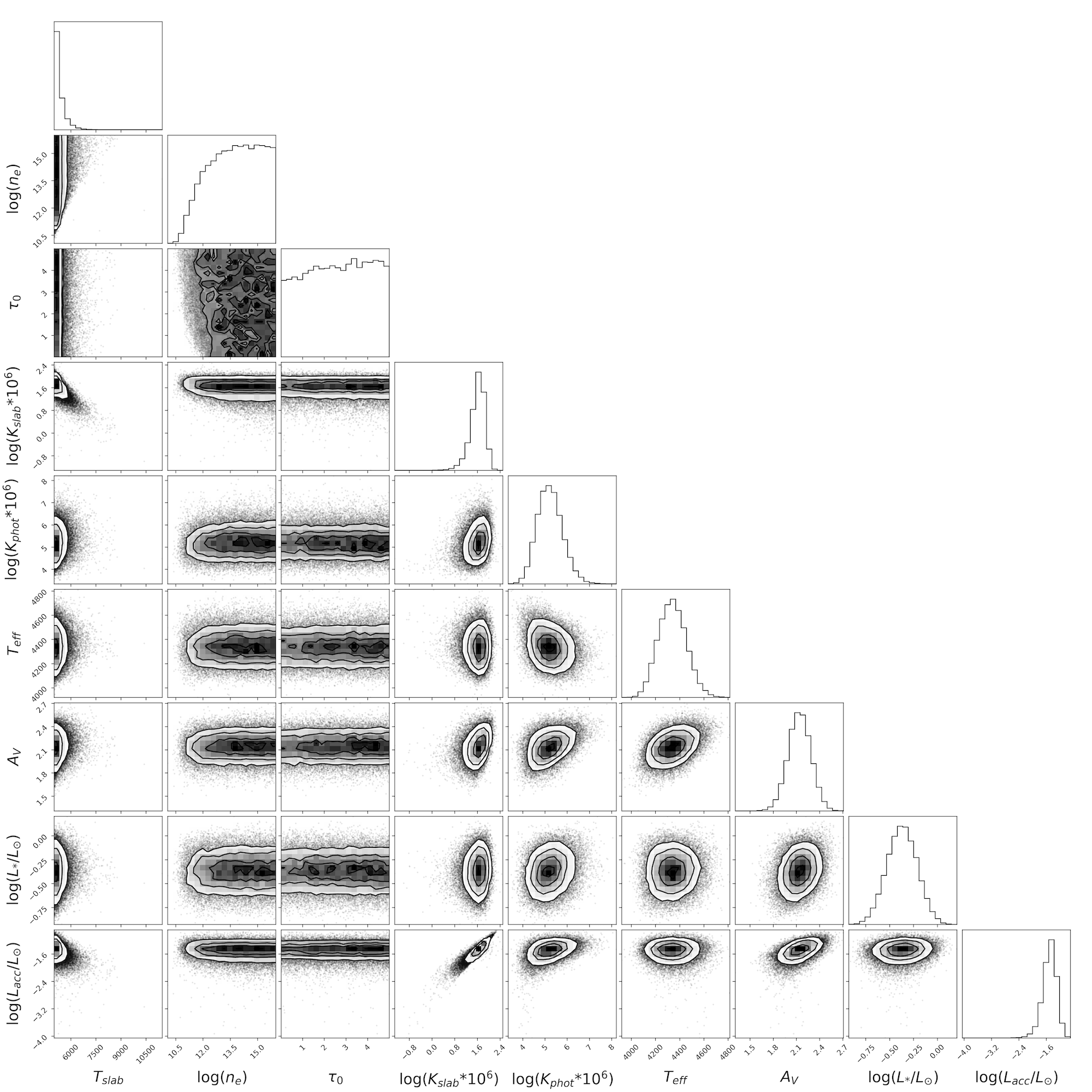}
\caption{The corner plot for Object 6, for model parameters and the log($L_{*}$) and log($L_{acc}$) posteriors.}
\end{figure*}

\begin{figure*} 
\centering
\includegraphics[scale=0.25]{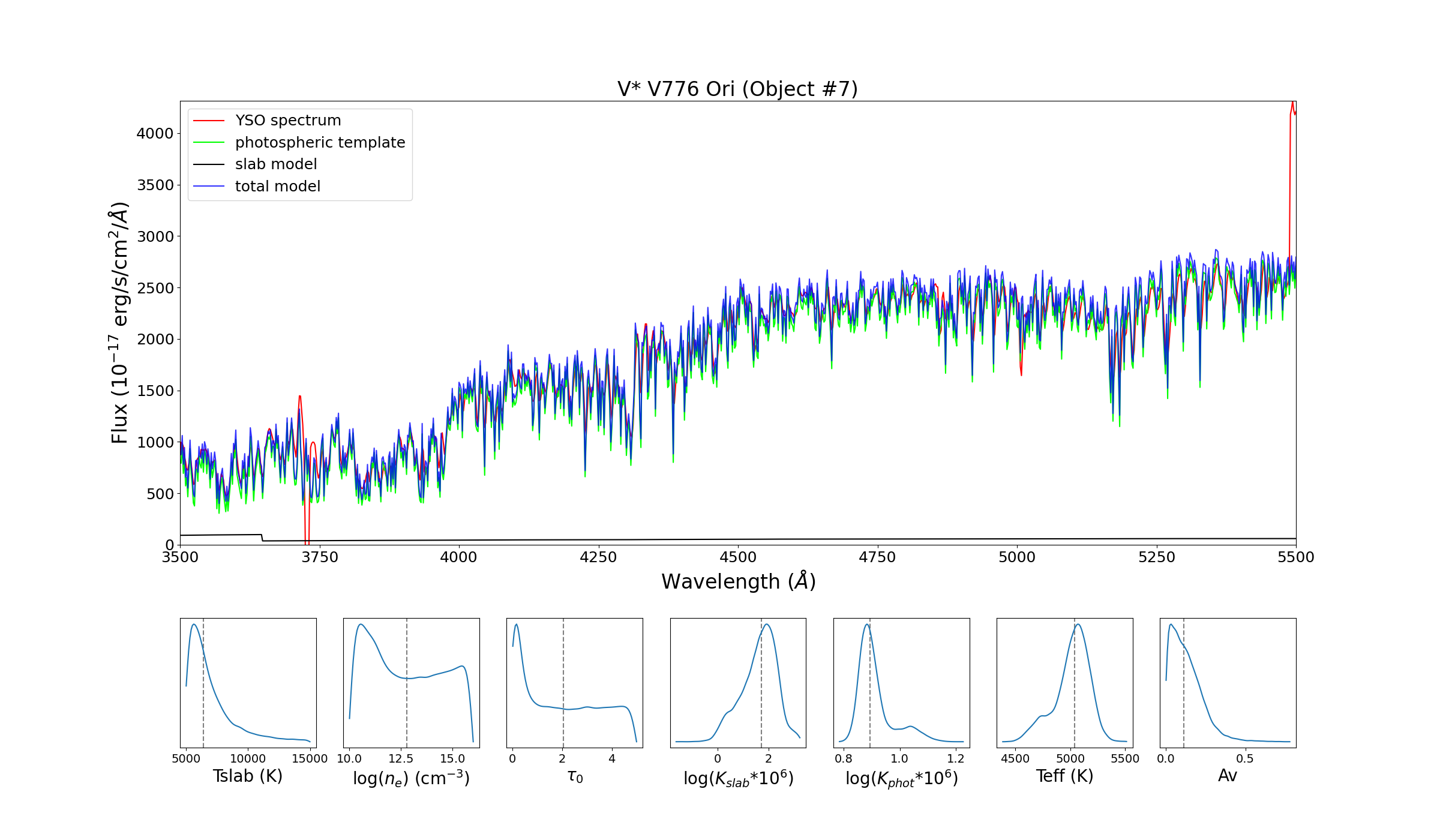}
\caption{The median model fit for Object 7 and the parameter posteriors with the same plotting convention as Figure \ref{fig:result_1}.}
\end{figure*}

\begin{figure*} 
\centering
\includegraphics[scale=0.66]{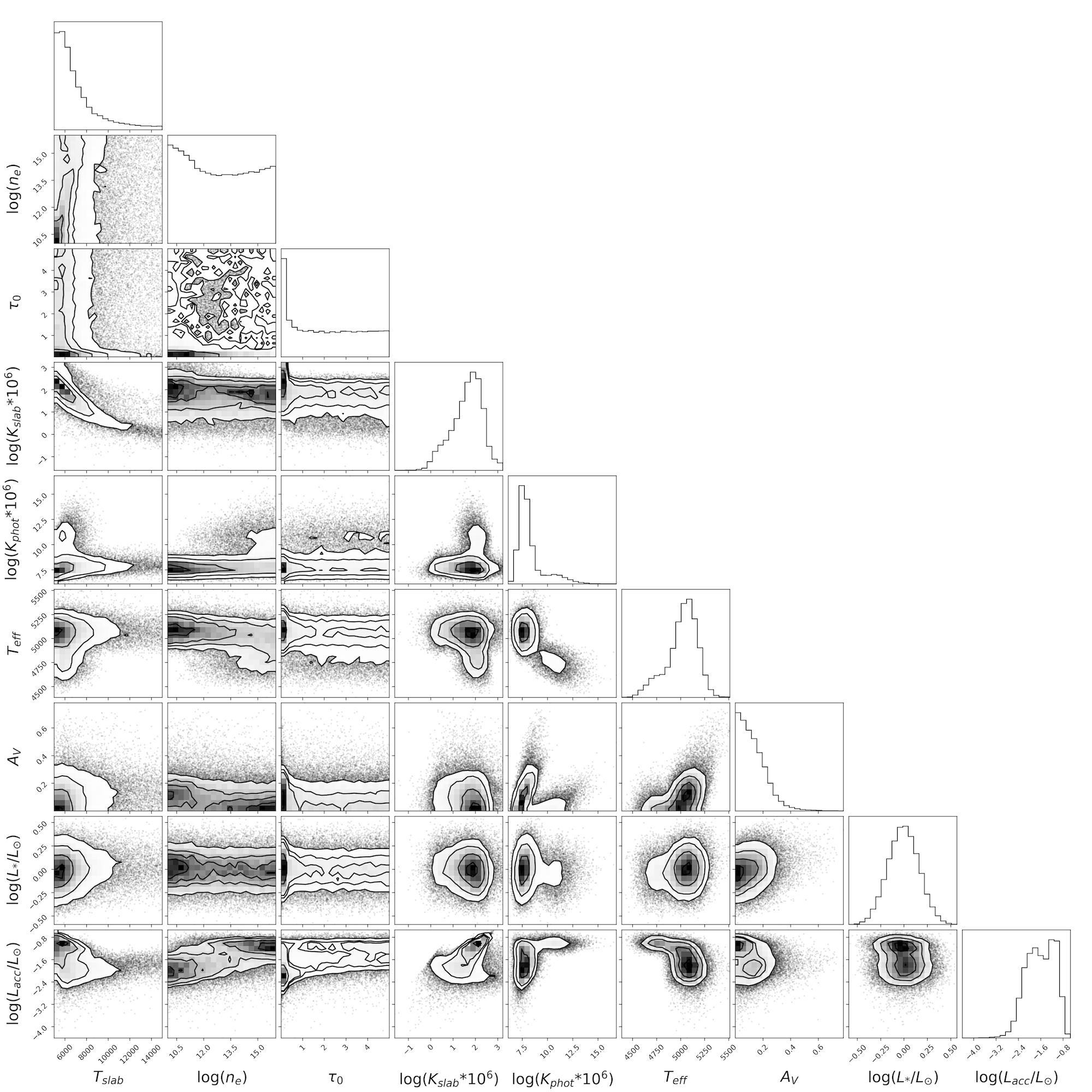}
\caption{The corner plot for Object 7, for model parameters and the log($L_{*}$) and log($L_{acc}$) posteriors.}
\end{figure*}

\begin{figure*} 
\centering
\includegraphics[scale=0.25]{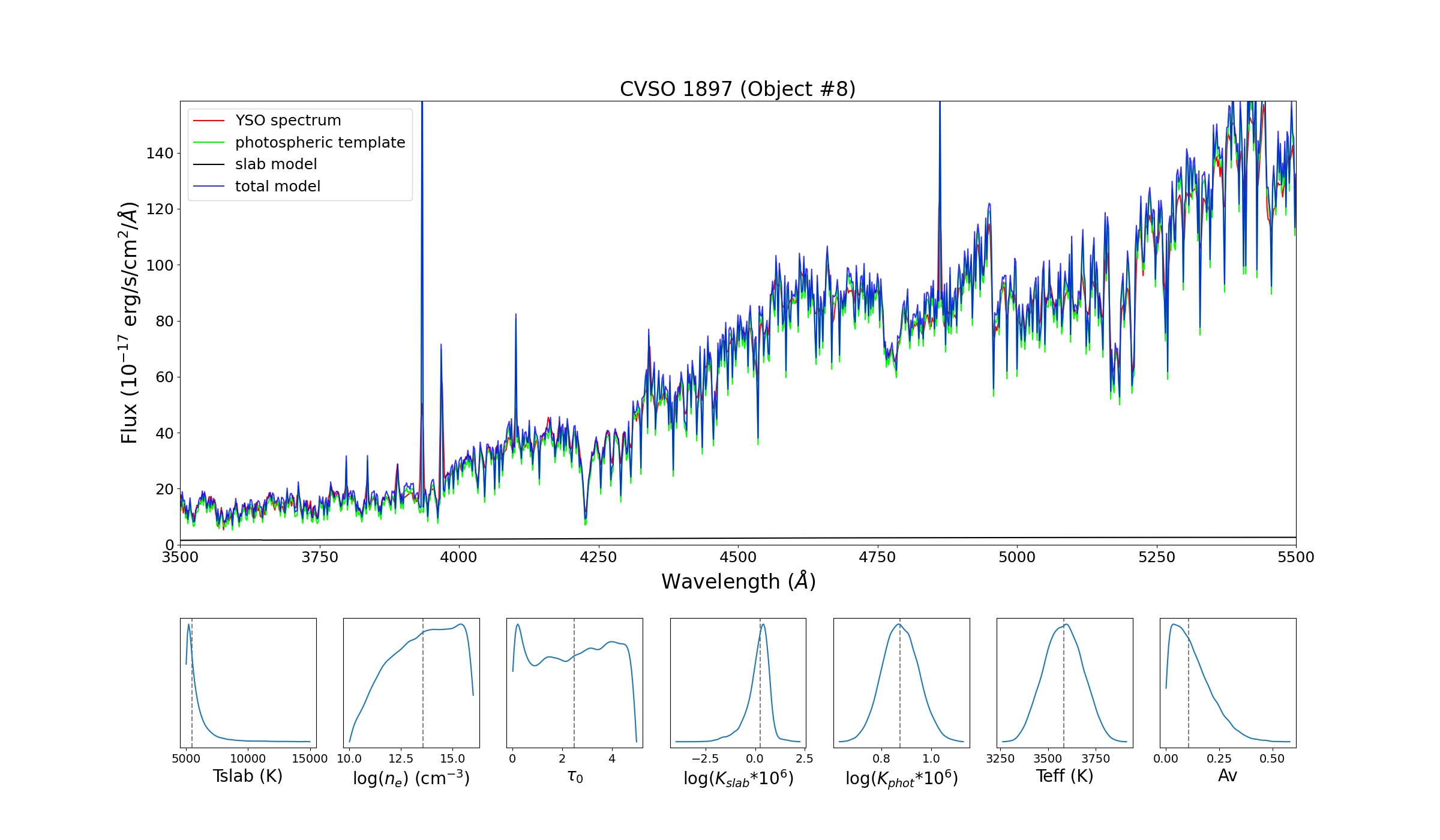}
\caption{The median model fit for Object 8 and the parameter posteriors with the same plotting convention as Figure \ref{fig:result_1}.}
\end{figure*}

\begin{figure*} 
\centering
\includegraphics[scale=0.66]{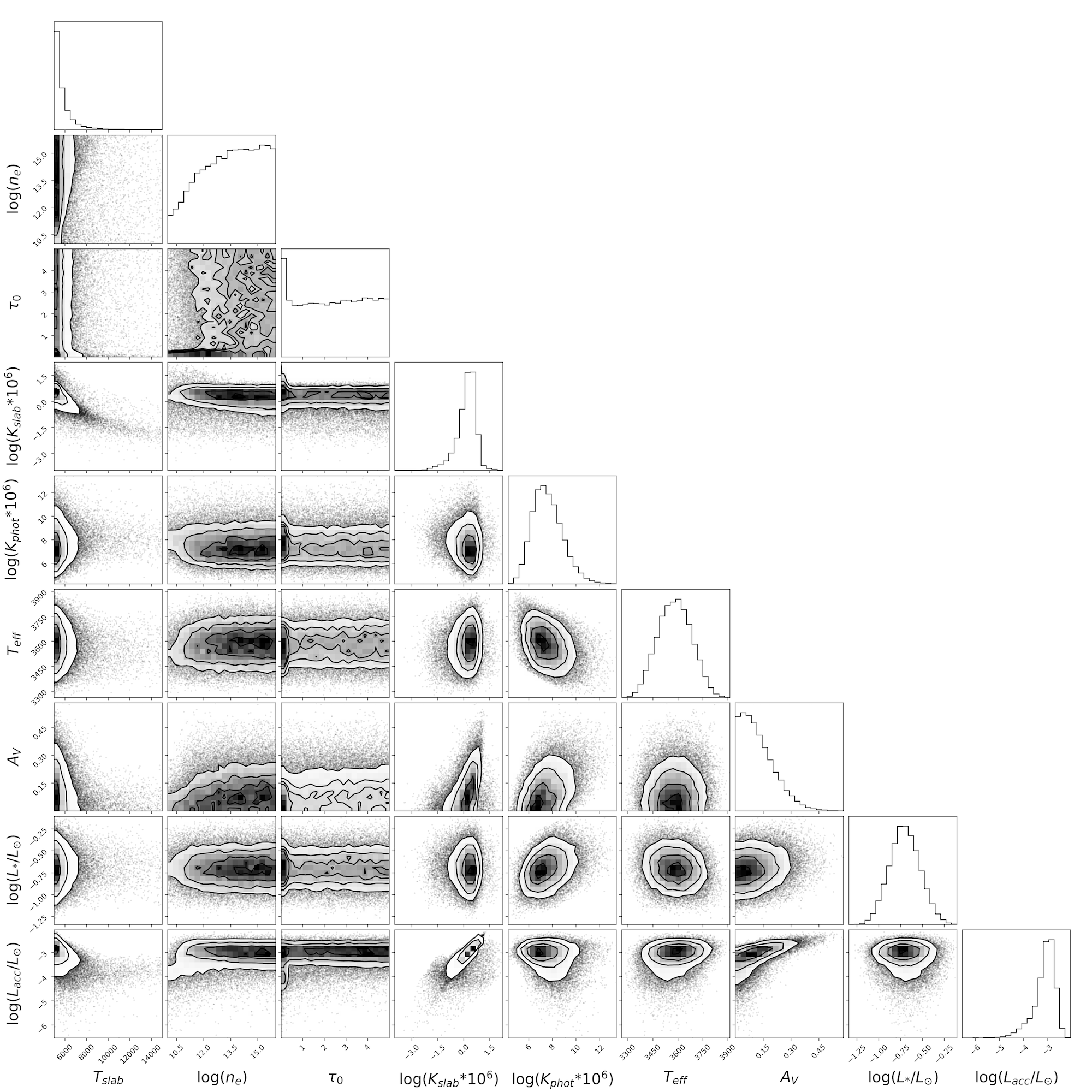}
\caption{The corner plot for Object 8, for model parameters and the log($L_{*}$) and log($L_{acc}$) posteriors.}
\end{figure*}

\begin{figure*} 
\centering
\includegraphics[scale=0.25]{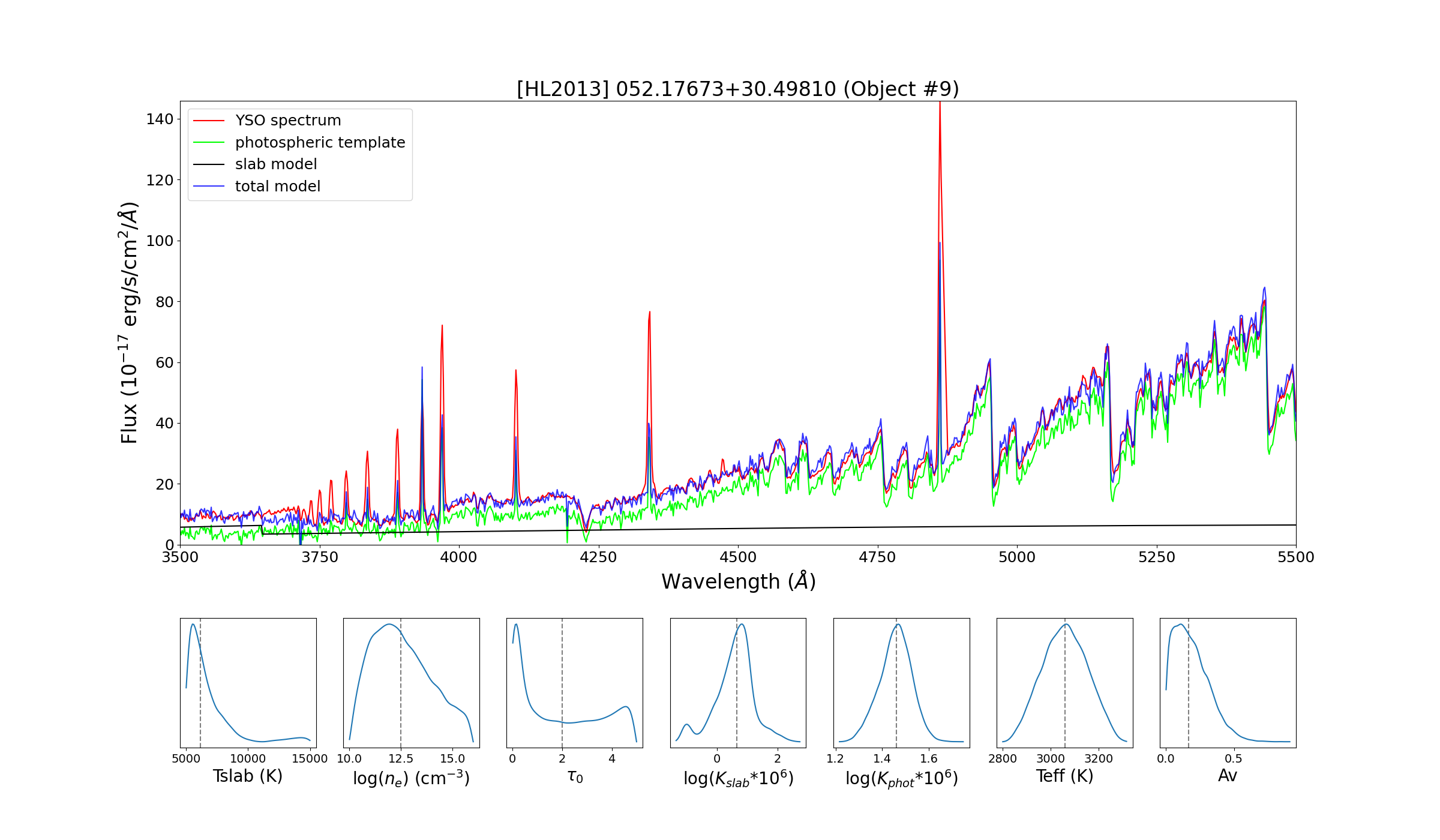}
\caption{The median model fit for Object 9 and the parameter posteriors with the same plotting convention as Figure \ref{fig:result_1}.}
\end{figure*}

\begin{figure*} 
\centering
\includegraphics[scale=0.66]{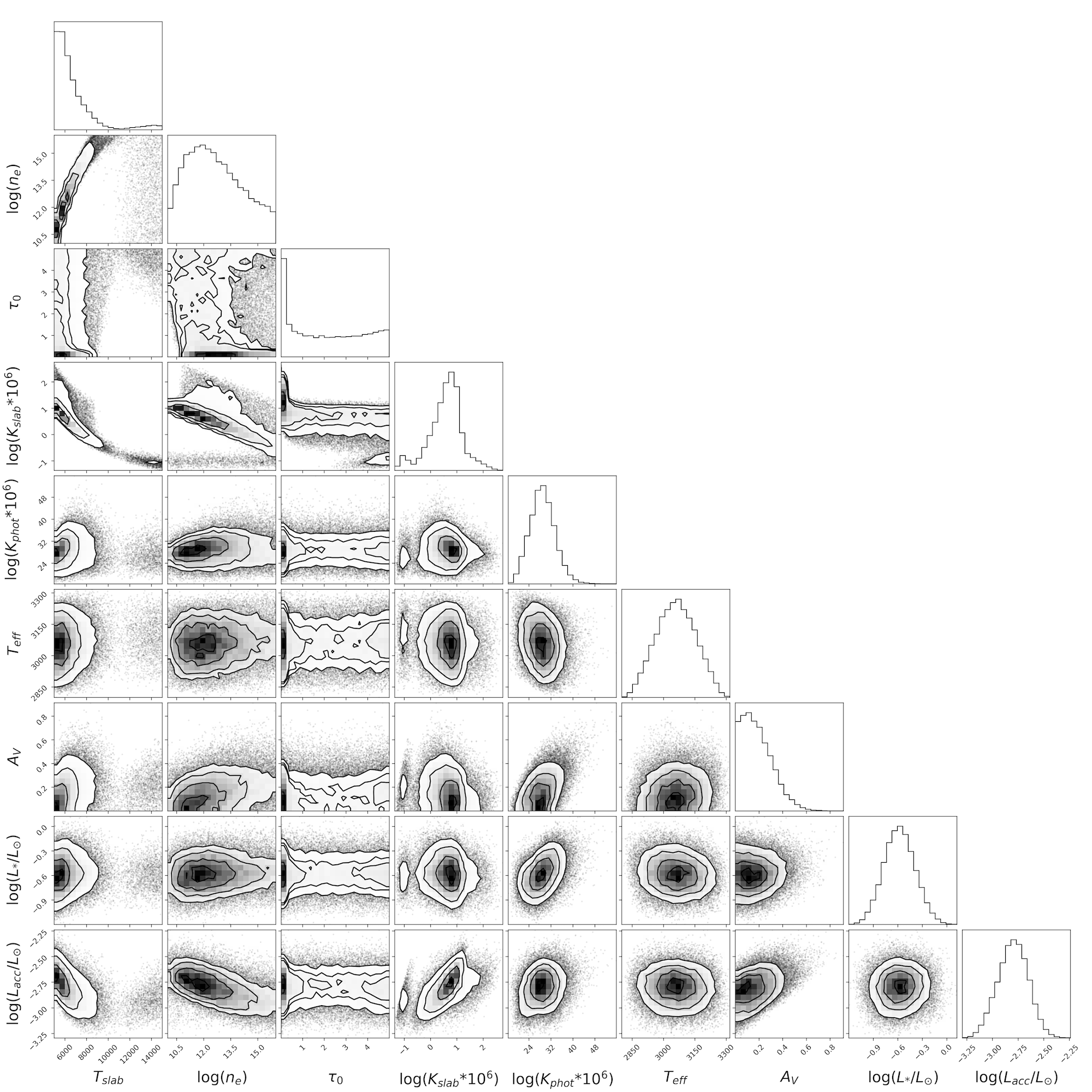}
\caption{The corner plot for Object 9, for model parameters and the log($L_{*}$) and log($L_{acc}$) posteriors.}
\end{figure*}

\begin{figure*}
\centering
\includegraphics[scale=0.25]{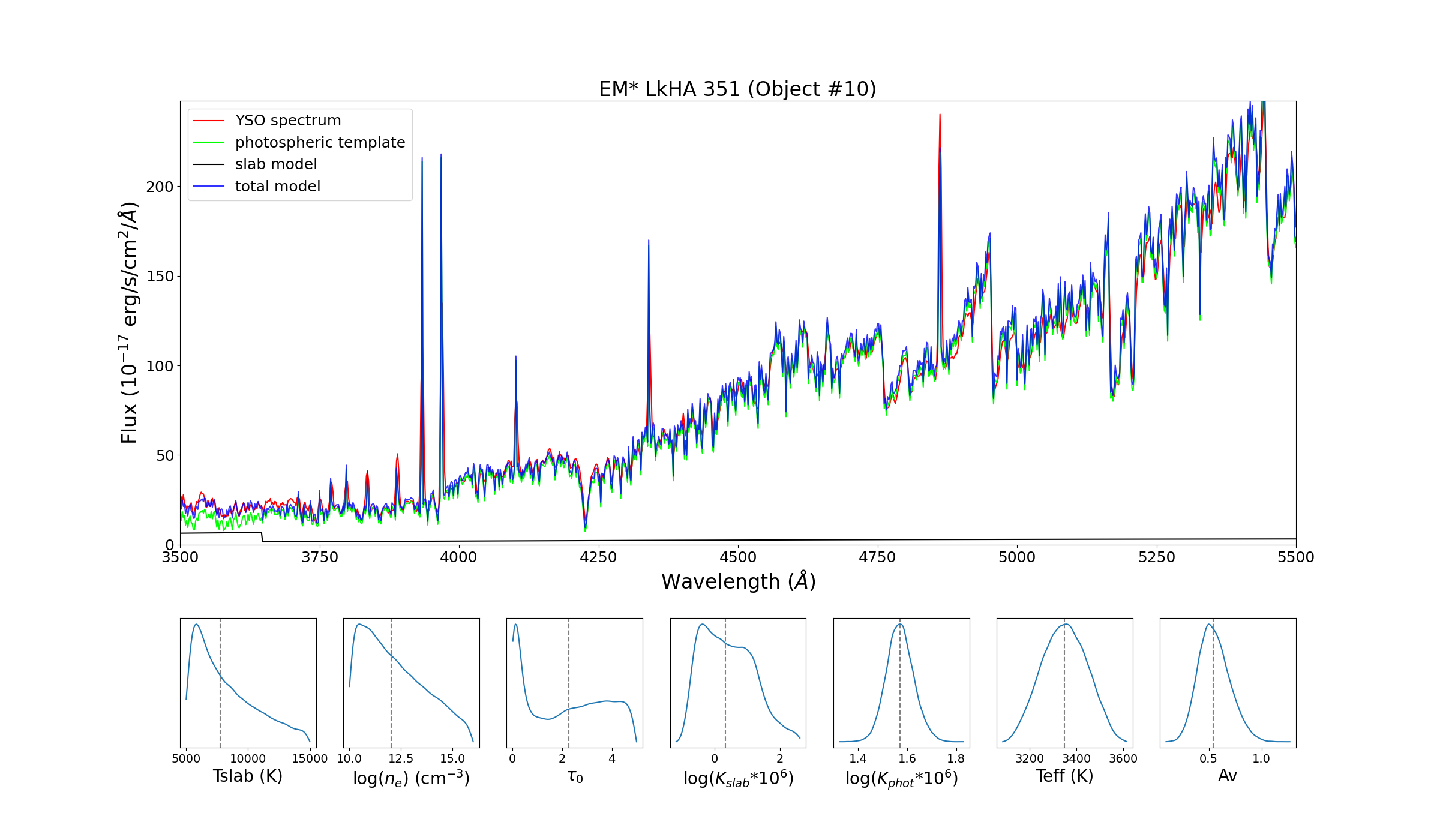}
\caption{The median model fit for Object 10 and the parameter posteriors with the same plotting convention as Figure \ref{fig:result_1}.}
\end{figure*}

\begin{figure*}
\centering
\includegraphics[scale=0.66]{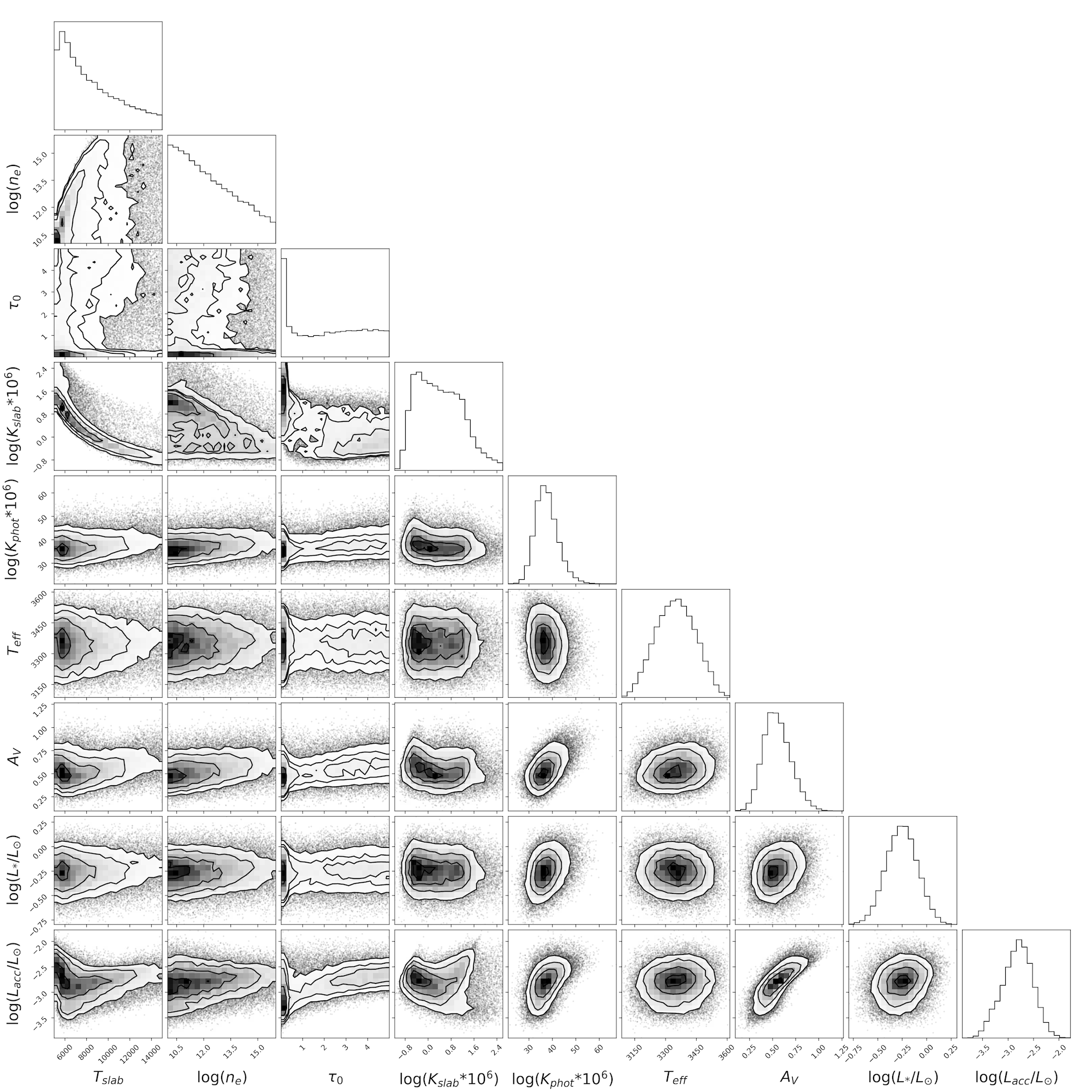}
\caption{The corner plot for Object 10, for model parameters and the log($L_{*}$) and log($L_{acc}$) posteriors.}
\end{figure*}

\begin{figure*} 
\centering
\includegraphics[scale=0.25]{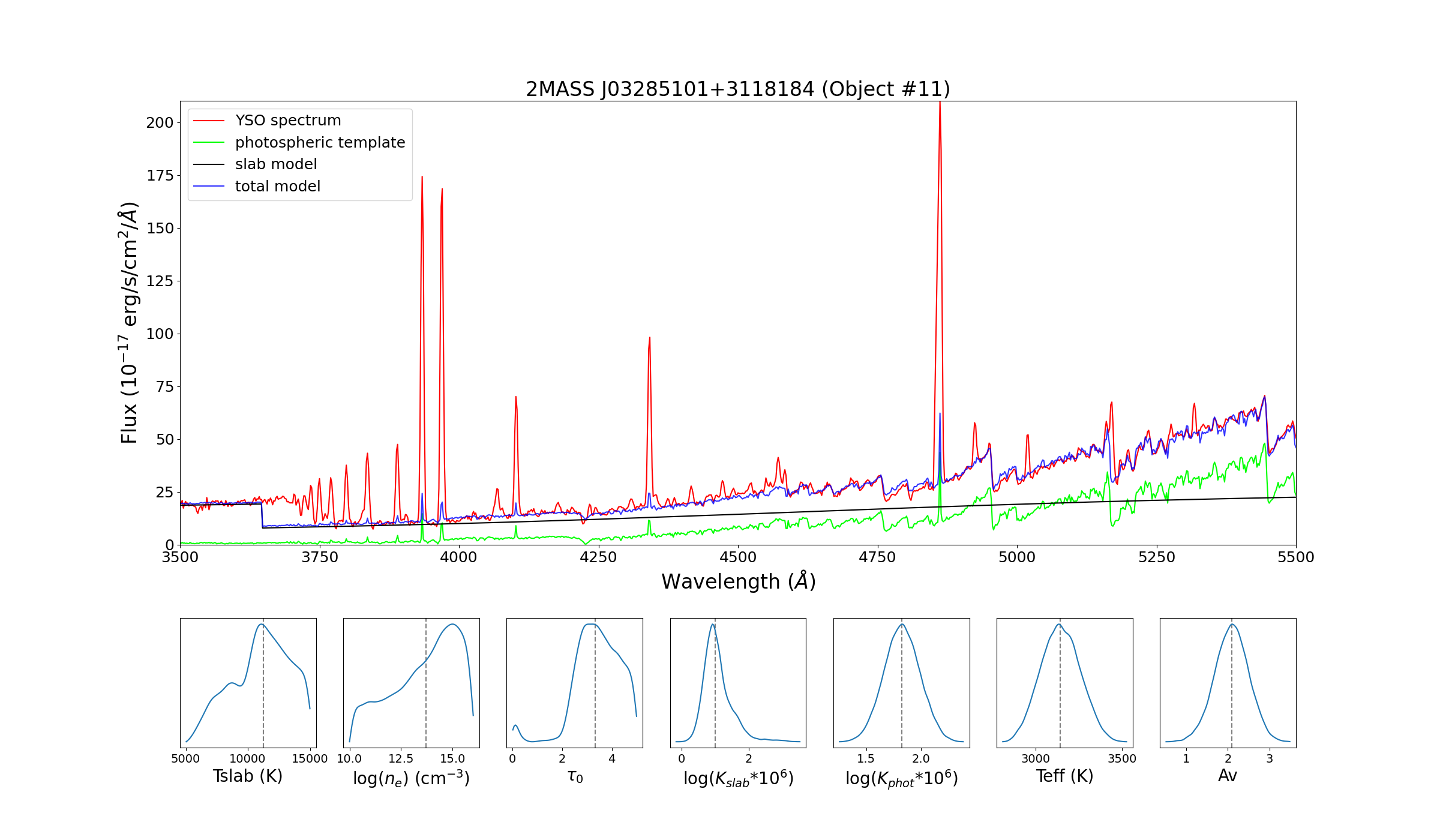}
\caption{The median model fit for Object 11 and the parameter posteriors with the same plotting convention as Figure \ref{fig:result_1}.}
\end{figure*}

\begin{figure*} 
\centering
\includegraphics[scale=0.66]{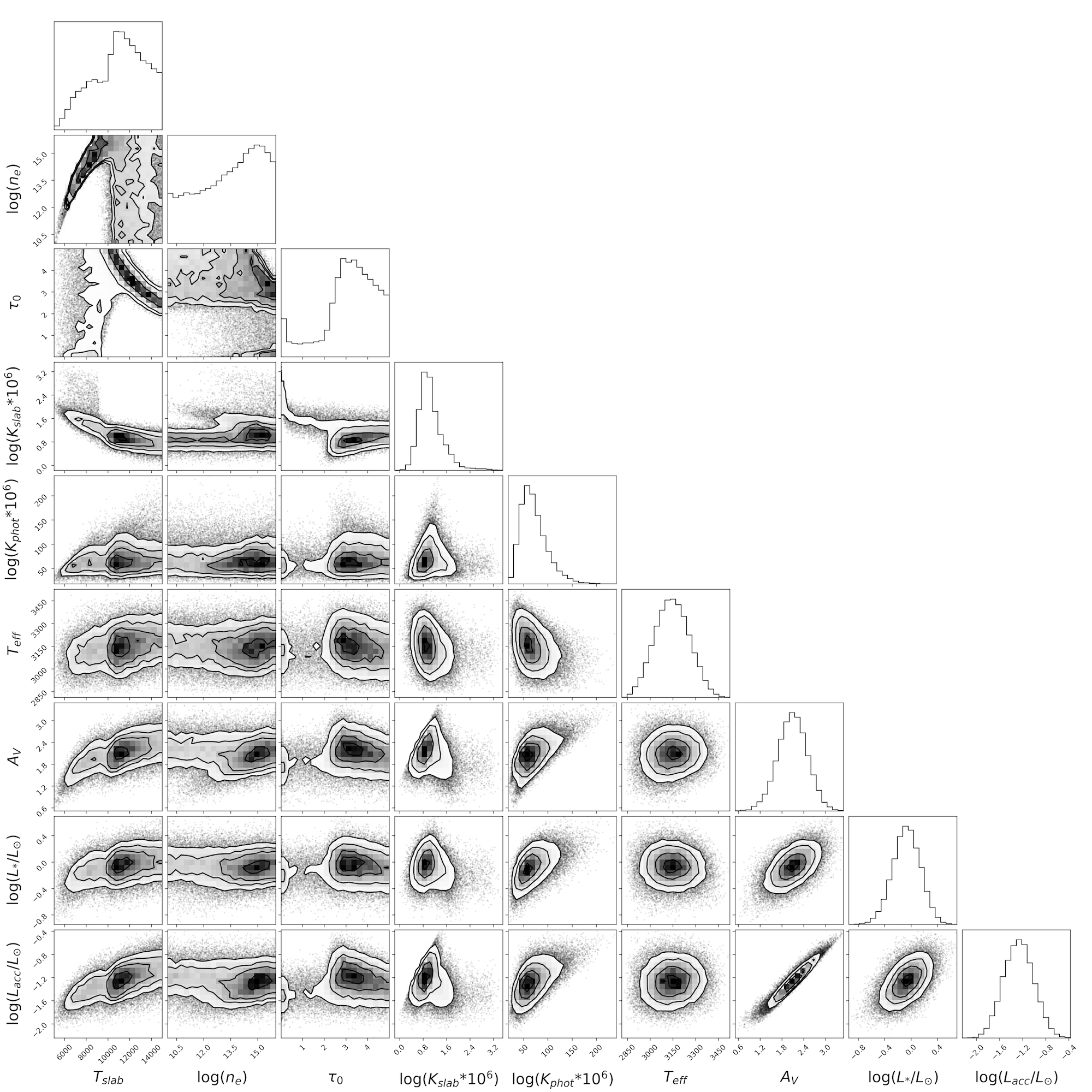}
\caption{The corner plot for Object 11, for model parameters and the log($L_{*}$) and log($L_{acc}$) posteriors.}
\end{figure*}

\begin{figure*} 
\centering
\includegraphics[scale=0.25]{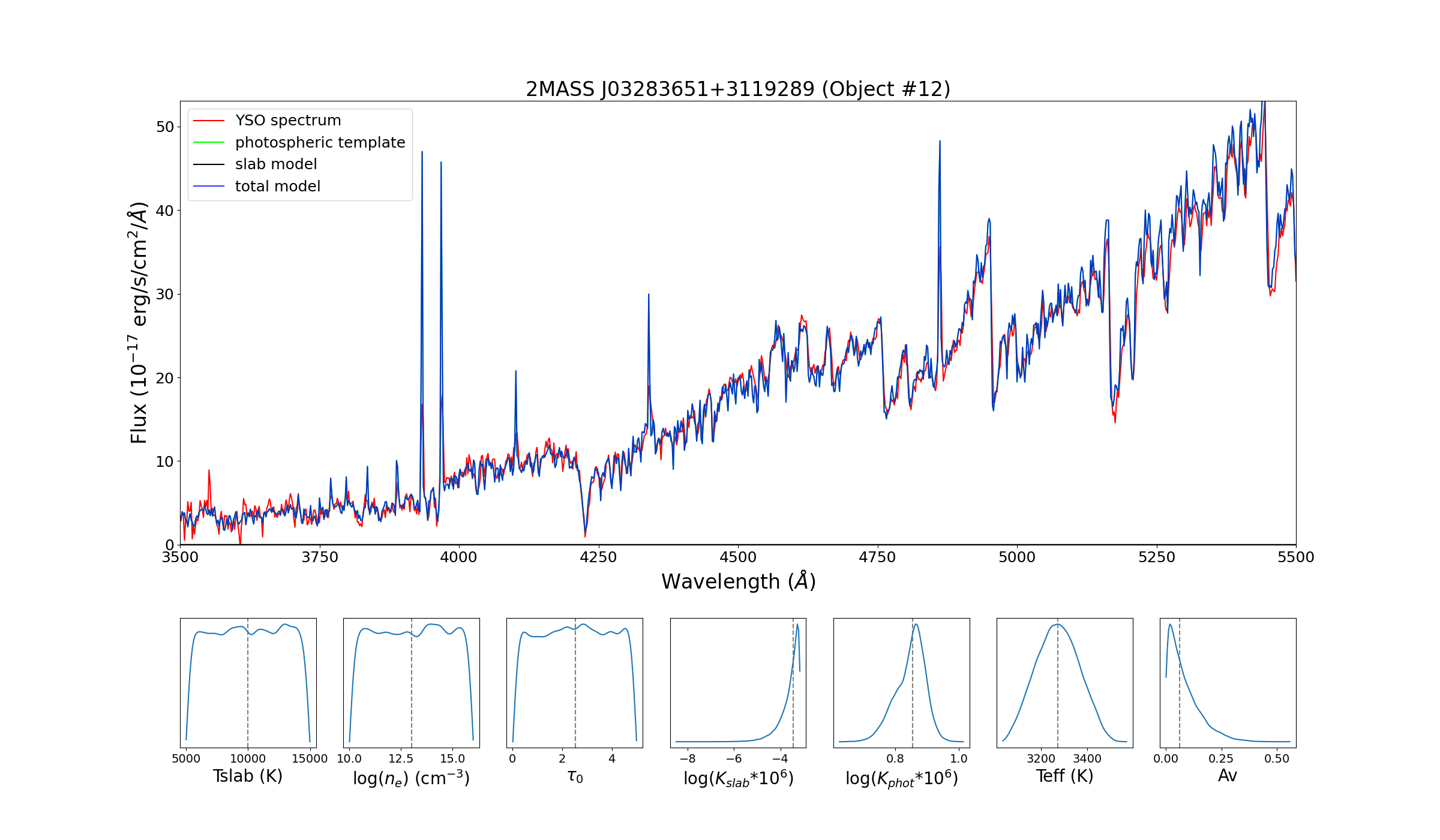}
\caption{The median model fit for Object 12 and the parameter posteriors with the same plotting convention as Figure \ref{fig:result_1}.}
\end{figure*}

\begin{figure*} 
\centering
\includegraphics[scale=0.66]{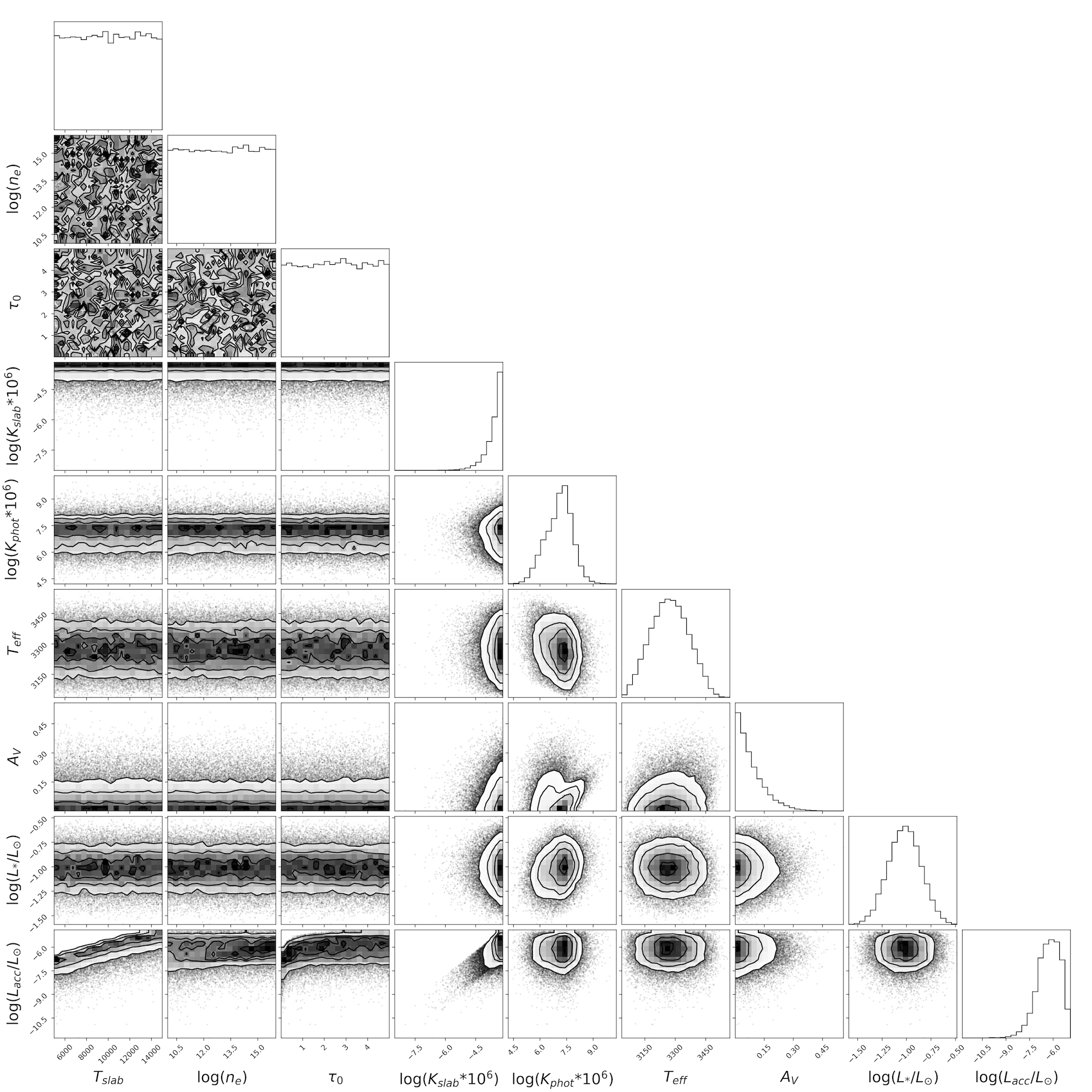}
\caption{The corner plot for Object 12, for model parameters and the log($L_{*}$) and log($L_{acc}$) posteriors.}
\end{figure*}

\begin{figure*} 
\centering
\includegraphics[scale=0.25]{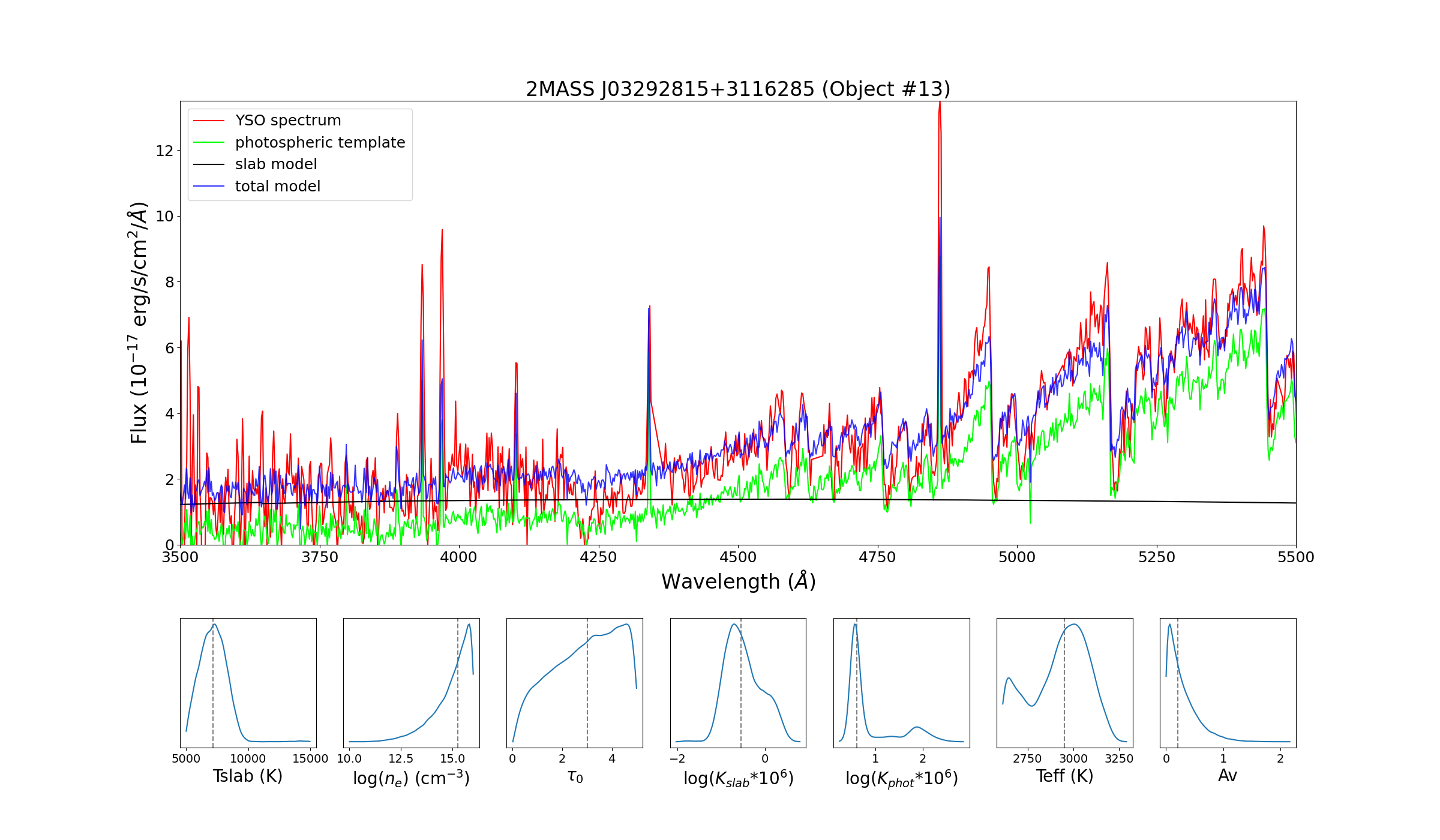}
\caption{The median model fit for Object 13 and the parameter posteriors with the same plotting convention as Figure \ref{fig:result_1}.}
\end{figure*}

\begin{figure*} 
\centering
\includegraphics[scale=0.66]{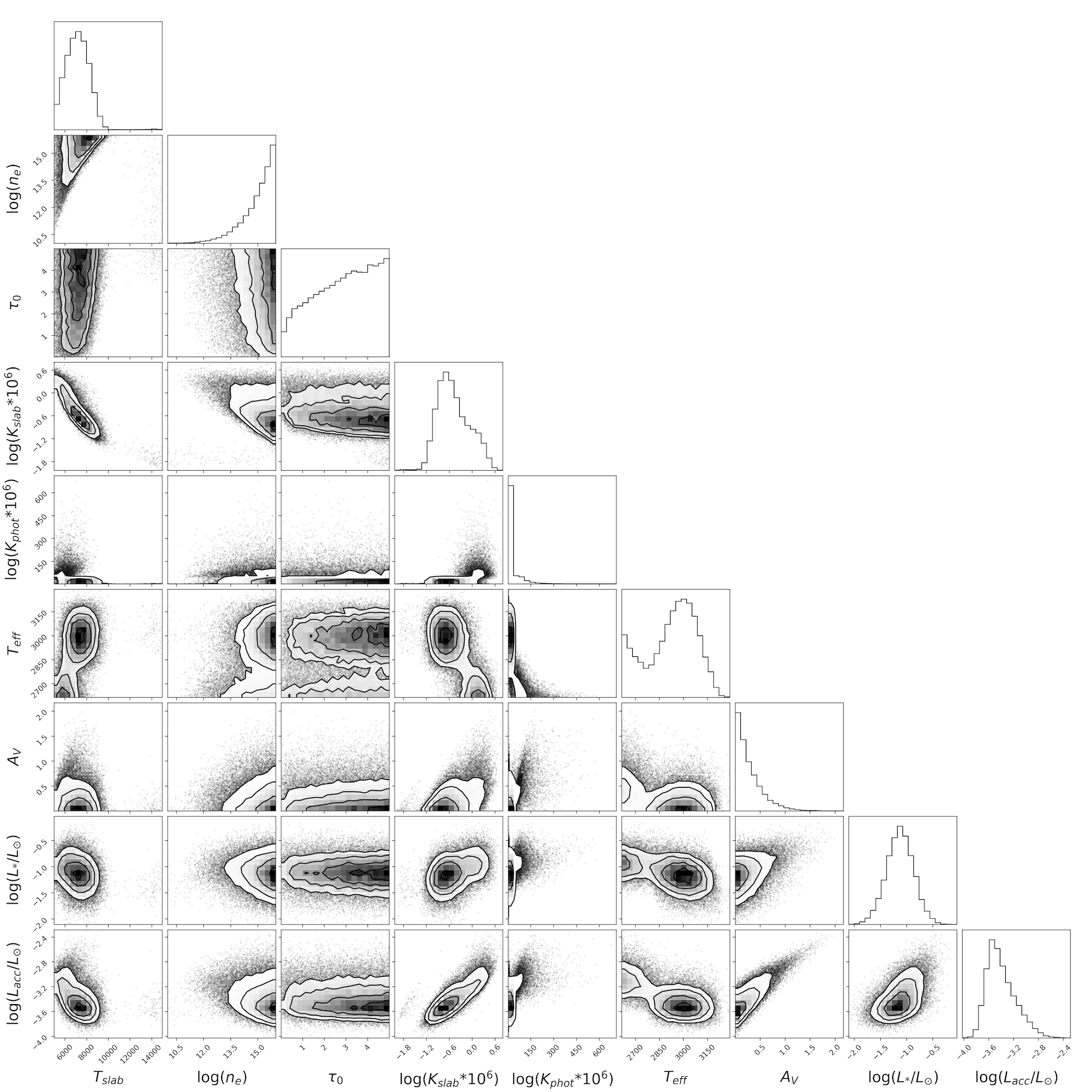}
\caption{The corner plot for Object 13, for model parameters and the log($L_{*}$) and log($L_{acc}$) posteriors.}
\end{figure*}

\begin{figure*} 
\centering
\includegraphics[scale=0.25]{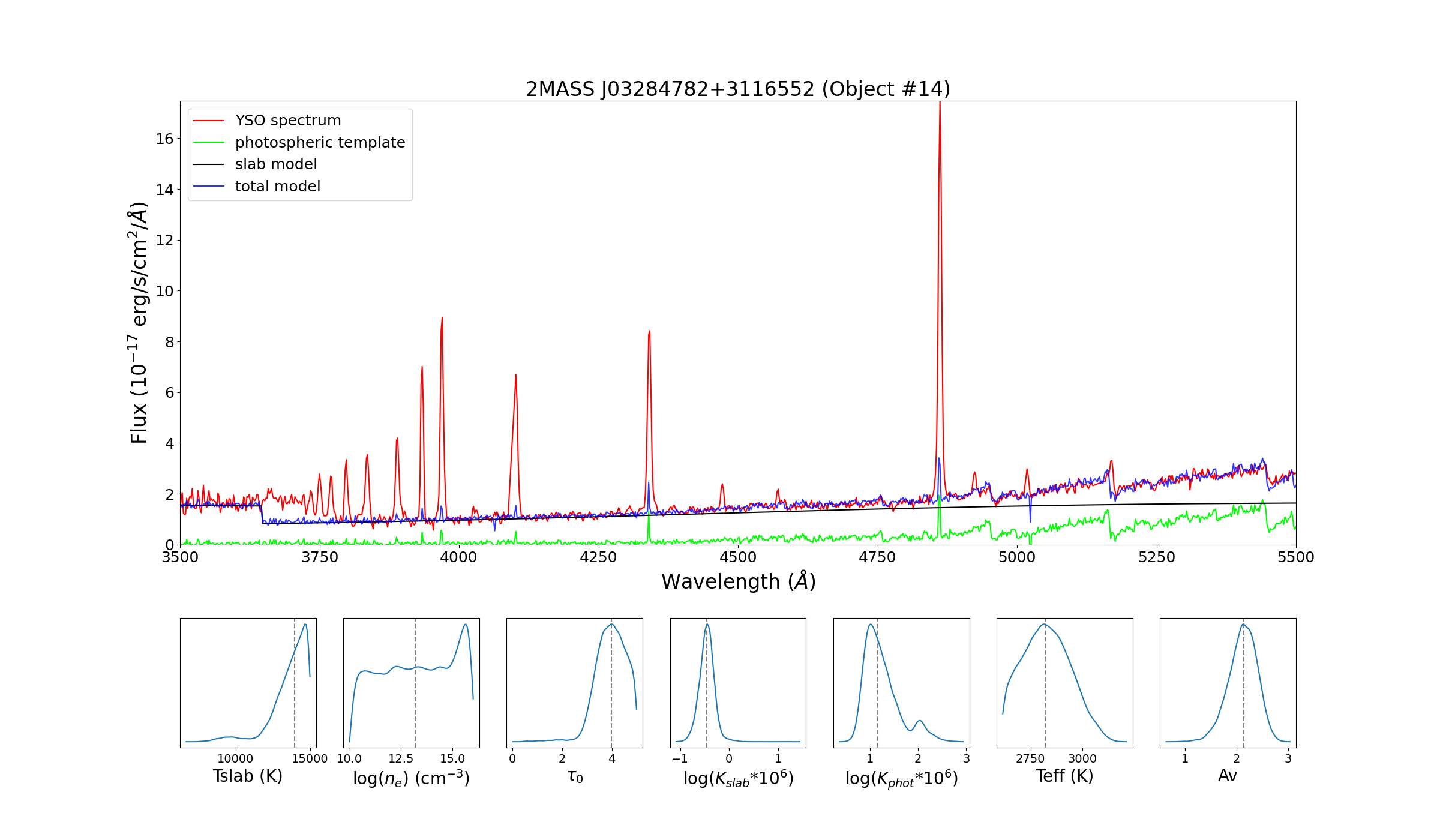}
\caption{The median model fit for Object 14 and the parameter posteriors with the same plotting convention as Figure \ref{fig:result_1}.}
\end{figure*}

\begin{figure*} 
\centering
\includegraphics[scale=0.66]{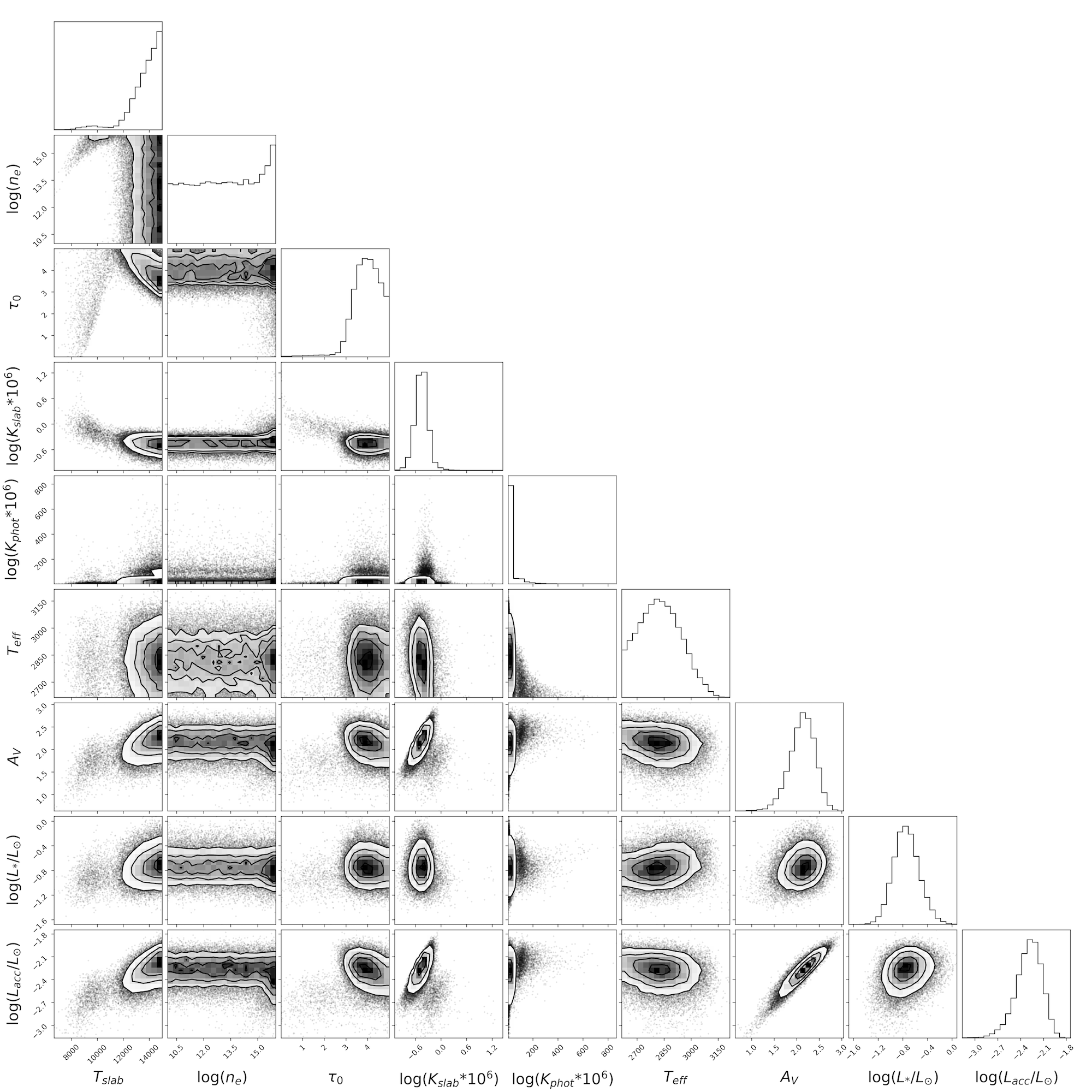}
\caption{The corner plot for Object 14, for model parameters and the log($L_{*}$) and log($L_{acc}$) posteriors.}
\end{figure*}

\begin{figure*} 
\centering
\includegraphics[scale=0.25]{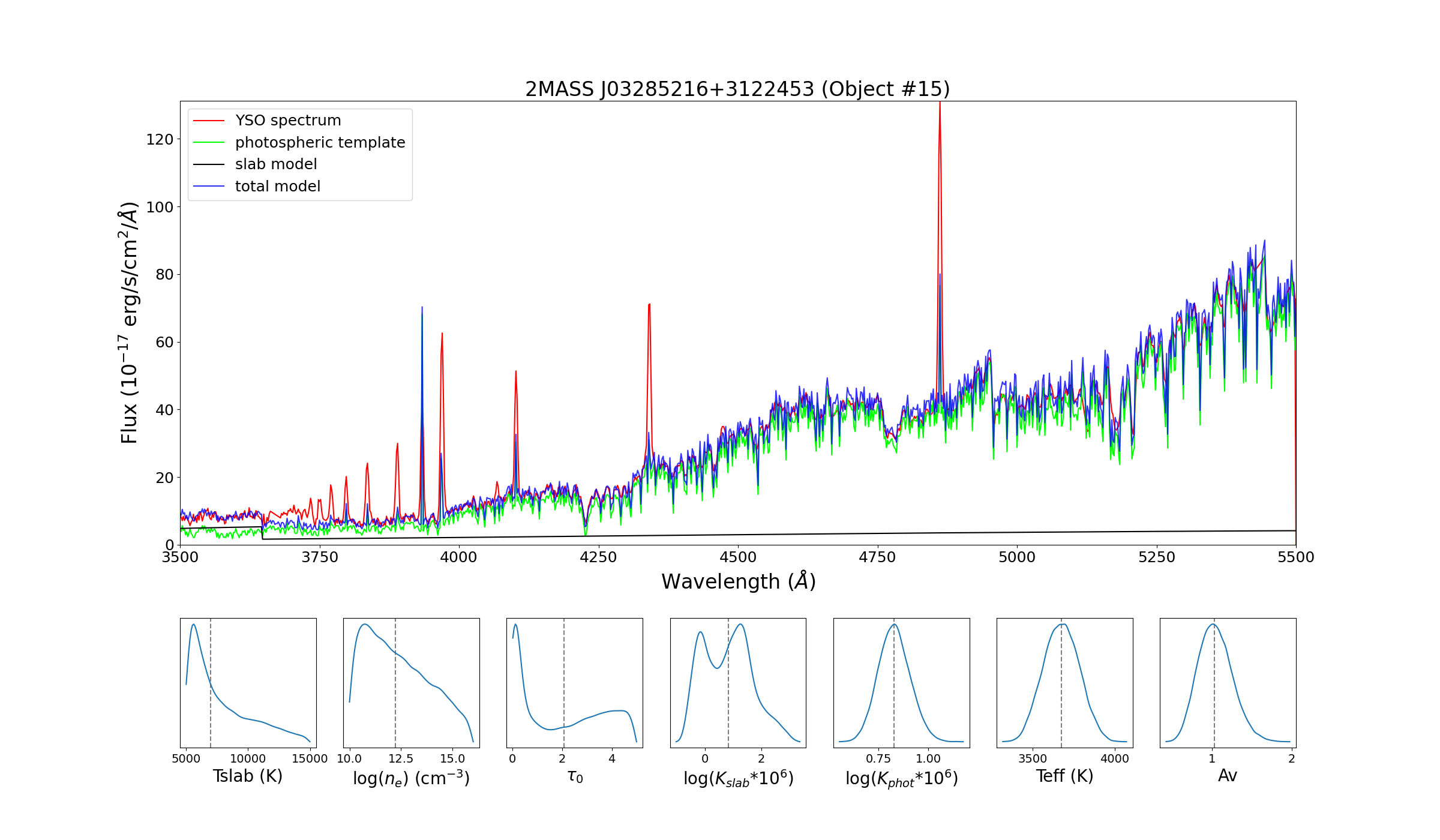}
\caption{The median model fit for Object 15 and the parameter posteriors with the same plotting convention as Figure \ref{fig:result_1}.}
\end{figure*}

\begin{figure*} 
\centering
\includegraphics[scale=0.66]{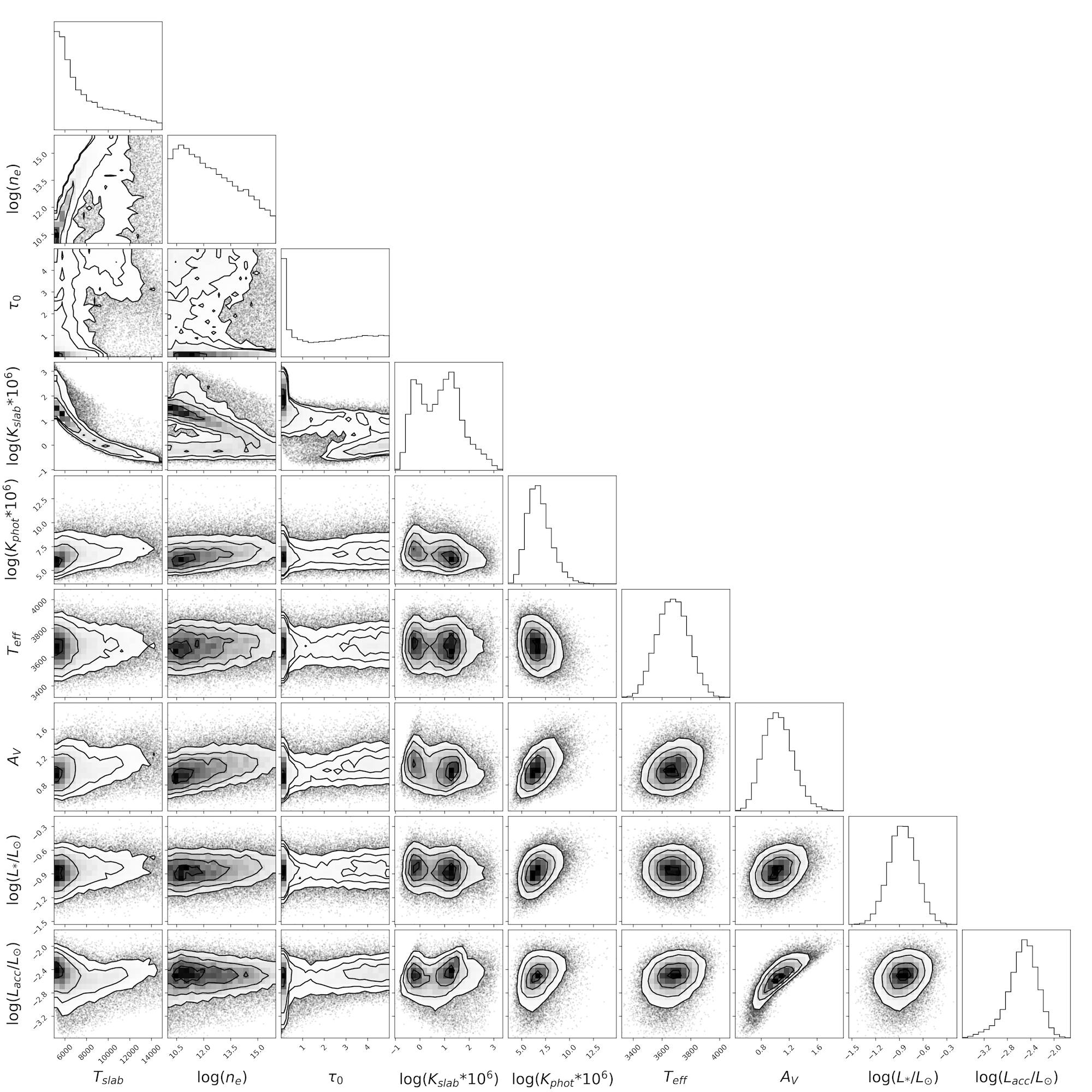}
\caption{The corner plot for Object 15, for model parameters and the log($L_{*}$) and log($L_{acc}$) posteriors.}
\end{figure*}

\end{document}